\documentclass[
 reprint,
 superscriptaddress,
 amsmath,
 amssymb,
 aps,
 nofootinbib
]{revtex4-2}

\usepackage[utf8]{inputenc}
\usepackage[T1]{fontenc}
\usepackage{graphicx}
\usepackage{dcolumn}
\usepackage{bm}
\usepackage{natbib}
\DeclareUnicodeCharacter{2299}{$_\odot$}
\usepackage{marvosym}
\usepackage[normalem]{ulem}

\usepackage[usenames]{color}

\usepackage{hyperref}
\hypersetup{colorlinks,linkcolor={black},citecolor={blue},urlcolor={blue}}

\usepackage{rotating}
\usepackage{booktabs}
\usepackage[detect-all]{siunitx}

\begin{document}

\title{Constraining neutrino-DM interactions with Milky Way dwarf spheroidals and supernova neutrinos}

\author{Sean Heston}
\email{seanh125@vt.edu}
\affiliation{Center for Neutrino Physics, Department of Physics, Virginia Tech, Blacksburg, Virginia 24061, USA.}

\author{Shunsaku Horiuchi}
\affiliation{Center for Neutrino Physics, Department of Physics, Virginia Tech, Blacksburg, Virginia 24061, USA.}
\affiliation{Kavli IPMU (WPI), UTIAS, The University of Tokyo, Kashiwa, Chiba 277-8583, Japan}

\author{Satoshi Shirai}
\affiliation{Kavli IPMU (WPI), UTIAS, The University of Tokyo, Kashiwa, Chiba 277-8583, Japan}

\date{\today}

\begin{abstract}
{We constrain the neutrino-dark matter cross section using properties of the dark matter density profiles of Milky Way dwarf spheroidal galaxies. The constraint arises from core-collapse supernova neutrinos scattering on dark matter as a form of energy injection, allowing the transformation of the dark matter density profile from a cusped profile to a flatter profile. We assume a standard cosmology of dark energy and cold, collisionless, and non-self-interacting dark matter. By requiring that the dark matter cores do not lose too much mass or overshoot constraints from stellar kinematics, we place an upper limit on the cross section of $\sigma_{\nu-\mathrm{DM}}(E_\nu=15 \, \mathrm{MeV}, m_\chi\lesssim130 \, \mathrm{GeV}) \approx 3.4 \times 10^{-23} \, \mathrm{cm^2}$ and $\sigma_{\nu-\mathrm{DM}}(E_\nu=15 \, \mathrm{MeV}, m_\chi\gtrsim130 \, \mathrm{GeV}) \approx 3.2 \times 10^{-27} \left( \frac{m_\chi}{1\,\mathrm{GeV}}\right)^2\, \mathrm{cm^2}$, which is stronger than previous bounds for these energies. Consideration of baryonic feedback or host galaxy effects on the dark matter profile can strengthen this constraint.}
\end{abstract}

\maketitle
\section{Introduction}\label{intro}

Constraining the interaction cross section between neutrinos and dark matter (DM), $\sigma_{\nu-\mathrm{DM}}$, is important to better understand both neutrinos and DM. Such an interaction is intriguing considering the fact that the neutrino properties we observe do not agree with the predictions from the Standard Model (SM), e.g., the origin of neutrino masses \cite{Asaka_2005, Ma_2006, Farzan_2012, deGouvea_2016, Escudero2017a, Escudero2017b} and oscillation experiment anomalies \cite{LSND_anomaly, MiniBooNE_anomaly}. However, given the weak nature of neutrino interactions, the cross section is only weakly tested, mostly from astrophysical neutrinos, e.g., SN1987A \cite{Mangano_2006}, higher energy astrophysical neutrinos observed at IceCube \cite{Choi_2019, Cline_2023, Cline:2023tkp, Ferrer_2023, Fujiwara_2023tidal}, or low energy relic neutrinos \cite{Wilkinson_2014, Akita2023constraints}. Other searches have looked for the effects such an interaction will have on Milky Way satellites \cite{Escudero_2018}, cosmology \cite{Mangano_2006, Boehm_2013, Mosbech_2021, Hooper_2022, Mosbech_2023}, astrophysics \cite{Fayet_2006, Koren_2019, Murase_2019, McMullen_2021, Carpio_2023}, boosted DM detection \cite{Farzan_2014, Das_2021, Chao_2021, Jho_2021, Zhang_2021, Ghosh_2022, Bardhan_2023, DeRomeri_2023, Lin_2023}, etc. 

If $\nu-$DM scattering can take place, it can also have effects on the small-scale structure of the Universe. For example, dark energy and cold dark matter ($\Lambda$CDM) by itself cannot reproduce the observed small-scale structures purely thought gravitational interactions (for a review see, e.g., Refs.~\cite{Bullock_2017, Salucci:2018hqu}). One such problem is that the DM only simulations predict subhalos inhabited by dwarf galaxies to have cuspy DM density profiles \cite{NFW, Wang_2020}, while observations suggest the presence of near constant density cores of DM \cite{Flores_1994, Moore_1994, Spergel_2000, Oh_2008, Donato_2009, Walker_2011, Salucci_2012, Oh_2015}. Many solutions to this discrepancy exist involving some form of feedback in order to cause a redistribution of the DM mass profile, e.g., Refs.~\cite{Navarro_1996, Penarrubia_2012, Pontzen_2012}. The question of how effective baryonic feedback should be is still an ongoing effort. The role of neutrinos, specifically those from core-collapse supernovae (CCSNe), as a source of feedback in dwarf galaxy-sized subhalos, is not well studied.

In this paper, we explore supernova neutrinos as a source of energy for feedback, in particular in light of beyond the standard model physics which allow larger interactions between neutrinos and DM. This is potentially a powerful probe since each core-collapse supernova (CCSN) emits $\mathcal{O}(10^{53})$ erg of energy in neutrinos \cite{Arnett_1966, Colgate&White_1966, Wilson_1982, Bethe&Wilson_1985, Bethe_1990, burrows1993theory, Kotake_2006, janka2012b, Mirizzi_2016}, so the total energy budget available for injection over the history of the dwarf galaxy is potentially very large. Since CCSN neutrinos have $\sim 10$ MeV energies, there are $\mathcal{O}(10^{58})$ neutrinos from each CCSN. Therefore, even small interaction cross sections may still have noticeable impacts on DM structures. 
This is because the scattering will inject some amount of energy into the DM, and this can change, e.g., an initially cuspy central density profile into a centrally cored profile. Since observations of stellar kinematics constrain the maximum size of such DM density cores, the amount of energy that can be injected into the subhalo is constrained which translates into a constraint on $\sigma_{\nu-\mathrm{DM}}$. 

In this work, we derive an upper bound on the cross section between neutrinos and DM. We use the fact that measured core sizes have estimated upper limits to constrain the maximum size of the interaction cross section. As we know that CCSNe have occurred in these dwarf spheroidals, there is an associated neutrino emission with typical energies of $3\times10^{53}$ erg released in neutrinos per supernova with an average neutrino energy of 15 MeV. 

In order to not overshoot estimated upper limits of DM core sizes or lose too much mass such that the cores cannot exist as we observe them today, we find that $\sigma_{\nu\mathrm{-DM}}(E_\nu=15 \, \mathrm{MeV}, m_\chi\lesssim130 \, \mathrm{GeV}) \lesssim 3.4 \times 10^{-23} \, \mathrm{cm^2}$ and $\sigma_{\nu\mathrm{-DM}}(E_\nu=15 \, \mathrm{MeV}, m_\chi\gtrsim130 \, \mathrm{GeV}) \lesssim 3.2 \times 10^{-27} \left( \frac{m_\chi}{1\,\mathrm{GeV}}\right)^2\, \mathrm{cm^2}$. This is slightly stronger than those at similar energies from SN1987A \cite{Mangano_2006}. Considering other forms of energy injection or feedback occurring would strengthen our constraints. 

\section{Dwarf Galaxy subhalo Properties}\label{Dwarf properties}

In this section, we cover the DM subhalo properties of the dwarf spheroidals (dSphs) studied in this work. These properties are estimated from stellar kinematic data of the stars in the dwarfs. We also cover how we model the amount of energy needed to transform a cusped profile into a cored profile, for which we follow the methods of Ref.~\cite{Penarrubia_2012}. 

\subsection{NFW parameters}\label{NFW parameters}

To be able to constrain the $\nu-$DM interaction, we assume that $\Lambda$CDM cosmology is correct and that the subhalos of DM that dSphs occupy initially have a Navarro-Frenk-White (NFW) profile \cite{NFW}. We first find the properties of the DM subhalo that each dwarf spheroidal occupies. 
 
We estimate the density distribution of the DM from the motions of stars in dwarf galaxies, based on the Jeans equation. This analysis closely mirrors previous work that investigated the deformation of the DM density distribution due to DM self-interactions; see Ref.~\cite{Hayashi:2020syu}. 

We model the subhalo DM density distribution for core radius $r_c$ with a modification of the NFW profile,
\begin{equation}\label{Modified NFW profile}
  \rho(r) = \frac{\rho_0 \, r_\mathrm s^3}{(r_\mathrm c+r)(r+r_\mathrm s)^2}, 
\end{equation}
where $\rho_0$ is the characteristic halo density and $r_\mathrm s$ is the scale radius.

We adopt the data on half-light radii, characterized using the Plummer profile, from Ref.~\cite{2018ApJ...860...66M}. 
Additionally, we utilize the stellar kinematics data of member stars in each dwarf galaxy, as reported in various sources \cite{Simon:2010ek,Kirby:2013isa,Koposov:2011zi,Simon:2007dq,DES:2015tfc,2016MNRAS.458L..59M,2017ApJ...838...83K,Kirby:2015ija,Walker:2015twz,DES:2019ncb,Koposov:2015jla,DES:2016fyd,DES:2016fyd,2015MNRAS.448.2717W,Walker:2008ji,Walker:2008ji,Walker:2008ax,2016ApJ...830..126F,2018AJ....156..257S,Mateo:2007xh,2017ApJ...836..202S}.
Furthermore, we apply the concentration-mass relation to the NFW model parameters, as described by Ref.~\cite{Moline:2016pbm}.

\begin{figure*}[t]
  \centering
  \includegraphics[width=2\columnwidth]{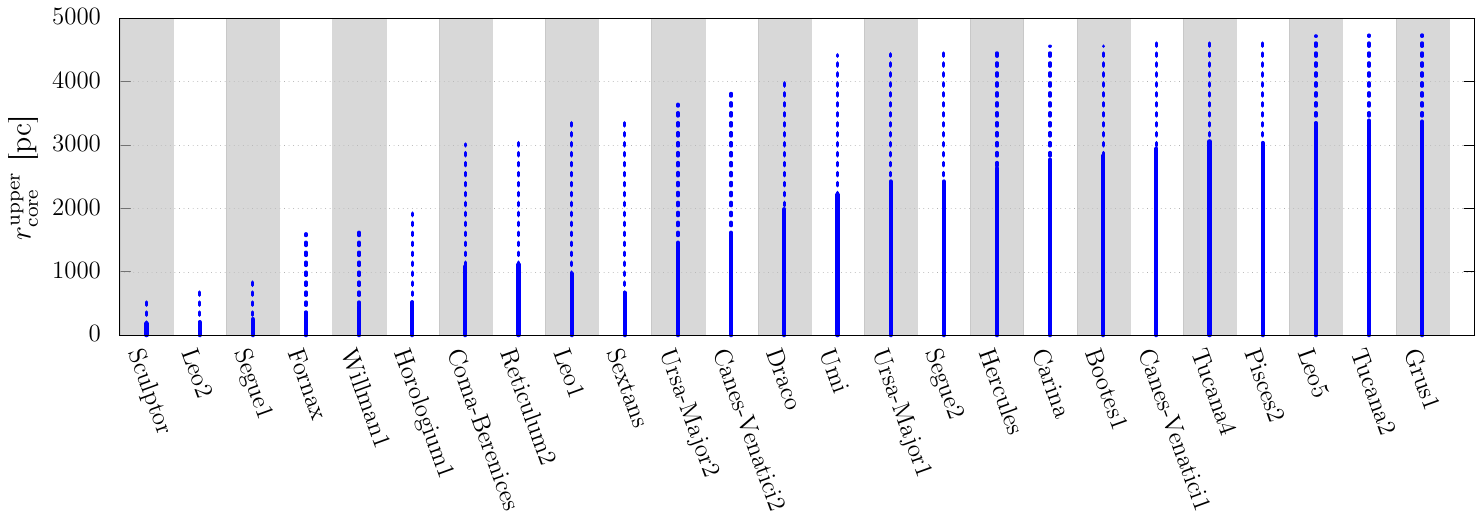}
  \caption{Upper limit of dwarf galaxy DM subhalo core size. Solid (dashed) lines lines correspond to to $1\sigma$ ($2\sigma$) bounds.}
  \label{Dwarf Core Size}
\end{figure*}

These allow us to estimate $r_\mathrm{c}$ for each dwarf galaxy we consider in this study. The upper limit for each is shown in Fig.~\ref{Dwarf Core Size}, where the solid (dashed) lines correspond to the $1\sigma$ ($2\sigma$) upper limit. These are the values of $r_\mathrm{c}$ that let us place an upper limit on the energy injection from $\nu-\mathrm{DM}$ interactions that is allowed. The NFW parameters assuming $r_\mathrm{c}=0$ are also estimated using Markov chain Monte Carlo techniques, the results of which are shown in Table \ref{Dwarf Galaxy Properties}.

\subsection{Virial radius and mass}\label{Virial radius and mass}

We assume the DM profile is virialized and in equilibrium before any CCSNe occur. Then, the neutrinos emitted from CCSNe within the dSphs interact with the DM subhalo, injecting energy via $\nu-$DM scatterings and causing the profile to become cored. In this process, we assume that DM is not lost such that the subhalo remains at a constant mass, and that the profiles revirialize and are equilibrated again into what we currently observe. 

We define the virial radius ($r_{200}$) and the virial mass ($M_{200}$) adopting $\Delta_\mathrm{crit}=200$, i.e., that the subhalo extends until the density of the DM is 200 times greater than the critical density of the Universe, 
\begin{equation}\label{Virial radius definition}
\rho(<r_{200})=\Delta_\mathrm{crit}\,\rho_\mathrm{crit}=200 \, \frac{3H(t)}{8\pi G_\mathrm{N}}, 
\end{equation}
where $H(t)$ is the Hubble constant and $G_\mathrm{N}$ is Newton's gravitational constant. 

Therefore, we can simply find the virial radius using the subhalo parameters found in Sec.~\ref{NFW parameters}. Using Eq.~(\ref{Modified NFW profile}) with $r_\mathrm{c}=0$ and the NFW parameters, we find at what radius the density satisfies the condition of Eq.~(\ref{Virial radius definition}). As we keep the virial mass constant and only the inner regions of the DM profile change with the energy injection, this should stay constant as the core forms. The specific values of $r_{200}$ for each dwarf galaxy are found in column 5 of Table \ref{Dwarf Galaxy Properties}. $M_{200}$ is then found easily as well. We take the volume integral of the mass density profile out to $r_{200}$. These masses are given in column 6 of Table \ref{Dwarf Galaxy Properties}.

\renewcommand{\arraystretch}{1.25}

\begin{table*}[t]
\caption{Dwarf galaxy DM properties from stellar kinematic observations. Included is a shorthand name for the dwarf, the scale density ($\rho_s$), the scale radius ($r_s$), the core radius and bounds ($r_c \pm 1\sigma)$, the virial radius ($r_{200}$), and the virial mass ($M_{200}$). Derivations of the DM subhalo parameters are covered in Sec. \ref{Dwarf properties}. The virial mass and radius are calculated from the mass profile assuming $\Delta_\mathrm{crit}=200$.}
\label{Dwarf Galaxy Properties}
\begin{ruledtabular}
  \begin{tabular}{ccccccc}
  Name & \multicolumn{1}{c}{$\rho_\mathrm{s}$ [$\mathrm{M}_\odot$/pc]\footnote{assuming a pure NFW profile}} & \multicolumn{1}{c}{$r_\mathrm{s} \; [\mathrm{pc}]^\mathrm{a}$} & \multicolumn{1}{c}{$(r_\mathrm{c}\pm1\sigma)\times10^2$ [pc]} & $r_{200} \; [\mathrm{pc}]^\mathrm{a}$ & $M_{200}$ [$\mathrm{M}_\odot$] & Refs. \\ \hline
  Seg1 &  $2.64 \times 10^{-2}$ &  $3.19 \times 10^{3}$ &  $1.24_{-1.00}^{+1.30}$ &  $2.93 \times 10^{4}$ &  $1.53 \times 10^{10}$ & \cite{Grcevich_2009, Simon_2011}\\
  Seg2 &  $1.14 \times 10^{-1}$ &  $1.56 \times 10^{2}$ &  $13.8_{-9.74}^{+10.5}$ &  $2.39 \times 10^{3}$ & $1.00 \times 10^{7}$ & \cite{Belokurov_2009}\\
  Boo1 &  $4.64 \times 10^{-2}$ &  $4.83 \times 10^{2}$ &  $19.6_{-12.2}^{+8.70}$ &  $5.40 \times 10^{3}$ & $1.04 \times 10^{8}$ & \cite{Grcevich_2009, Koposov_2011}\\
  Her  &  $3.87 \times 10^{-2}$ &  $5.63 \times 10^{2}$ &  $17.4_{-12.9}^{+8.92}$ &  $5.92 \times 10^{3}$ & $1.33 \times 10^{8}$ & \cite{Aden_2009a, Grcevich_2009}\\
  Com  &  $4.08 \times 10^{-2}$ &  $8.23 \times 10^{2}$ &  $6.56_{-4.38}^{+4.29}$ &  $8.80 \times 10^{3}$ & $4.41 \times 10^{8}$ & \cite{Simon_2007, Grcevich_2009}\\
  CVn1 &  $2.31 \times 10^{-2}$ &  $9.84 \times 10^{2}$ &  $18.6_{-16.3}^{+10.8}$ &  $8.60 \times 10^{3}$ & $3.81 \times 10^{8}$ & \cite{Simon_2007, Grcevich_2009}\\
  CVn2 &  $1.64 \times 10^{-2}$ &  $1.58 \times 10^{3}$ &  $10.0_{-6.41}^{+5.84}$ &  $1.22 \times 10^{4}$ & $1.05 \times 10^{9}$ & \cite{Simon_2007, Grcevich_2009}\\
  Leo5 &  $2.87 \times 10^{-2}$ &  $5.63 \times 10^{2}$ &  $23.6_{-15.8}^{+9.60}$ &  $5.33 \times 10^{3}$ & $9.32 \times 10^{7}$ & \cite{Grcevich_2009, Walker_2009a, Belokurov_2008}\\
  UMa1 &  $3.39 \times 10^{-2}$ &  $9.32 \times 10^{2}$ &  $15.5_{-12.8}^{+8.77}$ &  $9.34 \times 10^{3}$ & $5.14 \times 10^{8}$ & \cite{Simon_2007, Grcevich_2009}\\
  UMa2 &  $3.82 \times 10^{-2}$ &  $1.16 \times 10^{3}$ &  $8.45_{-5.80}^{+5.85}$ &  $1.21 \times 10^{4}$ & $1.14 \times 10^{9}$ & \cite{Simon_2007, Grcevich_2009}\\
  Ret2 &  $7.34 \times 10^{-2}$ &  $3.69 \times 10^{2}$ &  $7.02_{-4.78}^{+4.15}$ &  $4.85 \times 10^{3}$ & $7.95 \times 10^{7}$ & \cite{Koposov_2015a, Koposov_2015b, Muñoz_2018}\\
  Psc2 &  $4.46 \times 10^{-2}$ &  $3.54 \times 10^{2}$ &  $20.0_{-14.8}^{+10.4}$ &  $3.91 \times 10^{3}$ & $3.91 \times 10^{7}$ & \cite{McConnachie_2012}\\
  Gru1 &  $5.57 \times 10^{-2}$ &  $2.56 \times 10^{2}$ &  $23.6_{-15.5}^{+10.1}$ &  $3.06 \times 10^{3}$ & $1.93 \times 10^{7}$ & \cite{Muñoz_2018, Chiti_2022a}\\
  Hor1 &  $2.40 \times 10^{-2}$ &  $2.12 \times 10^{3}$ &  $2.64_{-2.06}^{+2.58}$ &  $1.88 \times 10^{4}$ & $3.97 \times 10^{9}$ & \cite{Muñoz_2018, Koposov_2015b}\\
  Tuc2 &  $7.29 \times 10^{-2}$ &  $2.77 \times 10^{2}$ &  $24.4_{-15.7}^{+9.44}$ &  $3.64 \times 10^{3}$ & $3.35 \times 10^{7}$ & \cite{Koposov_2015a, Chiti_2023}\\
  Tuc4 &  $1.05 \times 10^{-1}$ &  $1.97 \times 10^{2}$ &  $20.5_{-15.6}^{+10.1}$ &  $2.94 \times 10^{3}$ & $1.84 \times 10^{7}$ & \cite{Drlica-Wagner_2015, Simon_2020}\\
  Wil1 &  $3.50 \times 10^{-2}$ &  $1.42 \times 10^{3}$ &  $2.47_{-2.05}^{+2.05}$ &  $1.44 \times 10^{4}$ & $1.88 \times 10^{9}$ & \cite{Grcevich_2009, Martin_2007}\\
  Car  &  $8.76 \times 10^{-3}$ &  $2.76 \times 10^{3}$ &  $23.2_{-14.0}^{+4.50}$ &  $1.70 \times 10^{4}$ & $2.56 \times 10^{9}$ & \cite{Grcevich_2009, Walker_2008, Walker_2009b}\\
  Dra  &  $5.02 \times 10^{-3}$ &  $6.73 \times 10^{3}$ &  $12.5_{-5.30}^{+7.43}$ &  $3.37 \times 10^{4}$ &  $1.84 \times 10^{10}$ & \cite{Grcevich_2009, Wilkinson_2004, Walker_2007}\\
  For  &  $2.61 \times 10^{-2}$ &  $1.29 \times 10^{3}$ &  $2.97_{-2.10}^{+0.53}$ &  $1.18 \times 10^{4}$ & $9.94 \times 10^{8}$ & \cite{Grcevich_2009, Walker_2009b, Walker_2008, Bouchard_2006}\\
  Leo1 &  $1.50 \times 10^{-2}$ &  $1.96 \times 10^{3}$ &  $1.79_{-1.33}^{+7.93}$ &  $1.46 \times 10^{4}$ & $1.77 \times 10^{9}$ & \cite{Grcevich_2009, Mateo_2008}\\
  Leo2 &  $2.28 \times 10^{-2}$ &  $1.50 \times 10^{3}$ &  $1.05_{-0.75}^{+1.02}$ &  $1.30 \times 10^{4}$ & $1.32 \times 10^{9}$ & \cite{Grcevich_2009, Walker_2007}\\
  Scl  &  $2.42 \times 10^{-2}$ &  $1.42 \times 10^{3}$ &  $1.82_{-1.30}^{+0.04}$ &  $1.26 \times 10^{4}$ & $1.21 \times 10^{9}$ & \cite{Grcevich_2009, Walker_2008, Walker_2009b, Carignan_1998}\\
  Sex  &  $3.40 \times 10^{-2}$ &  $8.06 \times 10^{2}$ &  $2.51_{-1.64}^{+4.19}$ &  $8.09 \times 10^{3}$ & $3.34 \times 10^{8}$ & \cite{Grcevich_2009, Walker_2008, Walker_2009b}\\
  UMi  &  $3.12 \times 10^{-2}$ &  $1.09 \times 10^{3}$ &  $6.45_{-5.12}^{+15.7}$ &  $1.06 \times 10^{4}$ & $7.37 \times 10^{8}$ & \cite{Grcevich_2009, Wilkinson_2004, Walker_2009c}\\
  \end{tabular}
\end{ruledtabular}
\end{table*}

We assume that the only form of feedback that acts upon the DM subhalos is from neutrinos. This should make our upper bound a conservative limit, as we ignore other sources of feedback, e.g., gas blowout or tidal effects from the Milky Way DM halo, and other possible sources that can change the shape of the DM profile \cite{Navarro_1996, El-Zant_2001, Mayer_2001, Read&Gilmore_2005, Mashchenko_2008, Goerdt_2010, Governato_2010, Kazantzidis:2010cw, Cole_2011, Brooks:2012vi, Chang:2012rx, Pontzen_2012, DiCintio_2013, Garrison-Kimmel:2013yys, Kazantzidis:2013wi, Teyssier_2013, Nipoti:2014xha, Chan_2015, Read:2015sta, Dutton_2016, Tollet_2016, Wang:2016qol, Fitts_2017, Hiroshima:2018kfv, Freundlich_2019, Lazar_2020, Burger_2021}.

\section{Deriving the upper limit on the DM-neutrino cross section}\label{Upper limit constraining}

In this section, we first define the CCSN neutrino energy spectrum. We then use the dwarf galaxy properties to estimate how many CCSNe have gone off in each dwarf galaxy. Next, we get an upper limit on the amount of energy that can be injected into the DM subhalo by neutrino interactions using the estimated upper limit of the core radius found from stellar kinematics. We first place a bound on the interaction cross section for light DM that is accelerated above the escape velocity of the host subhalo. Next, we take the energy constraint and transform it into a constraint on the interaction cross section under certain assumptions of the interaction. These assumptions are that the interactions take place within the DM core region and that each interaction has maximal energy transfer.

\subsection{Neutrino emission}

In our phenomenological derivation of an upper limit on $\sigma_{\nu-\mathrm{DM}}$, we consider flavor-independent interactions. For the sake of simplicity, we therefore use a time-integrated, flavor-independent energy spectrum for CCSN neutrino emission from Ref.~\cite{PinchedSpectrum}, which when normalized is given by 
\begin{equation}\label{NeutrinoSpec}
  F_\nu(E_\nu)=\frac{E_{\nu, \mathrm{tot}} E_{\nu}^\alpha}{\langle E_\nu \rangle ^{2+\alpha}} \frac{(\alpha+1)^{(\alpha+1)}}{\Gamma(\alpha+1)} \, \mathrm{Exp}\left[-(\alpha+1)\frac{E_{\nu}}{\langle E_\nu \rangle} \right],
\end{equation}
where $E_{\nu, \mathrm{tot}}$ is the total energy emitted in neutrinos, $\langle E_\nu \rangle$ is the average neutrino energy, $\alpha$ is the pinching parameter, and $\Gamma$ is the gamma function. The normalization of $F_\nu$ is such that $\int F_\nu \, dE_\nu=\mathcal{N}_\nu$, where $\mathcal{N}_\nu$ is the total number of neutrinos emitted. The bounds of the integral are the energy range of the neutrinos, which throughout this work we choose to be 0 to 1 GeV. The specific values of the spectrum parameters change over time during core collapse due to the different stages of collapse, asymmetries, dependence upon the stellar progenitor, and remnant. To reduce complexity, we consider only the time-integrated spectrum with averaged values motivated by simulations of CCSNe \cite{PinchedSpectrum, Totani_1998, Thompson_2003, Sumiyoshi_2005, Nakazato_2013, Suwa_2019, Bollig_2021}: $E_{\nu, \mathrm{tot}} = 3\times10^{53}$ erg, $\langle E_\nu \rangle = 15$ MeV, and $\alpha=2.3$ (Fermi-Dirac). This is reasonable given the duration of neutrino emission for each CCSN, which is only $\sim 10$ seconds and much shorter than the time of the DM profile adjusting. 

We then find the total energy budget of CCSN neutrinos for each dwarf galaxy. For this, we take the observed stellar masses of the dwarf and use them as the total historical stellar mass. This assumption and the fact that in order to find the stellar mass, it is assumed that $\mathrm{M}_\odot / L_\odot=1$, should lead to a minimum estimate for the total historical stellar mass and therefore a conservative estimate on the total neutrino energy budget. Next, we assume an initial mass function of Ref.~\cite{Kroupa_2002},
\begin{equation}\label{IMF}
  \xi(m)\propto
  \begin{cases}
 m^{-0.3}, \quad \mathrm{if} \; m\leq0.08\,\mathrm{M_\odot}, \\
 m^{-1.3}, \quad \mathrm{if} \; 0.08\,\mathrm{M_\odot}<m\leq0.5\,\mathrm{M_\odot}, \\
 m^{-2.3}, \quad \mathrm{if} \; 0.5\,\mathrm{M_\odot}<m,
  \end{cases}
\end{equation}
normalized such that $\int^{100 \, \mathrm{M}_\odot}_{0.1 \, \mathrm{M}_\odot} \, \xi(m) \, dm = 1$. We can then estimate the approximate number of massive stars that used to exist, and therefore the number of CCSNe ($\mathcal{N}_\mathrm{CCSNe}$) that have occurred in each dwarf galaxy using the following equation,
\begin{equation}\label{Number of massive stars}
  \mathcal{N}_\mathrm{CCSNe}= M_* \, \frac{\int_{8\,\mathrm{M}_\odot}^{100\,\mathrm{M}_\odot} \xi(m) \, dm}{\int_{0.1\,\mathrm{M}_\odot}^{100\,\mathrm{M}_\odot} m \, \xi(m) \, dm},
\end{equation}
which builds on the observation that the minimum mass threshold for core collapse is $8 \, \mathrm{M}_\odot$ \cite{Smartt_2009, Smartt_2015, Diaz-Rodríguez_2018, Diaz-Rodriguez_2021}. Here, $M_*$ is the stellar mass of the dwarf galaxy which sets the overall normalization. The calculation of this for each dwarf galaxy is shown in the third column of Table \ref{Dwarf Galaxy Energetics}, kept as floating point numbers. Then, we simply assume that each CCSN emits $3\times10^{53}$ erg of energy in neutrinos in order to get the total $\nu-$DM energy budget as seen in column 4 of Table \ref{Dwarf Galaxy Energetics}, i.e., $E_\mathrm{\nu, budget} = \mathcal{N}_\mathrm{CCSNe} \times (3\times10^{53} \, \mathrm{erg})$.

\subsection{Constraining the energy injected into DM}\label{Energy constraint}

In order to place a constraint on the energy injection, we must first find out the energy needed to transform the DM profile. For this, we follow Ref.~\cite{Penarrubia_2012}. We assume a $\Lambda$CDM cosmology with $H_0 = 71 \, \mathrm{km \, s^{-1} \, Mpc^{-1}}$, $\Omega_\mathrm m$=0.268, $\Omega_\Lambda=1-\Omega_\mathrm m$, and $\Delta_\mathrm{crit} = 200$ ($r_\mathrm{vir} = r_{200}$). With these parameters and assuming constant subhalo mass and full equilibration for the initial cusped profile and the final cored profile we observe today, the virial theorem can be applied to find the energy needed for the transformation to occur. This change in energy required is
\begin{equation}\label{Virial energy change}
  \Delta E = \frac{W_\mathrm{core} - W_\mathrm{cusp}}{2},
\end{equation}
where the work, $W$, is given by 
\begin{equation}\label{Work}
  W=-4\pi \,G_N \int_0^{r_{200}} r \, \rho(r) \, M(r) \,dr, 
\end{equation}
with $M(r)$ being the halo mass profile. In the case of $W_\mathrm{cusp}$, $r_c=0$\,pc so Eq.~\ref{Modified NFW profile} becomes the normal NFW profile. The halo mass profile is found by integrating the density profile over volume and has the analytic form of
\begin{equation}\label{Halo mass profile}
 \frac{M(r)}{M_0}= 
 \begin{cases}
 \mathrm{ln}(1+\tilde{r}) - \frac{\tilde{r}\,(2+3\tilde{r})}{2(1+\tilde{r})^2}, \, x=1 \\
 \frac{x^2\,\mathrm{ln}(1+\tilde{r}/x)+(1-2x)\mathrm{ln}(1+\tilde{r})}{(1-x)^2}-\frac{\tilde{r}}{(1+\tilde{r})(1-x)}, \, x\neq1 \\
 \end{cases}
\end{equation}
where $M_0$ is the total mass of the halo, $\tilde{r}\equiv r / r_\mathrm s$, and $x\equiv r / r_\mathrm c$. 

Using Eqs.~(\ref{Modified NFW profile})--(\ref{Halo mass profile}) and the dwarf galaxy properties shown in Table \ref{Dwarf Galaxy Properties}, we calculate the energy required for this transformation between the $\pm1\sigma$ bounds of the core radius estimates, which are the last two columns in Table \ref{Dwarf Galaxy Energetics}. 

\begin{figure}[t]
  \centering
  \includegraphics[width=\columnwidth]{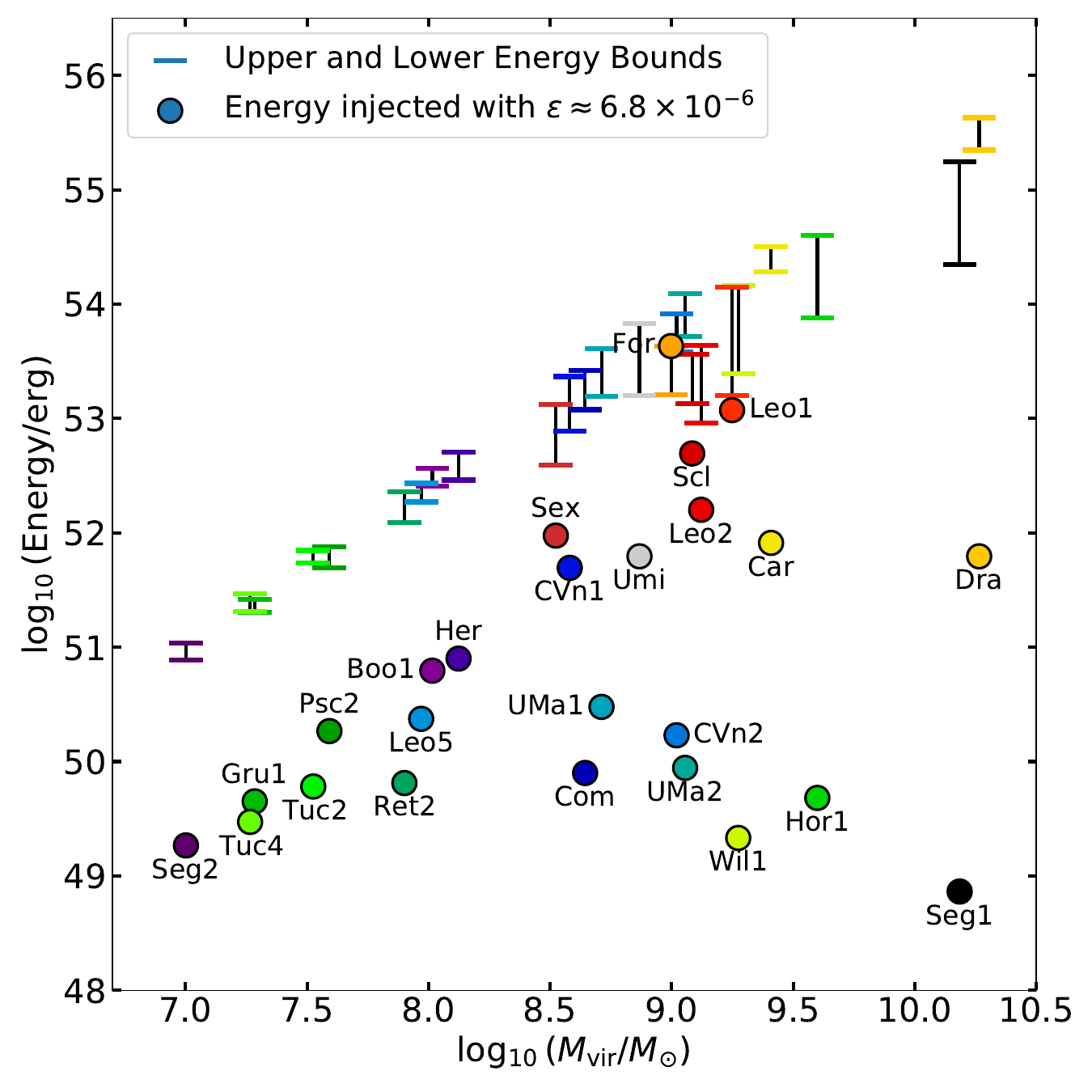}
  \caption{Scatter plot of the energy injected into the DM profiles of Milky Way dSph galaxies: the lines represent the upper and lower bounds such that the DM core radii are not larger than the $1\sigma$ upper limit from Fig.~\ref{Dwarf Core Size}, and the circles represent the energy injected from CCSN neutrinos assuming $\varepsilon\approx6.8\times10^{-6}$ -- this is the value when the energy injected from neutrinos is just small enough to not conflict with DM core radii limits.}
  \label{Upper Core Limit Energy Plot}
\end{figure}

We constrain the amount of energy that can be injected into the DM subhalo of each dwarf galaxy using Eq.~(\ref{Virial energy change}). We do this by requiring that the energy injected must not be too large such that the core size becomes larger than the $r_\mathrm{c}^\mathrm{upper} \equiv r_\mathrm{c}+1\sigma$ estimation. For this, we allow the fraction of energy injected, 
\begin{equation}
\varepsilon \equiv \frac{E_\mathrm{inj}}{ E_\mathrm{\nu, budget}},
\end{equation}
to be freely changed until it satisfies our virial energy change constraint for a dwarf galaxy that is well studied. This is shown in Fig.~\ref{Upper Core Limit Energy Plot}, where  the horizontal bars are the virial energy change found from Eq.~(\ref{Virial energy change}) for the upper $1\sigma$ core radius ($E_\mathrm{upper}$) and the lower $1\sigma$ core radius ($E_\mathrm{lower}$). The circles correspond to the energy injected calculated using $E_\mathrm{inj}=\varepsilon\times E_\mathrm{\nu, tot}$. The smallest upper limit on the energy injection fraction we get is $\varepsilon\approx 6.8\times10^{-6}$ such that the requirement of $r_\mathrm{c} \leq r_\mathrm{c,upper}$ is maintained, which occurs for Fornax. We can see this limit in Fig.~\ref{Upper Core Limit Energy Plot} as the circle for Fornax's energy injection is overlaying the upper energy bound (top bar). This energy injection constraint is carried out for each dSph. However, this is only valid in a regime where the DM is not accelerated above the host subhalo escape velocity (i.e., non-escaping). 

\renewcommand{\arraystretch}{1.25}

\begin{table*}[t]
\caption{Dwarf galaxy energetic properties. The second column shows the estimated stellar mass ($M_*$). The number of massive stars ($\mathcal{N}_\mathrm{CCSNe}$) is found from Eq.~(\ref{Number of massive stars}). The total $\nu$ energy budget ($E_{\nu,\mathrm{budget}}$) is found by assuming each CCSN emits $3\times10^{53}$ erg across all flavors, and sets the maximum energy that can be injected to the DM profile. For the energy injected ($E_\mathrm{inj}$), an overall energy transfer fraction between CCSN neutrinos and DM of $\varepsilon=E_{\mathrm{inj}}/E_{\nu, \mathrm{tot}}= 6.8\times10^{-6}$ is assumed, which is the smallest energy injection limit we find (Fornax). The energy bounds ($E_\mathrm{lower}$ and $E_\mathrm{upper})$ are calculated using Eq.~(\ref{Virial energy change}) for the different core radii estimates.}
\label{Dwarf Galaxy Energetics}
\begin{ruledtabular}
  \begin{tabular}{ccccccc}
  Name & \multicolumn{1}{c}{$M_*$ [$\mathrm{M}_\odot$]\footnote{assuming $\mathrm{M}_\odot / L_\odot = 1$ from Refs.~\cite{McConnachie_2012, Hayashi_2023}}} & \multicolumn{1}{c}{$\mathcal{N}_\mathrm{CCSNe}$} & \multicolumn{1}{c}{$E_{\nu,\mathrm{budget}}$ [erg]} & \multicolumn{1}{c}{$E_\mathrm{inj}$ [erg]} & \multicolumn{1}{c}{$E_\mathrm{lower}$ [erg]} & \multicolumn{1}{c}{$E_\mathrm{upper}$ [erg]} \\
  \hline
Seg1 &  $3.40 \times 10^{2}$ &  $3.56 \times 10^{0}$ & $1.07 \times 10^{54}$ &  $7.31 \times 10^{48}$ &   $2.23 \times 10^{54}$ &   $1.75 \times 10^{55}$ \\
Seg2 &  $8.60 \times 10^{2}$ &  $9.01 \times 10^{0}$ & $2.70 \times 10^{54}$ &  $1.85 \times 10^{49}$ &   $7.73 \times 10^{50}$ &   $1.09 \times 10^{51}$ \\
Boo1 &  $2.90 \times 10^{4}$ &  $3.04 \times 10^{2}$ & $9.12 \times 10^{55}$ &  $6.24 \times 10^{50}$ &   $2.57 \times 10^{52}$ &   $3.66 \times 10^{52}$ \\
Her  &  $3.70 \times 10^{4}$ &  $3.88 \times 10^{2}$ & $1.16 \times 10^{56}$ &  $7.96 \times 10^{50}$ &   $2.90 \times 10^{52}$ &   $5.07 \times 10^{52}$ \\
Com  &  $3.70 \times 10^{3}$ &  $3.88 \times 10^{1}$ & $1.16 \times 10^{55}$ &  $7.96 \times 10^{49}$ &   $1.19 \times 10^{53}$ &   $2.64 \times 10^{53}$ \\
CVn1 &  $2.30 \times 10^{5}$ &  $2.41 \times 10^{3}$ & $7.23 \times 10^{56}$ &  $4.95 \times 10^{51}$ &   $7.77 \times 10^{52}$ &   $2.33 \times 10^{53}$ \\
CVn2 &  $7.90 \times 10^{3}$ &  $8.28 \times 10^{1}$ & $2.48 \times 10^{55}$ &  $1.70 \times 10^{50}$ &   $3.83 \times 10^{53}$ &   $8.20 \times 10^{53}$ \\
Leo5 &  $1.10 \times 10^{4}$ &  $1.15 \times 10^{2}$ & $3.46 \times 10^{55}$ &  $2.37 \times 10^{50}$ &   $1.86 \times 10^{52}$ &   $2.72 \times 10^{52}$ \\
UMa1 &  $1.40 \times 10^{4}$ &  $1.47 \times 10^{2}$ & $4.40 \times 10^{55}$ &  $3.01 \times 10^{50}$ &   $1.57 \times 10^{53}$ &   $4.08 \times 10^{53}$ \\
UMa2 &  $4.10 \times 10^{3}$ &  $4.30 \times 10^{1}$ & $1.29 \times 10^{55}$ &  $8.82 \times 10^{49}$ &   $5.21 \times 10^{53}$ &   $1.23 \times 10^{54}$ \\
Ret2 &  $3.02 \times 10^{3}$ &  $3.16 \times 10^{1}$ & $9.49 \times 10^{54}$ &  $6.50 \times 10^{49}$ &   $1.24 \times 10^{52}$ &   $2.28 \times 10^{52}$ \\
Psc2 &  $8.60 \times 10^{3}$ &  $9.01 \times 10^{1}$ & $2.70 \times 10^{55}$ &  $1.85 \times 10^{50}$ &   $4.92 \times 10^{51}$ &   $7.59 \times 10^{51}$ \\
Gru1 &  $2.09 \times 10^{3}$ &  $2.19 \times 10^{1}$ & $6.57 \times 10^{54}$ &  $4.49 \times 10^{49}$ &   $2.03 \times 10^{51}$ &   $2.61 \times 10^{51}$ \\
Hor1 &  $2.24 \times 10^{3}$ &  $2.35 \times 10^{1}$ & $7.04 \times 10^{54}$ &  $4.82 \times 10^{49}$ &   $7.55 \times 10^{53}$ &   $4.01 \times 10^{54}$ \\
Tuc2 &  $2.82 \times 10^{3}$ &  $2.95 \times 10^{1}$ & $8.86 \times 10^{54}$ &  $6.06 \times 10^{49}$ &   $5.45 \times 10^{51}$ &   $7.01 \times 10^{51}$ \\
Tuc4 &  $1.38 \times 10^{3}$ &  $1.45 \times 10^{1}$ & $4.34 \times 10^{54}$ &  $2.97 \times 10^{49}$ &   $2.05 \times 10^{51}$ &   $2.93 \times 10^{51}$ \\
Wil1 &  $1.00 \times 10^{3}$ &  $1.05 \times 10^{1}$ & $3.14 \times 10^{54}$ &  $2.15 \times 10^{49}$ &   $2.48 \times 10^{53}$ &   $1.45 \times 10^{54}$ \\
Car  &  $3.80 \times 10^{5}$ &  $3.98 \times 10^{3}$ & $1.19 \times 10^{57}$ &  $8.17 \times 10^{51}$ &   $1.91 \times 10^{54}$ &   $3.18 \times 10^{54}$ \\
Dra  &  $2.90 \times 10^{5}$ &  $3.04 \times 10^{3}$ & $9.12 \times 10^{56}$ &  $6.24 \times 10^{51}$ &   $2.22 \times 10^{55}$ &   $4.27 \times 10^{55}$ \\
For  &  $2.00 \times 10^{7}$ &  $2.10 \times 10^{5}$ & $6.29 \times 10^{58}$ &  $4.30 \times 10^{53}$ &   $1.62 \times 10^{53}$ &   $4.30 \times 10^{53}$ \\
Leo1 &  $5.50 \times 10^{6}$ &  $5.76 \times 10^{4}$ & $1.73 \times 10^{58}$ &  $1.18 \times 10^{53}$ &   $1.59 \times 10^{53}$ &   $1.41 \times 10^{54}$ \\
Leo2 &  $7.40 \times 10^{5}$ &  $7.75 \times 10^{3}$ & $2.33 \times 10^{57}$ &  $1.59 \times 10^{52}$ &   $9.09 \times 10^{52}$ &   $4.34 \times 10^{53}$ \\
Scl  &  $2.30 \times 10^{6}$ &  $2.41 \times 10^{4}$ & $7.23 \times 10^{57}$ &  $4.95 \times 10^{52}$ &   $1.35 \times 10^{53}$ &   $3.66 \times 10^{53}$ \\
Sex  &  $4.40 \times 10^{5}$ &  $4.61 \times 10^{3}$ & $1.38 \times 10^{57}$ &  $9.46 \times 10^{51}$ &   $3.94 \times 10^{52}$ &   $1.32 \times 10^{53}$ \\
UMi  &  $2.90 \times 10^{5}$ &  $3.04 \times 10^{3}$ & $9.12 \times 10^{56}$ &  $6.24 \times 10^{51}$ &   $1.59 \times 10^{53}$ &   $6.79 \times 10^{53}$ \\
  \end{tabular}
\end{ruledtabular}
\end{table*}

\subsection{Mass loss cross section bound}

In order to find the valid DM mass regime in which the energy injected by neutrinos does not cause the DM to become gravitationally unbound from the host subhalo (i.e., non-escaping), we must first find the escape velocity. This is done by setting the work it takes to unbind the DM as its kinetic energy. Assuming a NFW profile, the escape velocity, $v_\mathrm{esc}$, is given by

\begin{equation}\label{Escape Velocity}
    v_\mathrm{esc}^2 = \int_0^\infty \frac{2 G_N M(r)}{r^2} dr = 8 \pi G_N \rho_\mathrm{s} r_\mathrm{s}^2,
\end{equation}
\noindent where $M(r)$ is the mass interior to radius $r$, given by Eq.~(\ref{Halo mass profile}). With this, we can find the minimum mass at which the neutrinos do not give enough energy to the DM to cause it to escape by setting the kinetic energy equal to the energy transfer per interaction. We assume the energy transfered is what is maximally allowed from relativistic collisions, $\Delta E_\mathrm{int} = \frac{2E_\nu^2}{2E_\nu+m_\chi}$ where $m_\chi$ is the DM mass, and that the DM is initially at rest. We then average over the neutrino energy spectrum, i.e.,
\begin{equation}
    \frac{1}{2}m_\chi v_\mathrm{esc}^2 = \frac{\int^{1\,\mathrm{GeV}}_{0\,\mathrm{GeV}} F_\nu(E_\nu) \, 2 E_\nu^2/(2 E_\nu + m_\chi)\, dE_\nu}{\int^{1\,\mathrm{GeV}}_{0\,\mathrm{GeV}} F_\nu(E_\nu) dE_\nu},
\end{equation}
With the dSphs we study in this work, the escape velocities are in between $\sim20-200 \, \mathrm{km \, s^{-1}}$, which translates to DM with $m_\chi \lesssim 50-500 \, \mathrm{GeV}$ becoming gravitationally unbound from the $\nu$-DM interaction. In order to place a constraint on the interaction cross section in this region, we invoke a mass loss bound in which the subhalo cannot lose too much mass within its core region. We set the limit on the mass lost by the $\nu$-DM interaction to be when the subhalo core region loses enough mass such that it cannot form the core we see today, i.e.,
\begin{equation}
    \Delta M_\mathrm{lim} = M_\mathrm{NFW}(r_\mathrm{c}) - M_\mathrm{cored}(r_\mathrm{c}). 
\end{equation}
This is then transformed to a limit on the cross section first by getting the fraction of neutrinos that must interact to cause this mass loss,
\begin{equation}
    \eta\equiv \frac{\mathcal{N}_\mathrm{int}}{\mathcal{N}_{\nu,\mathrm{tot}}}=\frac{\Delta M_\mathrm{lim}}{m_\chi \, \mathcal{N}_{\nu,\mathrm{tot}}},
\end{equation}
\noindent where $\mathcal{N}_\mathrm{int}$ is the number of interactions and $\mathcal{N}_{\nu,\mathrm{tot}}=\mathcal{N}_\mathrm{CCSNe}\times\mathcal{N}_\mathrm{\nu}$ is the total number of neutrinos emitted for all CCSNe. The next component that is needed to place a limit on the cross section is the column number density of DM that a neutrino would pass through in the core region,
\begin{equation}\label{Column number density}
  \Sigma_\mathrm{DM} = \frac{\int^{r_\mathrm{c}}_0 \rho_\mathrm{NFW}(r)\,dr}{m_\chi},
\end{equation} 
\noindent where we choose to integrate out to the best fit core radius, as that is where the interactions need to occur for the mass loss. Changing the upper limit of the integration to be the $1\sigma$ core radius upper limit does not appreciably change the column number density ($\sim0.5\%$). The cross section is then given by

\begin{equation}\label{Cross Section Defn}
\begin{split}
        \langle \sigma_{\nu\mathrm{-DM}}(m_\chi {\leq} m_{\chi,\mathrm{lim}}) \rangle & = \frac{\eta}{\Sigma_\mathrm{DM}} = \frac{\Delta M_\mathrm{lim}}{m_\chi \, \Sigma_\mathrm{DM} \, \mathcal{N}_{\nu,\mathrm{tot}}}, \\
        & = \frac{\Delta M_\mathrm{lim}}{\mathcal{N}_{\nu,\mathrm{tot}} \int_0^{r_\mathrm{c}} \rho_\mathrm{NFW}(r) dr,}
\end{split}
\end{equation}
\noindent where we include $\langle \rangle$ to denote that this is averaged over the neutrino energy spectrum. Note, that this limit is a constant for each dSph and not a function of $m_\chi$. This comes from the number of interactions needed to remove the same amount of mass increasing at the same rate as the number density of DM.

Calculating this bound, the strongest limit comes from Fornax, which has $m_{\chi,\mathrm{lim}}\approx130\,\mathrm{GeV}$. The bound from Fornax is the strongest as the subhalo has a large escape velocity compared to other dSphs, and it has a large number of massive stars so the total number of CCSN neutrinos is large. The limit comes out to be $\langle\sigma_{\nu\mathrm{-DM}}\rangle\approx3.4\times10^{-23}\,\mathrm{cm}^{-2}$.

\subsection{Energy injection cross section bound}

With the constraint on $\varepsilon$, we now convert that into a constraint on the cross section for DM that is not gravitationally unbound. For this, we assume that the fraction of energy transferred between the neutrinos and the DM in their scattering interaction is what is maximally allowed in relativistic collisions, and that fraction is given by,
\begin{equation}\label{Kinematic factor}
  f_\mathrm{max}=\frac{2E_\nu}{m_\chi+2E_\nu},
\end{equation}

\noindent In this energy injection limit regime, we can then relate $\eta$ to the energy injection limit $\varepsilon$ and the fraction of energy injected $f$ by the following equation: 
\begin{equation}\label{Interaction fraction}
\begin{split}
  \eta\equiv\frac{\mathcal{N}_\mathrm{int}}{\mathcal{N}_{\nu,\mathrm{tot}}}=\frac{\mathcal{N}_\mathrm{int}}{\mathcal{N}_{\nu,\mathrm{tot}}} \times \frac{\langle E_\nu \rangle}{\langle E_\nu \rangle} = \frac{E_\mathrm{inj} / f}{E_{\nu,\mathrm{budget}}}=
  \frac{\varepsilon}{f_\mathrm{max}}. 
\end{split}
\end{equation}

We estimate the limit on the cross section again by using Eq.~(\ref{Cross Section Defn}). We still integrate $\Sigma_\mathrm{DM}$ out to the upper limit of the core radius. As we assume an initial cusped NFW profile a majority of the interactions should occur near the center of the subhalo, which is where it needs to happen for core formation. Integrating out to $r_{200}$ instead only increases $\Sigma_\mathrm{DM}$ by a few percent.
This yields the limit,
\begin{equation}\label{Cross section upper bound}
  \sigma_{\nu\mathrm{-DM}}(m_\chi {>} m_{\chi,\mathrm{lim}})= \frac{\varepsilon}{f_\mathrm{max} \, \Sigma_\mathrm{DM}}, 
\end{equation}
where $\sigma_0$ is the value of the cross section assuming that all neutrinos are emitted at a single energy. However, as the neutrino spectrum from CCSNe is not monoenergetic, we take into account the spectrum energy dependence. To do so, we average the value of the cross section over the neutrino energy spectrum
\begin{equation}\label{More realistic cross section upper bound}
  \langle \sigma_{\nu\mathrm{-DM}}(m_\chi{>}m_{\chi,\mathrm{lim}}) \rangle = \frac{\int_{\mathrm{0 \, GeV}}^{\mathrm{1 \, GeV}} F_\nu(E_\nu) \, \sigma_{\nu\mathrm{-DM}}(E_\nu) \, dE_\nu}{\int_{\mathrm{0 \, GeV}}^{\mathrm{1 \, GeV}} F_\nu(E_\nu) \, dE_\nu},
\end{equation}
where $\sigma_{\nu\mathrm{-DM}}$ has mass and energy dependence from $\Sigma_\mathrm{DM}$ and $f_\mathrm{max}$. 

\begin{figure}[t]
  \centering
  \includegraphics[width=\columnwidth]{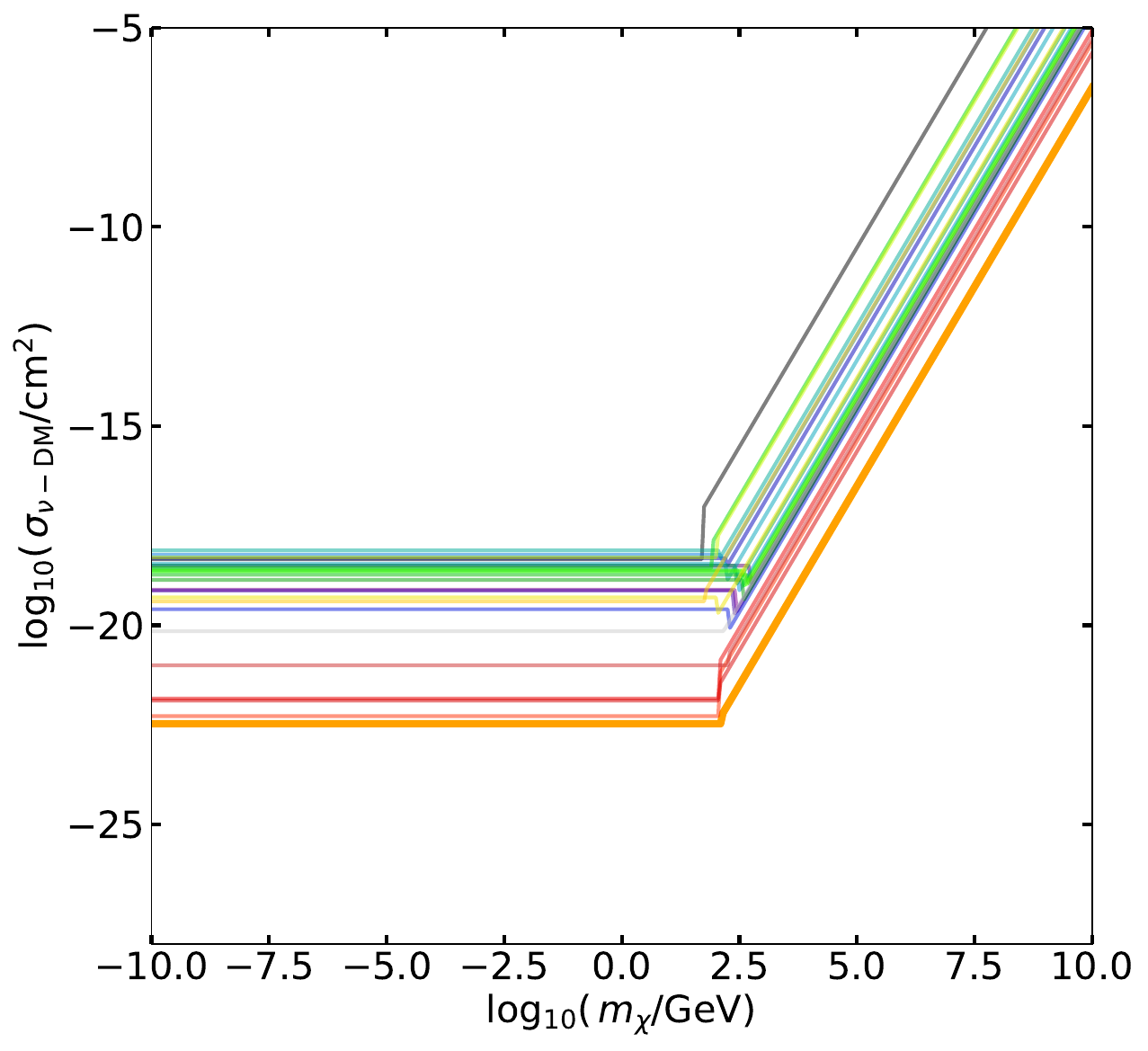}
  \caption{Dependence of the phenomenological cross section on the DM mass, where the colors represent the bounds from different dSphs, same colors used as Fig.~\ref{Upper Core Limit Energy Plot}. Fornax has the strongest constraint for all DM masses. For small masses the bound comes from a mass loss argument, and for large masses the bound comes from an energy injection argument.}
  \label{Cross Section DM Mass Dependence}
\end{figure}

\begin{figure}[t]
  \centering
  \includegraphics[width=\columnwidth]{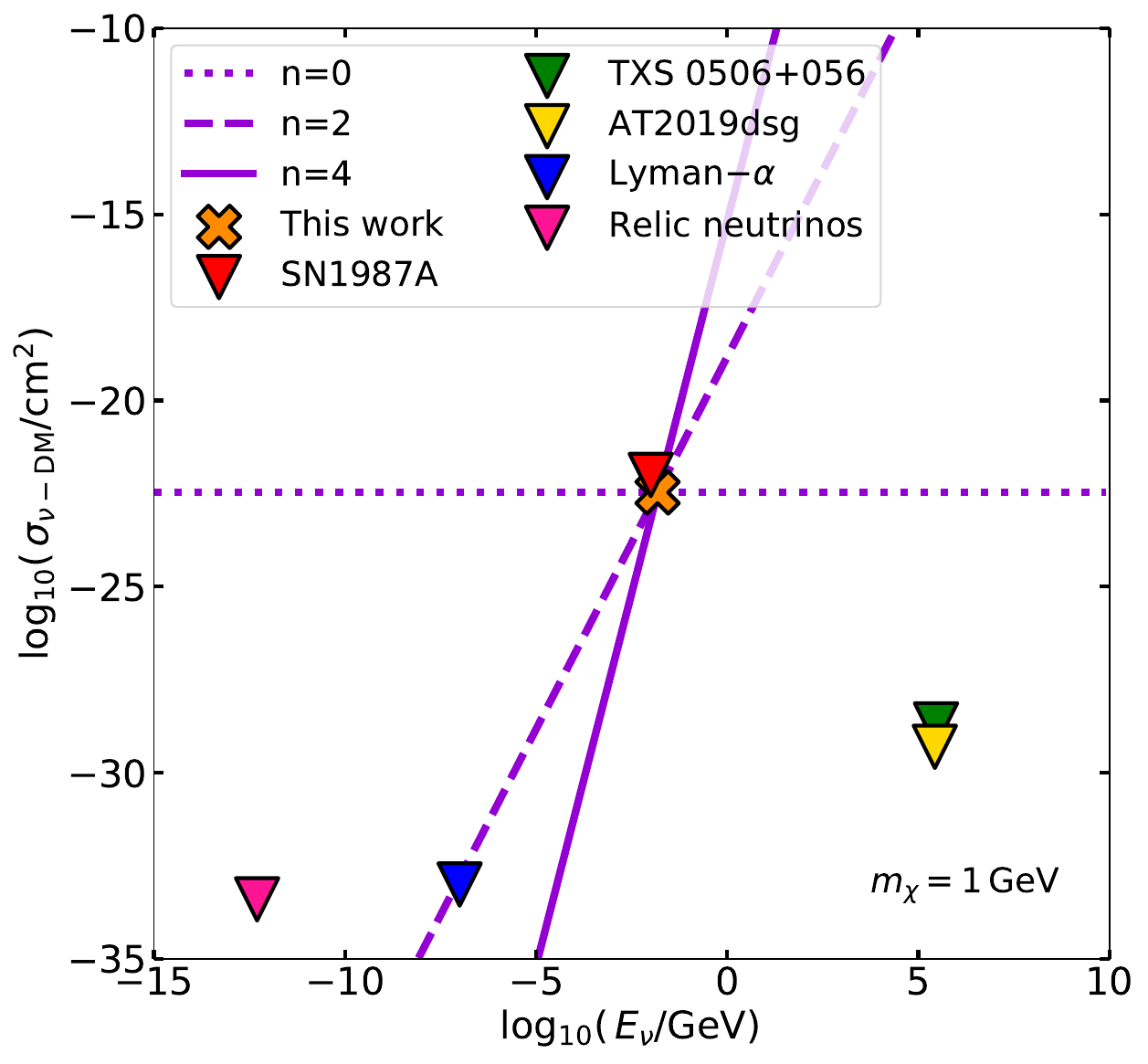}
  \caption{Upper bound assuming a cross section with a simple power law energy dependence with index $n$; see Eq.~(\ref{Simple power law cross section}). This does not take into account the changing $m_{\chi,\mathrm{lim}}$ for differing $E_\nu$.} All forms are normalized such that they fit the upper bound found in this work (orange cross). Shown for comparison are bounds from SN1987A \cite{Mangano_2006} (red triangle), TXS 0506+056 \cite{Cline_2023, Ferrer_2023} (green triangle), AT2019dsg \cite{Fujiwara_2023tidal} (yellow triangle), Lyman$-\alpha$ forest \cite{Wilkinson_2014} (blue triangle), and relic neutrinos \cite{Akita2023constraints} (pink triangle).
  \label{Power Law Cross Section Plot}
\end{figure}

The resulting $\langle \sigma_{\nu\mathrm{-DM}} \rangle$ only has mass dependence. We plot the cross section dependence on the DM mass in Fig.~\ref{Cross Section DM Mass Dependence}, showing both the mass loss bound and the energy injection bound. We can approximate the strongest bound from Fornax in the two $m_\chi$ regions numerically as 
\begin{align}
  \langle \sigma_{\nu\mathrm{-DM}} \rangle \approx 
  \begin{cases}
  3.4 \times 10^{-23} \,  \, \mathrm{cm^2},  &m_\chi \leq 130\,\mathrm{GeV}, \\
  3.2 \times 10^{-27} \, \left( \frac{m_\chi}{1 \, \mathrm{GeV}}\right)^2 \, \mathrm{cm^2}, &m_\chi > 130\,\mathrm{GeV}.
  \end{cases}
\end{align}
If we assume a power law form of the energy dependence of the cross section, then our upper bound is,
\begin{equation}\label{Simple power law cross section}
  \sigma_{\nu\mathrm{-DM}}(E_\nu)=\langle \sigma_{\nu\mathrm{-DM}} \rangle \left( \frac{E_\nu}{15\,\mathrm{MeV}}\right)^n ,
\end{equation}
where the specific value of $\langle \sigma_{\nu\mathrm{-DM}} \rangle$ depends on the choice of $m_\chi$ and $n$ determines the specific energy dependence. We plot our upper limit in its naive power law form along with other constraints from similar works in Fig.~{\ref{Power Law Cross Section Plot}} where in all cases $m_\chi=1$ GeV is fixed. We see that in the energy range of CCSN neutrinos, our upper limit is slightly stronger than those placed by SN1987A \cite{Mangano_2006}. 

\section{Discussion}

In this section we discuss our resulting upper limit and compare it to other limits on the $\nu-$DM cross section. We consider a simple particle DM model and work out how to use our energy constraint to place constraints on the coupling constants. We also go over the different uncertainties that are in our analysis.

\subsection{Comparison to limits from previous works}

There are several previous works that constrain the interaction cross section between neutrinos and DM, of which we have chosen a few to compare to. The bounds from the similar studies are also plotted in Fig.~\ref{Power Law Cross Section Plot}, represented by the upside-down triangles. The works we chose to compare to are spread out in terms of the neutrino energy, going from ${\sim}$0.5 meV up to 290 TeV.  

The most relevant comparison is to the bounds from SN1987A, which remains the only supernova we have measured neutrinos from. Reference \cite{Mangano_2006} looks at elastic scattering of neutrinos on DM and uses the fact that the theoretical neutrino flux from SN1987A agrees with the observed neutrino flux \cite{1987A1, 1987A2, 1987A3} to retrieve a bound of $\sigma\sim10^{-22}$ cm$^2$, which is just the column number density for a 1 GeV DM candidate (red triangle). Our bound for a 1 GeV DM particle is slightly stronger than this, by around a factor of 3. We are only slightly stronger as this falls within the region where the DM is accelerated above the escape velocity, so we cannot use the strong energy injection constraint. If we extend to larger (smaller) $m_\chi$, our limit becomes stronger (weaker) in comparison.

At higher energies, works have used high energy neutrino events observed by the IceCube experiment, e.g., Ref.~\cite{Choi_2019}, which looks at the 290 TeV neutrino from the flaring blazar TXS 0506+056 \cite{IceCube2018}. More stringent constraints have also been made using TXS 0506+056, by assuming DM spike profiles near the supermassive black hole which are model dependent \cite{Cline_2023, Ferrer_2023} (green triangle). There are also bounds from a high energy neutrino at 270 TeV thought to be associated with the tidal disruption event AT2019dsg \cite{Stein:2020xhk} also assuming a dark matter spike model \cite{Fujiwara_2023tidal} (yellow triangle).

Low energy limits come from looking at data from the cosmic microwave background, baryon acoustic oscillations, matter power spectrum, and Lyman-$\alpha$ data \cite{Mangano_2006, Wilkinson_2014, Olivares-Del-Campo_2018, Palanque-Delabrouille_2020, Mosbech_2021, Hooper_2022}. There are also constraints of relic neutrinos interacting with Milky Way satellites \cite{Akita2023constraints}. These limits have a smaller upper bound for the cross section, and are stronger than the limit found in this paper if the energy dependence scales as $n\lesssim2$. 

\subsection{\ensuremath{Z'} model} \label{Z' model}

As an illustrative example, we discuss a simple model for a Dirac fermion DM candidate that scatters with neutrinos, the $Z'$ model. It is an interaction that has a vector boson exchange ($Z'$) with mass $m_{Z'}$, and two couplings, $g_\nu$ and $g_\chi$. For simplicity, $g_\nu$ is assumed to take the same value for all flavors and for particles and antiparticles. This is the same model as discussed in Refs.~\cite{Cline_2023, Cline:2023tkp, Fujiwara_2023tidal}. 

We place a constraint on the product of the couplings by computing the ratio between the upper limit we find and the full form of the cross section with dependence on the DM mass, mediator mass, and neutrino energy (we assume $E\nu=15$\,MeV for simplicity) . The full form of the cross section, from Ref.~\cite{Arguelles_2017}, is
\begin{widetext}
\begin{equation}
\begin{split}
    \sigma_{\nu\mathrm{-DM}} = &\frac{(g_\nu g_\chi)^2}{16 \pi E_\nu^2 m_\chi^2} \left[ (m_{Z'}^2 + m_\chi^2 + 2E_\nu m_\chi)\mathrm{log}\left(\frac{m_{Z'}^2(2E_\nu+m_\chi)}{m_\chi(4E_\nu^2+m_{Z'}^2) + 2E_\nu m_{Z'}^2}\right) \right.\\ 
    & \; + \left. 4E_\nu^2\left(1 + \frac{m_\chi^2}{m_{Z'}^2} - \frac{2E_\nu(4E_\nu^2 m_\chi + E_\nu(m_\chi^2 + 2m_{Z'}^2) + m_\chi m_{Z'}^2)}{(2E_\nu+m_\chi)(m_\chi(4E_\nu^2+m_{Z'}^2)+2E_\nu m_{Z'}^2}\right) \right] 
\end{split}
\end{equation}
\end{widetext}

\begin{figure}[t]
  \centering
  \includegraphics[width=\columnwidth]{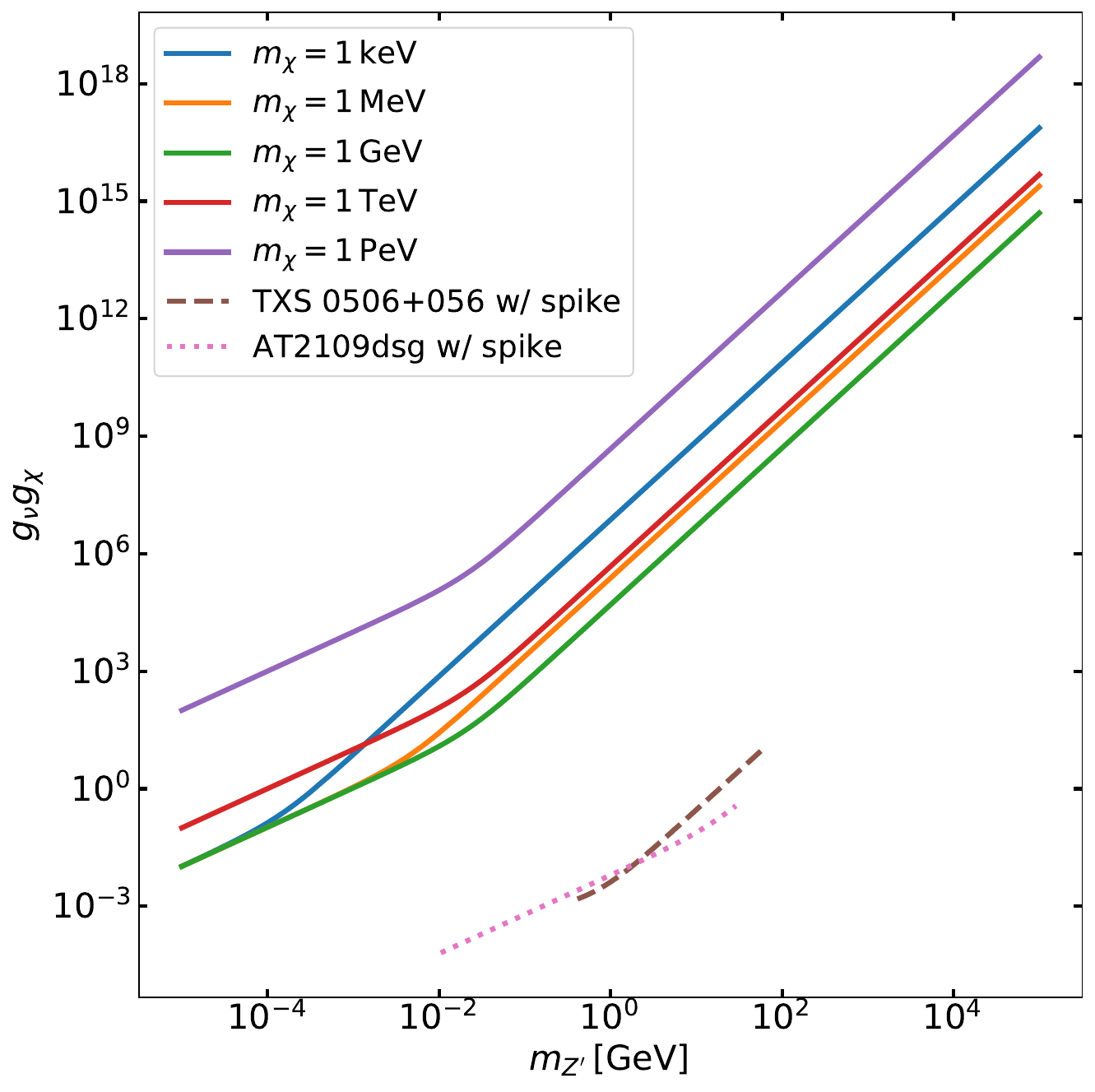}
  \caption{Upper limits from Fornax on the product of the neutrino coupling and DM coupling as a function of the vector boson mediator mass, plotted for different DM mass values of 1 keV (blue), 1 MeV (orange), 1 GeV (green), 1 TeV (red) and 1 PeV (purple). Included are the most stringent bounds on the $Z'$ model from works using high energy neutrinos propagating through enhanced DM spikes, TXS 0506+056 \cite{Cline_2023} with $m_\chi=1$ keV (brown dashed) and AT2019dsg \cite{Fujiwara_2023tidal} with $m_\chi=1$ MeV (pink dotted).}
  \label{Coupling vs Zprime mass plot}
\end{figure}

\begin{figure}[t]
  \centering
  \includegraphics[width=\columnwidth]{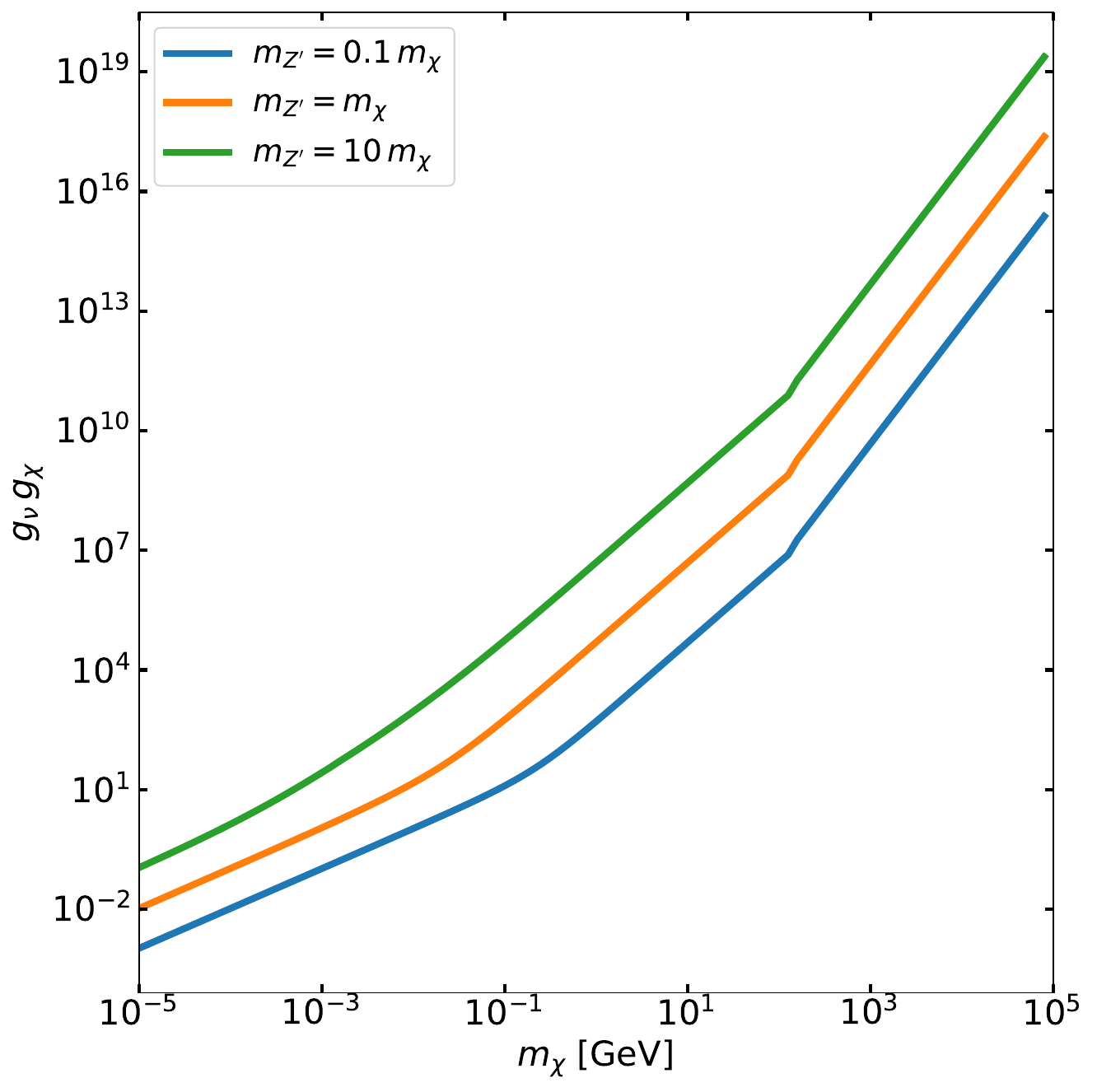}
  \caption{Upper limits from Fornax on the product of the neutrino coupling and DM coupling as a function of the DM mass for varying mediator masses that are some factor times the DM mass. The values for the multiplication factor we choose are 0.1 (blue), 1 (orange), and 10 (green).}
  \label{Coupling vs DM mass plot}
\end{figure}

We plot this calculation using the cross section limit from Fornax both as a function of mediator mass (see Fig.~\ref{Coupling vs Zprime mass plot}) and as a function of DM mass (see Fig.~\ref{Coupling vs DM mass plot}). In Fig.~\ref{Coupling vs Zprime mass plot}, we also show bounds that are calculated using enhanced DM spikes around supermassive black holes with neutrinos from TXS 0506+056 \cite{Cline_2023} with $m_\chi= 1\,\mathrm{keV}$ and AT2019dsg \cite{Fujiwara_2023tidal} with $m_\chi=1\,\mathrm{MeV}$. Those bounds on the product of the couplings are much stronger than ours for any DM mass and/or mediator mass.

\subsection{Uncertainties}

There are various sources of uncertainty that we do not include in our derivation of an upper bound in the $\nu$-DM cross section. Uncertainties with the stellar kinematics and stellar groupings can have large effects on the inferred subhalo properties and observed stellar masses. In order to try and minimize the impact of uncertainties, we make assumptions that result in a more conservative $\sigma_{\nu-\mathrm{DM}}$ when possible. For example, we chose to use the upper limit for the core radius such that the cross section estimation was maximal. For the same reasoning, we used a minimum estimation for the total dSph stellar mass, which results in the minimum estimate of the number of CCSNe, taking it directly from the observed luminosity with a conservative mass-to-light ratio $\mathrm{M}_\odot/L_\odot=1$.  

We also ignore any effects other than $\nu-$DM interactions that could contribute to DM core formation. For example, baryonic feedback which results from supernova explosion energy \cite{Governato_2010, Mashchenko_2008, Brooks:2012vi, DiCintio_2013, Garrison-Kimmel:2013yys, Teyssier_2013, Chan_2015, Read:2015sta, Dutton_2016, Tollet_2016, Freundlich_2019, Burger_2021} (note that this is the explosion energy, which is $\sim1$\% of the total neutrino energy) and tidal effects from the Milky Way halo \cite{Mayer_2001, Kazantzidis:2010cw, Brooks:2012vi, Chang:2012rx, Kazantzidis:2013wi, Read:2015sta, Wang:2016qol, Hiroshima:2018kfv} are both neglected. If any of these were included, they would contribute to DM core formation, yielding stronger constraints on the energy injection and $\nu-$DM interaction.  

Other effects we do not consider come from specific DM particle models, such as flavor dependence, that we do not consider as we solely look at the phenomenological upper bound. This can also change the energy dependence of the cross section to not be a simple power law. The mediator of the interaction also affects the cross section which depends on the specific particle model that is considered. 

Uncertainties in our derivation itself come from neglecting the distribution of energy transfer, that all interactions occur near the center of the subhalo, and neglecting time dependence of CCSN occurrence. The energy transfer distribution is a highly model dependent quantity, so we only consider one example. For assuming most interactions happen within the central cusp/core region, one can expect most of the interactions to happen within the DM cusp as the number density of DM is much larger compared to other regions of the subhalo, e.g., for Fornax the column number density only increases by ${\sim}3\%$ when changing the upper integration bound from $r_c^\mathrm{upper}$ to $r_{200}$; thus most of the DM lies in the central region. Even when an initial core begins to form, the number density is still larger within the core, so interactions are more favorable within the inner volume. There is also an uncertainty in the pinching/anti-pinching of the neutrino spectrum, but changing this parameter has only a small impact on $\sigma_{\nu-\mathrm{DM}}$. Finally, CCSNe are transient injections of energy, and a numerical simulation is desirable to explore beyond our treatment based on Ref.~\cite{Penarrubia_2012}. For example, the impact of energy injection from supernova feedback is dependent on the number and size of so-called blowouts, with a single large blowout influencing the DM profile more strongly than multiple weaker blowouts \cite{Garrison-Kimmel:2013yys}. While the star-forming activity of Milky Way satellites often occurs in prominent bursts \cite{Weisz:2014qra}, it would still be interesting to explore the impacts of time-dependent energy injections from CCSN neutrinos. 

\section{Summary}

In this work, we find a phenomenological upper bound on the interaction cross section between neutrinos and DM using the condition that CCSN neutrinos cannot have too many interactions with DM particles inside the subhalos of Milky Way dwarf spheroidals such that their DM cores are less massive or become too large when compared to observed estimations. Using stellar kinematic data, we estimate the subhalo properties for different dSphs. We then estimate the amount of energy that needs to be injected in order to transform an initial NFW cusped DM profile into the cored profile that we observe today. Under the assumptions that $\Lambda$CDM is correct, the $\nu-$DM interactions are the only source of energy injection and feedback, and that the energy transfer between the neutrinos and the DM particle is maximal, we find an upper bound of $\sigma_{\nu\mathrm{-DM}}(E_\nu=15 \, \mathrm{MeV}, m_\chi\lesssim130 \, \mathrm{GeV}) \lesssim 3.4 \times 10^{-23} \, \mathrm{cm^2}$ and $\sigma_{\nu\mathrm{-DM}}(E_\nu=15 \, \mathrm{MeV}, m_\chi\gtrsim130 \, \mathrm{GeV}) \lesssim 3.2 \times 10^{-27} \left( \frac{m_\chi}{1\,\mathrm{GeV}}\right)^2\, \mathrm{cm^2}$. This upper limit we find is slightly stronger than previous limits using supernova neutrino data \cite{Mangano_2006} for $\mathcal{O}(10)$ MeV neutrinos. 

There are several sources of uncertainty in our derivation, e.g., constant energy transfer, stellar kinematics, and assumptions of core formation. Wherever possible we adopt assumptions within the calculation that make the upper limit more conservative. Therefore, our limit is conservative, and inclusion of additional effects will likely result in stronger constraints. 

Additional uncertainties that can affect our result are model dependent. While we remain model agnostic throughout the derivation, we briefly discuss a simple model of particle DM that can interact with neutrinos that is flavor independent (Sec.~\ref{Z' model}). We then use the derivation of our cross section upper limit to place constraints on parameters of the model itself, specifically the product of the neutrino and DM couplings to the mediator. For other models the same can be done, however one needs to remain careful about the flavor dependence of the interaction(s) as well as the kinematics.  

Many different astrophysical neutrino sources have been used to constrain the cross section of neutrinos and DM. At MeV energies, our limits remain competitive compared to those from SN1987A neutrinos \cite{Mangano_2006}. Looking to the future, the detection prospects of the diffuse supernova neutrino background are very promising \cite{Li_2022, Ekanger_2023:DSNB}, which can help in constraining $\sigma_{\nu-\mathrm{DM}}$ further.

Future missions like the Thirty Meter Telescope \cite{TMT:2015pvw} and the Vera C. Rubin Observatory with the Legacy Survey of Space and Time \cite{LSST} will be able to help reduce the uncertainties. First and foremost, they will be able to get more precise measurements of the proper motions, reducing uncertainties on the stellar kinematics. Another important future effort, as mentioned in Refs.~\cite{deMartino_2022,Guerra_2023}, would be to reduce the uncertainties in confirming more stars associated in these dwarf galaxies. Together, the DM profiles of dwarfs would be determined better, which will result in more stringent constraints for neutrino-DM interactions.

\begin{acknowledgments}
We thank Kohei Hayashi for providing us the stellar kinematics data used in the analysis. We thank Gonzalo Herrera, R. Andrew Gustafson, Nirmal Raj, Akash Kumar Saha, Ranjan Laha, and Biplob Bhattacherjee for useful discussions. 
S.~Heston is supported by U.S.~Department of Energy Office of Science under Award No.~DE-SC0020262 and NSF Grant No.~PHY-2209420. The work of S.~Horiuchi is supported by the U.S.~Department of Energy Office of Science under Award No.~DE-SC0020262, NSF Grant No.~AST1908960 and No.~PHY-2209420, JSPS KAKENHI Grants No.~JP22K03630 and No.~JP23H04899, and the Julian Schwinger Foundation. 
The work of S.~S.~is supported by Grant-in-Aid for Scientific Research from the Ministry of Education, Culture, Sports, Science, and Technology (MEXT), Japan, 20H01895, 20H05860, and 21H00067.
This work was supported by World Premier International Research Center Initiative (WPI Initiative), MEXT, Japan. 
\end{acknowledgments}

\bibliography{research.bib}

\providecommand{\noopsort}[1]{}\providecommand{\singleletter}[1]{#1}
\begin{thebibliography}{159}%
\makeatletter
\providecommand \@ifxundefined [1]{%
 \@ifx{#1\undefined}
}%
\providecommand \@ifnum [1]{%
 \ifnum #1\expandafter \@firstoftwo
 \else \expandafter \@secondoftwo
 \fi
}%
\providecommand \@ifx [1]{%
 \ifx #1\expandafter \@firstoftwo
 \else \expandafter \@secondoftwo
 \fi
}%
\providecommand \natexlab [1]{#1}%
\providecommand \enquote  [1]{``#1''}%
\providecommand \bibnamefont  [1]{#1}%
\providecommand \bibfnamefont [1]{#1}%
\providecommand \citenamefont [1]{#1}%
\providecommand \href@noop [0]{\@secondoftwo}%
\providecommand \href [0]{\begingroup \@sanitize@url \@href}%
\providecommand \@href[1]{\@@startlink{#1}\@@href}%
\providecommand \@@href[1]{\endgroup#1\@@endlink}%
\providecommand \@sanitize@url [0]{\catcode `\\12\catcode `\$12\catcode
  `\&12\catcode `\#12\catcode `\^12\catcode `\_12\catcode `\%12\relax}%
\providecommand \@@startlink[1]{}%
\providecommand \@@endlink[0]{}%
\providecommand \url  [0]{\begingroup\@sanitize@url \@url }%
\providecommand \@url [1]{\endgroup\@href {#1}{\urlprefix }}%
\providecommand \urlprefix  [0]{URL }%
\providecommand \Eprint [0]{\href }%
\providecommand \doibase [0]{https://doi.org/}%
\providecommand \selectlanguage [0]{\@gobble}%
\providecommand \bibinfo  [0]{\@secondoftwo}%
\providecommand \bibfield  [0]{\@secondoftwo}%
\providecommand \translation [1]{[#1]}%
\providecommand \BibitemOpen [0]{}%
\providecommand \bibitemStop [0]{}%
\providecommand \bibitemNoStop [0]{.\EOS\space}%
\providecommand \EOS [0]{\spacefactor3000\relax}%
\providecommand \BibitemShut  [1]{\csname bibitem#1\endcsname}%
\let\auto@bib@innerbib\@empty
\bibitem [{\citenamefont {Asaka}\ \emph {et~al.}(2005)\citenamefont {Asaka},
  \citenamefont {Blanchet},\ and\ \citenamefont {Shaposhnikov}}]{Asaka_2005}%
  \BibitemOpen
  \bibfield  {author} {\bibinfo {author} {\bibfnamefont {T.}~\bibnamefont
  {Asaka}}, \bibinfo {author} {\bibfnamefont {S.}~\bibnamefont {Blanchet}},\
  and\ \bibinfo {author} {\bibfnamefont {M.}~\bibnamefont {Shaposhnikov}},\
  }\bibfield  {title} {\bibinfo {title} {The $\nu${MSM},dark matter and
  neutrino masses},\ }\href {https://doi.org/10.1016/j.physletb.2005.09.070}
  {\bibfield  {journal} {\bibinfo  {journal} {Phys. Lett. B}\ }\textbf
  {\bibinfo {volume} {631}},\ \bibinfo {pages} {151} (\bibinfo {year}
  {2005})}\BibitemShut {NoStop}%
\bibitem [{\citenamefont {Ma}(2006)}]{Ma_2006}%
  \BibitemOpen
  \bibfield  {author} {\bibinfo {author} {\bibfnamefont {E.}~\bibnamefont
  {Ma}},\ }\bibfield  {title} {\bibinfo {title} {{Verifiable radiative seesaw
  mechanism of neutrino mass and dark matter}},\ }\href
  {https://doi.org/10.1103/PhysRevD.73.077301} {\bibfield  {journal} {\bibinfo
  {journal} {Phys. Rev. D}\ }\textbf {\bibinfo {volume} {73}},\ \bibinfo
  {pages} {077301} (\bibinfo {year} {2006})}\BibitemShut {NoStop}%
\bibitem [{\citenamefont {Farzan}\ and\ \citenamefont
  {Ma}(2012)}]{Farzan_2012}%
  \BibitemOpen
  \bibfield  {author} {\bibinfo {author} {\bibfnamefont {Y.}~\bibnamefont
  {Farzan}}\ and\ \bibinfo {author} {\bibfnamefont {E.}~\bibnamefont {Ma}},\
  }\bibfield  {title} {\bibinfo {title} {Dirac neutrino mass generation from
  dark matter},\ }\href {https://doi.org/10.1103/PhysRevD.86.033007} {\bibfield
   {journal} {\bibinfo  {journal} {Phys. Rev. D}\ }\textbf {\bibinfo {volume}
  {86}},\ \bibinfo {pages} {033007} (\bibinfo {year} {2012})}\BibitemShut
  {NoStop}%
\bibitem [{\citenamefont {de~Gouv\^{e}a}(2016)}]{deGouvea_2016}%
  \BibitemOpen
  \bibfield  {author} {\bibinfo {author} {\bibfnamefont {A.}~\bibnamefont
  {de~Gouv\^{e}a}},\ }\bibfield  {title} {\bibinfo {title} {Neutrino mass
  models},\ }\href {https://doi.org/10.1146/annurev-nucl-102115-044600}
  {\bibfield  {journal} {\bibinfo  {journal} {Annu. Rev. Nucl. Part. Sci.}\
  }\textbf {\bibinfo {volume} {66}},\ \bibinfo {pages} {197} (\bibinfo {year}
  {2016})}\BibitemShut {NoStop}%
\bibitem [{\citenamefont {Escudero}\ \emph
  {et~al.}(2017{\natexlab{a}})\citenamefont {Escudero}, \citenamefont {Rius},\
  and\ \citenamefont {Sanz}}]{Escudero2017a}%
  \BibitemOpen
  \bibfield  {author} {\bibinfo {author} {\bibfnamefont {M.}~\bibnamefont
  {Escudero}}, \bibinfo {author} {\bibfnamefont {N.}~\bibnamefont {Rius}},\
  and\ \bibinfo {author} {\bibfnamefont {V.}~\bibnamefont {Sanz}},\ }\bibfield
  {title} {\bibinfo {title} {Sterile neutrino portal to dark matter {I}: {The}
  ${U}(1)_{B-L}$ case},\ }\href {https://doi.org/10.1007/JHEP02(2017)045}
  {\bibfield  {journal} {\bibinfo  {journal} {J. High Energy Phys}\ }\textbf
  {\bibinfo {volume} {2017}},\ \bibinfo {pages} {45} (\bibinfo {year}
  {2017}{\natexlab{a}})}\BibitemShut {NoStop}%
\bibitem [{\citenamefont {Escudero}\ \emph
  {et~al.}(2017{\natexlab{b}})\citenamefont {Escudero}, \citenamefont {Rius},\
  and\ \citenamefont {Sanz}}]{Escudero2017b}%
  \BibitemOpen
  \bibfield  {author} {\bibinfo {author} {\bibfnamefont {M.}~\bibnamefont
  {Escudero}}, \bibinfo {author} {\bibfnamefont {N.}~\bibnamefont {Rius}},\
  and\ \bibinfo {author} {\bibfnamefont {V.}~\bibnamefont {Sanz}},\ }\bibfield
  {title} {\bibinfo {title} {Sterile neutrino portal to dark matter {II}:
  {Exact} dark symmetry},\ }\href
  {https://doi.org/10.1140/epjc/s10052-017-4963-x} {\bibfield  {journal}
  {\bibinfo  {journal} {Eur. Phys. J.}\ }\textbf {\bibinfo {volume} {77}},\
  \bibinfo {pages} {397} (\bibinfo {year} {2017}{\natexlab{b}})}\BibitemShut
  {NoStop}%
\bibitem [{\citenamefont {Aguilar}\ \emph {et~al.}(2001)\citenamefont {Aguilar}
  \emph {et~al.}}]{LSND_anomaly}%
  \BibitemOpen
  \bibfield  {author} {\bibinfo {author} {\bibfnamefont {A.}~\bibnamefont
  {Aguilar}} \emph {et~al.} (\bibinfo {collaboration} {LSND Collaboration}),\
  }\bibfield  {title} {\bibinfo {title} {Evidence for neutrino oscillations
  from the observation of ${\overline{\nu}}_{e}$ appearance in a
  ${\overline{\nu}}_{\mu}$ beam},\ }\href
  {https://doi.org/10.1103/PhysRevD.64.112007} {\bibfield  {journal} {\bibinfo
  {journal} {Phys. Rev. D}\ }\textbf {\bibinfo {volume} {64}},\ \bibinfo
  {pages} {112007} (\bibinfo {year} {2001})}\BibitemShut {NoStop}%
\bibitem [{\citenamefont {Aguilar-Arevalo}\ \emph {et~al.}(2021)\citenamefont
  {Aguilar-Arevalo} \emph {et~al.}}]{MiniBooNE_anomaly}%
  \BibitemOpen
  \bibfield  {author} {\bibinfo {author} {\bibfnamefont {A.~A.}\ \bibnamefont
  {Aguilar-Arevalo}} \emph {et~al.} (\bibinfo {collaboration} {MiniBooNE
  Collaboration}),\ }\bibfield  {title} {\bibinfo {title} {Updated {MiniBooNE}
  neutrino oscillation results with increased data and new background
  studies},\ }\href {https://doi.org/10.1103/PhysRevD.103.052002} {\bibfield
  {journal} {\bibinfo  {journal} {Phys. Rev. D}\ }\textbf {\bibinfo {volume}
  {103}},\ \bibinfo {pages} {052002} (\bibinfo {year} {2021})}\BibitemShut
  {NoStop}%
\bibitem [{\citenamefont {Mangano}\ \emph {et~al.}(2006)\citenamefont
  {Mangano}, \citenamefont {Melchiorri}, \citenamefont {Serra}, \citenamefont
  {Cooray},\ and\ \citenamefont {Kamionkowski}}]{Mangano_2006}%
  \BibitemOpen
  \bibfield  {author} {\bibinfo {author} {\bibfnamefont {G.}~\bibnamefont
  {Mangano}}, \bibinfo {author} {\bibfnamefont {A.}~\bibnamefont {Melchiorri}},
  \bibinfo {author} {\bibfnamefont {P.}~\bibnamefont {Serra}}, \bibinfo
  {author} {\bibfnamefont {A.}~\bibnamefont {Cooray}},\ and\ \bibinfo {author}
  {\bibfnamefont {M.}~\bibnamefont {Kamionkowski}},\ }\bibfield  {title}
  {\bibinfo {title} {Cosmological bounds on dark-matter-neutrino
  interactions},\ }\bibfield  {journal} {\bibinfo  {journal} {Phys. Rev. D}\
  }\textbf {\bibinfo {volume} {74}},\ \href
  {https://doi.org/10.1103/physrevd.74.043517} {10.1103/physrevd.74.043517}
  (\bibinfo {year} {2006})\BibitemShut {NoStop}%
\bibitem [{\citenamefont {Choi}\ \emph {et~al.}(2019)\citenamefont {Choi},
  \citenamefont {Kim},\ and\ \citenamefont {Rott}}]{Choi_2019}%
  \BibitemOpen
  \bibfield  {author} {\bibinfo {author} {\bibfnamefont {K.-Y.}\ \bibnamefont
  {Choi}}, \bibinfo {author} {\bibfnamefont {J.}~\bibnamefont {Kim}},\ and\
  \bibinfo {author} {\bibfnamefont {C.}~\bibnamefont {Rott}},\ }\bibfield
  {title} {\bibinfo {title} {Constraining dark matter-neutrino interactions
  with {IceCube}-170922a},\ }\bibfield  {journal} {\bibinfo  {journal} {Phys.
  Rev. D}\ }\textbf {\bibinfo {volume} {99}},\ \href
  {https://doi.org/10.1103/physrevd.99.083018} {10.1103/physrevd.99.083018}
  (\bibinfo {year} {2019})\BibitemShut {NoStop}%
\bibitem [{\citenamefont {Cline}\ \emph {et~al.}(2023)\citenamefont {Cline},
  \citenamefont {Gao}, \citenamefont {Guo}, \citenamefont {Lin}, \citenamefont
  {Liu}, \citenamefont {Puel}, \citenamefont {Todd},\ and\ \citenamefont
  {Xiao}}]{Cline_2023}%
  \BibitemOpen
  \bibfield  {author} {\bibinfo {author} {\bibfnamefont {J.~M.}\ \bibnamefont
  {Cline}}, \bibinfo {author} {\bibfnamefont {S.}~\bibnamefont {Gao}}, \bibinfo
  {author} {\bibfnamefont {F.}~\bibnamefont {Guo}}, \bibinfo {author}
  {\bibfnamefont {Z.}~\bibnamefont {Lin}}, \bibinfo {author} {\bibfnamefont
  {S.}~\bibnamefont {Liu}}, \bibinfo {author} {\bibfnamefont {M.}~\bibnamefont
  {Puel}}, \bibinfo {author} {\bibfnamefont {P.}~\bibnamefont {Todd}},\ and\
  \bibinfo {author} {\bibfnamefont {T.}~\bibnamefont {Xiao}},\ }\bibfield
  {title} {\bibinfo {title} {Blazar constraints on neutrino-dark matter
  scattering},\ }\href {https://doi.org/10.1103/PhysRevLett.130.091402}
  {\bibfield  {journal} {\bibinfo  {journal} {Phys. Rev. Lett.}\ }\textbf
  {\bibinfo {volume} {130}},\ \bibinfo {pages} {091402} (\bibinfo {year}
  {2023})}\BibitemShut {NoStop}%
\bibitem [{\citenamefont {Cline}\ and\ \citenamefont
  {Puel}(2023)}]{Cline:2023tkp}%
  \BibitemOpen
  \bibfield  {author} {\bibinfo {author} {\bibfnamefont {J.~M.}\ \bibnamefont
  {Cline}}\ and\ \bibinfo {author} {\bibfnamefont {M.}~\bibnamefont {Puel}},\
  }\bibfield  {title} {\bibinfo {title} {Ngc 1068 constraints on neutrino-dark
  matter scattering},\ }\href {https://doi.org/10.1088/1475-7516/2023/06/004}
  {\bibfield  {journal} {\bibinfo  {journal} {J. Cosmol. Astropart. Phys.}\
  }\textbf {\bibinfo {volume} {2023}}\bibinfo  {number} { (06)},\ \bibinfo
  {pages} {004}}\BibitemShut {NoStop}%
\bibitem [{\citenamefont {Ferrer}\ \emph {et~al.}(2023)\citenamefont {Ferrer},
  \citenamefont {Herrera},\ and\ \citenamefont {Ibarra}}]{Ferrer_2023}%
  \BibitemOpen
\bibfield  {number} {  }\bibfield  {author} {\bibinfo {author} {\bibfnamefont
  {F.}~\bibnamefont {Ferrer}}, \bibinfo {author} {\bibfnamefont
  {G.}~\bibnamefont {Herrera}},\ and\ \bibinfo {author} {\bibfnamefont
  {A.}~\bibnamefont {Ibarra}},\ }\bibfield  {title} {\bibinfo {title} {New
  constraints on the dark matter-neutrino and dark matter-photon scattering
  cross sections from {TXS} 0506+056},\ }\href
  {https://doi.org/10.1088/1475-7516/2023/05/057} {\bibfield  {journal}
  {\bibinfo  {journal} {J. Cosmol. Astropart. Phys.}\ }\textbf {\bibinfo
  {volume} {2023}}\bibinfo  {number} { (05)},\ \bibinfo {pages}
  {057}}\BibitemShut {NoStop}%
\bibitem [{\citenamefont {Fujiwara}\ and\ \citenamefont
  {Herrera}(2024)}]{Fujiwara_2023tidal}%
  \BibitemOpen
\bibfield  {number} {  }\bibfield  {author} {\bibinfo {author} {\bibfnamefont
  {M.}~\bibnamefont {Fujiwara}}\ and\ \bibinfo {author} {\bibfnamefont
  {G.}~\bibnamefont {Herrera}},\ }\bibfield  {title} {\bibinfo {title} {{Tidal
  disruption events and dark matter scatterings with neutrinos and photons}},\
  }\href {https://doi.org/10.1016/j.physletb.2024.138573} {\bibfield  {journal}
  {\bibinfo  {journal} {Phys. Lett. B}\ }\textbf {\bibinfo {volume} {851}},\
  \bibinfo {pages} {138573} (\bibinfo {year} {2024})}\BibitemShut {NoStop}%
\bibitem [{\citenamefont {Wilkinson}\ \emph {et~al.}(2014)\citenamefont
  {Wilkinson}, \citenamefont {B{\oe}hm},\ and\ \citenamefont
  {Lesgourgues}}]{Wilkinson_2014}%
  \BibitemOpen
  \bibfield  {author} {\bibinfo {author} {\bibfnamefont {R.~J.}\ \bibnamefont
  {Wilkinson}}, \bibinfo {author} {\bibfnamefont {C.}~\bibnamefont
  {B{\oe}hm}},\ and\ \bibinfo {author} {\bibfnamefont {J.}~\bibnamefont
  {Lesgourgues}},\ }\bibfield  {title} {\bibinfo {title} {Constraining dark
  matter-neutrino interactions using the {CMB} and large-scale structure},\
  }\href {https://doi.org/10.1088/1475-7516/2014/05/011} {\bibfield  {journal}
  {\bibinfo  {journal} {J. Cosmol. Astropart. Phys.}\ }\textbf {\bibinfo
  {volume} {2014}}\bibinfo  {number} { (05)},\ \bibinfo {pages}
  {011}}\BibitemShut {NoStop}%
\bibitem [{\citenamefont {Akita}\ and\ \citenamefont
  {Ando}(2023)}]{Akita2023constraints}%
  \BibitemOpen
\bibfield  {number} {  }\bibfield  {author} {\bibinfo {author} {\bibfnamefont
  {K.}~\bibnamefont {Akita}}\ and\ \bibinfo {author} {\bibfnamefont
  {S.}~\bibnamefont {Ando}},\ }\bibfield  {title} {\bibinfo {title}
  {Constraints on dark matter-neutrino scattering from the milky-way satellites
  and subhalo modeling for dark acoustic oscillations},\ }\href
  {https://doi.org/10.1088/1475-7516/2023/11/037} {\bibfield  {journal}
  {\bibinfo  {journal} {J. Cosmol. Astropart. Phys.}\ }\textbf {\bibinfo
  {volume} {2023}}\bibinfo  {number} { (11)},\ \bibinfo {pages}
  {037}}\BibitemShut {NoStop}%
\bibitem [{\citenamefont {Escudero}\ \emph {et~al.}(2018)\citenamefont
  {Escudero}, \citenamefont {Lopez-Honorez}, \citenamefont {Mena},
  \citenamefont {Palomares-Ruiz},\ and\ \citenamefont
  {Villanueva-Domingo}}]{Escudero_2018}%
  \BibitemOpen
\bibfield  {number} {  }\bibfield  {author} {\bibinfo {author} {\bibfnamefont
  {M.}~\bibnamefont {Escudero}}, \bibinfo {author} {\bibfnamefont
  {L.}~\bibnamefont {Lopez-Honorez}}, \bibinfo {author} {\bibfnamefont
  {O.}~\bibnamefont {Mena}}, \bibinfo {author} {\bibfnamefont {S.}~\bibnamefont
  {Palomares-Ruiz}},\ and\ \bibinfo {author} {\bibfnamefont {P.}~\bibnamefont
  {Villanueva-Domingo}},\ }\bibfield  {title} {\bibinfo {title} {A fresh look
  into the interacting dark matter scenario},\ }\href
  {https://doi.org/10.1088/1475-7516/2018/06/007} {\bibfield  {journal}
  {\bibinfo  {journal} {J. Cosmol. Astropart. Phys.}\ }\textbf {\bibinfo
  {volume} {2018}}\bibinfo  {number} { (06)},\ \bibinfo {pages}
  {007}}\BibitemShut {NoStop}%
\bibitem [{\citenamefont {B{\oe}hm}\ \emph {et~al.}(2013)\citenamefont
  {B{\oe}hm}, \citenamefont {Dolan},\ and\ \citenamefont
  {McCabe}}]{Boehm_2013}%
  \BibitemOpen
\bibfield  {number} {  }\bibfield  {author} {\bibinfo {author} {\bibfnamefont
  {C.}~\bibnamefont {B{\oe}hm}}, \bibinfo {author} {\bibfnamefont {M.~J.}\
  \bibnamefont {Dolan}},\ and\ \bibinfo {author} {\bibfnamefont
  {C.}~\bibnamefont {McCabe}},\ }\bibfield  {title} {\bibinfo {title} {A lower
  bound on the mass of cold thermal dark matter from planck},\ }\href
  {https://doi.org/10.1088/1475-7516/2013/08/041} {\bibfield  {journal}
  {\bibinfo  {journal} {J. Cosmol. Astropart. Phys.}\ }\textbf {\bibinfo
  {volume} {2013}}\bibinfo  {number} { (08)},\ \bibinfo {pages}
  {041}}\BibitemShut {NoStop}%
\bibitem [{\citenamefont {Mosbech}\ \emph {et~al.}(2021)\citenamefont
  {Mosbech}, \citenamefont {Boehm}, \citenamefont {Hannestad}, \citenamefont
  {Mena}, \citenamefont {Stadler},\ and\ \citenamefont {Wong}}]{Mosbech_2021}%
  \BibitemOpen
\bibfield  {number} {  }\bibfield  {author} {\bibinfo {author} {\bibfnamefont
  {M.~R.}\ \bibnamefont {Mosbech}}, \bibinfo {author} {\bibfnamefont
  {C.}~\bibnamefont {Boehm}}, \bibinfo {author} {\bibfnamefont
  {S.}~\bibnamefont {Hannestad}}, \bibinfo {author} {\bibfnamefont
  {O.}~\bibnamefont {Mena}}, \bibinfo {author} {\bibfnamefont {J.}~\bibnamefont
  {Stadler}},\ and\ \bibinfo {author} {\bibfnamefont {Y.~Y.}\ \bibnamefont
  {Wong}},\ }\bibfield  {title} {\bibinfo {title} {The full {Boltzmann}
  hierarchy for dark matter-massive neutrino interactions},\ }\href
  {https://doi.org/10.1088/1475-7516/2021/03/066} {\bibfield  {journal}
  {\bibinfo  {journal} {J. Cosmol. Astropart. Phys.}\ }\textbf {\bibinfo
  {volume} {2021}}\bibinfo  {number} { (03)},\ \bibinfo {pages}
  {066}}\BibitemShut {NoStop}%
\bibitem [{\citenamefont {Hooper}\ and\ \citenamefont
  {Lucca}(2022)}]{Hooper_2022}%
  \BibitemOpen
\bibfield  {number} {  }\bibfield  {author} {\bibinfo {author} {\bibfnamefont
  {D.~C.}\ \bibnamefont {Hooper}}\ and\ \bibinfo {author} {\bibfnamefont
  {M.}~\bibnamefont {Lucca}},\ }\bibfield  {title} {\bibinfo {title} {Hints of
  dark matter-neutrino interactions in {Lyman}-$\alpha$ data},\ }\bibfield
  {journal} {\bibinfo  {journal} {Phys. Rev. D}\ }\textbf {\bibinfo {volume}
  {105}},\ \href {https://doi.org/10.1103/physrevd.105.103504}
  {10.1103/physrevd.105.103504} (\bibinfo {year} {2022})\BibitemShut {NoStop}%
\bibitem [{\citenamefont {Mosbech}\ \emph {et~al.}(2023)\citenamefont
  {Mosbech}, \citenamefont {Boehm},\ and\ \citenamefont {Wong}}]{Mosbech_2023}%
  \BibitemOpen
  \bibfield  {author} {\bibinfo {author} {\bibfnamefont {M.~R.}\ \bibnamefont
  {Mosbech}}, \bibinfo {author} {\bibfnamefont {C.}~\bibnamefont {Boehm}},\
  and\ \bibinfo {author} {\bibfnamefont {Y.~Y.}\ \bibnamefont {Wong}},\
  }\bibfield  {title} {\bibinfo {title} {Probing dark matter interactions with
  21cm observations},\ }\href {https://doi.org/10.1088/1475-7516/2023/03/047}
  {\bibfield  {journal} {\bibinfo  {journal} {J. Cosmol. Astropart. Phys.}\
  }\textbf {\bibinfo {volume} {2023}}\bibinfo  {number} { (03)},\ \bibinfo
  {pages} {047}}\BibitemShut {NoStop}%
\bibitem [{\citenamefont {Fayet}\ \emph {et~al.}(2006)\citenamefont {Fayet},
  \citenamefont {Hooper},\ and\ \citenamefont {Sigl}}]{Fayet_2006}%
  \BibitemOpen
\bibfield  {number} {  }\bibfield  {author} {\bibinfo {author} {\bibfnamefont
  {P.}~\bibnamefont {Fayet}}, \bibinfo {author} {\bibfnamefont
  {D.}~\bibnamefont {Hooper}},\ and\ \bibinfo {author} {\bibfnamefont
  {G.}~\bibnamefont {Sigl}},\ }\bibfield  {title} {\bibinfo {title}
  {Constraints on light dark matter from core-collapse supernovae},\ }\href
  {https://doi.org/10.1103/PhysRevLett.96.211302} {\bibfield  {journal}
  {\bibinfo  {journal} {Phys. Rev. Lett.}\ }\textbf {\bibinfo {volume} {96}},\
  \bibinfo {pages} {211302} (\bibinfo {year} {2006})}\BibitemShut {NoStop}%
\bibitem [{\citenamefont {Koren}(2019)}]{Koren_2019}%
  \BibitemOpen
  \bibfield  {author} {\bibinfo {author} {\bibfnamefont {S.}~\bibnamefont
  {Koren}},\ }\bibfield  {title} {\bibinfo {title} {Neutrino-dark matter
  scattering and coincident detections of {UHE} neutrinos with {EM} sources},\
  }\href {https://doi.org/10.1088/1475-7516/2019/09/013} {\bibfield  {journal}
  {\bibinfo  {journal} {J. Cosmol. Astropart. Phys.}\ }\textbf {\bibinfo
  {volume} {2019}}\bibinfo  {number} { (09)},\ \bibinfo {pages}
  {013}}\BibitemShut {NoStop}%
\bibitem [{\citenamefont {Murase}\ and\ \citenamefont
  {Shoemaker}(2019)}]{Murase_2019}%
  \BibitemOpen
\bibfield  {number} {  }\bibfield  {author} {\bibinfo {author} {\bibfnamefont
  {K.}~\bibnamefont {Murase}}\ and\ \bibinfo {author} {\bibfnamefont {I.~M.}\
  \bibnamefont {Shoemaker}},\ }\bibfield  {title} {\bibinfo {title} {Neutrino
  echoes from multimessenger transient sources},\ }\bibfield  {journal}
  {\bibinfo  {journal} {Phys. Rev. Lett.}\ }\textbf {\bibinfo {volume} {123}},\
  \href {https://doi.org/10.1103/physrevlett.123.241102}
  {10.1103/physrevlett.123.241102} (\bibinfo {year} {2019})\BibitemShut
  {NoStop}%
\bibitem [{\citenamefont {McMullen}\ \emph {et~al.}(2021)\citenamefont
  {McMullen}, \citenamefont {Vincent}, \citenamefont {Arguelles},\ and\
  \citenamefont {Schneider}}]{McMullen_2021}%
  \BibitemOpen
  \bibfield  {author} {\bibinfo {author} {\bibfnamefont {A.}~\bibnamefont
  {McMullen}}, \bibinfo {author} {\bibfnamefont {A.}~\bibnamefont {Vincent}},
  \bibinfo {author} {\bibfnamefont {C.}~\bibnamefont {Arguelles}},\ and\
  \bibinfo {author} {\bibfnamefont {A.}~\bibnamefont {Schneider}} (\bibinfo
  {collaboration} {IceCube Collaboration}),\ }\bibfield  {title} {\bibinfo
  {title} {{Dark matter neutrino scattering in the galactic centre with
  IceCube}},\ }\href {https://doi.org/10.1088/1748-0221/16/08/C08001}
  {\bibfield  {journal} {\bibinfo  {journal} {J. Instrum.}\ }\textbf {\bibinfo
  {volume} {16}}\bibinfo  {number} { (08)},\ \bibinfo {pages}
  {C08001}}\BibitemShut {NoStop}%
\bibitem [{\citenamefont {Carpio}\ \emph {et~al.}(2023)\citenamefont {Carpio},
  \citenamefont {Kheirandish},\ and\ \citenamefont {Murase}}]{Carpio_2023}%
  \BibitemOpen
\bibfield  {number} {  }\bibfield  {author} {\bibinfo {author} {\bibfnamefont
  {J.~A.}\ \bibnamefont {Carpio}}, \bibinfo {author} {\bibfnamefont
  {A.}~\bibnamefont {Kheirandish}},\ and\ \bibinfo {author} {\bibfnamefont
  {K.}~\bibnamefont {Murase}},\ }\bibfield  {title} {\bibinfo {title}
  {Time-delayed neutrino emission from supernovae as a probe of dark
  matter-neutrino interactions},\ }\href
  {https://doi.org/10.1088/1475-7516/2023/04/019} {\bibfield  {journal}
  {\bibinfo  {journal} {J. Cosmol. Astropart. Phys.}\ }\textbf {\bibinfo
  {volume} {2023}}\bibinfo  {number} { (04)},\ \bibinfo {pages}
  {019}}\BibitemShut {NoStop}%
\bibitem [{\citenamefont {Farzan}\ and\ \citenamefont
  {Palomares-Ruiz}(2014)}]{Farzan_2014}%
  \BibitemOpen
\bibfield  {number} {  }\bibfield  {author} {\bibinfo {author} {\bibfnamefont
  {Y.}~\bibnamefont {Farzan}}\ and\ \bibinfo {author} {\bibfnamefont
  {S.}~\bibnamefont {Palomares-Ruiz}},\ }\bibfield  {title} {\bibinfo {title}
  {Dips in the diffuse supernova neutrino background},\ }\href
  {https://doi.org/10.1088/1475-7516/2014/06/014} {\bibfield  {journal}
  {\bibinfo  {journal} {J. Cosmol. Astropart. Phys.}\ }\textbf {\bibinfo
  {volume} {2014}}\bibinfo  {number} { (06)},\ \bibinfo {pages}
  {014}}\BibitemShut {NoStop}%
\bibitem [{\citenamefont {Das}\ and\ \citenamefont {Sen}(2021)}]{Das_2021}%
  \BibitemOpen
\bibfield  {number} {  }\bibfield  {author} {\bibinfo {author} {\bibfnamefont
  {A.}~\bibnamefont {Das}}\ and\ \bibinfo {author} {\bibfnamefont
  {M.}~\bibnamefont {Sen}},\ }\bibfield  {title} {\bibinfo {title} {Boosted
  dark matter from diffuse supernova neutrinos},\ }\href
  {https://doi.org/10.1103/PhysRevD.104.075029} {\bibfield  {journal} {\bibinfo
   {journal} {Phys. Rev. D}\ }\textbf {\bibinfo {volume} {104}},\ \bibinfo
  {pages} {075029} (\bibinfo {year} {2021})}\BibitemShut {NoStop}%
\bibitem [{\citenamefont {Chao}\ \emph {et~al.}(2021)\citenamefont {Chao},
  \citenamefont {Li},\ and\ \citenamefont {Liao}}]{Chao_2021}%
  \BibitemOpen
  \bibfield  {author} {\bibinfo {author} {\bibfnamefont {W.}~\bibnamefont
  {Chao}}, \bibinfo {author} {\bibfnamefont {T.}~\bibnamefont {Li}},\ and\
  \bibinfo {author} {\bibfnamefont {J.}~\bibnamefont {Liao}},\ }\href@noop {}
  {\bibinfo {title} {Connecting primordial black hole to boosted sub-gev dark
  matter through neutrino}} (\bibinfo {year} {2021}),\ \Eprint
  {https://arxiv.org/abs/2108.05608} {arXiv:2108.05608 [hep-ph]} \BibitemShut
  {NoStop}%
\bibitem [{\citenamefont {Jho}\ \emph {et~al.}(2021)\citenamefont {Jho},
  \citenamefont {Park}, \citenamefont {Park},\ and\ \citenamefont
  {Tseng}}]{Jho_2021}%
  \BibitemOpen
  \bibfield  {author} {\bibinfo {author} {\bibfnamefont {Y.}~\bibnamefont
  {Jho}}, \bibinfo {author} {\bibfnamefont {J.-C.}\ \bibnamefont {Park}},
  \bibinfo {author} {\bibfnamefont {S.~C.}\ \bibnamefont {Park}},\ and\
  \bibinfo {author} {\bibfnamefont {P.-Y.}\ \bibnamefont {Tseng}},\ }\href@noop
  {} {\bibinfo {title} {Cosmic-neutrino-boosted dark matter ($\nu$bdm)}}
  (\bibinfo {year} {2021}),\ \Eprint {https://arxiv.org/abs/2101.11262}
  {arXiv:2101.11262 [hep-ph]} \BibitemShut {NoStop}%
\bibitem [{\citenamefont {Zhang}(2021)}]{Zhang_2021}%
  \BibitemOpen
  \bibfield  {author} {\bibinfo {author} {\bibfnamefont {Y.}~\bibnamefont
  {Zhang}},\ }\bibfield  {title} {\bibinfo {title} {Speeding up dark matter
  with solar neutrinos},\ }\bibfield  {journal} {\bibinfo  {journal} {Progress
  of Theoretical and Experimental Physics}\ }\textbf {\bibinfo {volume}
  {2022}},\ \href {https://doi.org/10.1093/ptep/ptab156} {10.1093/ptep/ptab156}
  (\bibinfo {year} {2021})\BibitemShut {NoStop}%
\bibitem [{\citenamefont {Ghosh}\ \emph {et~al.}(2022)\citenamefont {Ghosh},
  \citenamefont {Guha},\ and\ \citenamefont {Sachdeva}}]{Ghosh_2022}%
  \BibitemOpen
  \bibfield  {author} {\bibinfo {author} {\bibfnamefont {D.}~\bibnamefont
  {Ghosh}}, \bibinfo {author} {\bibfnamefont {A.}~\bibnamefont {Guha}},\ and\
  \bibinfo {author} {\bibfnamefont {D.}~\bibnamefont {Sachdeva}},\ }\bibfield
  {title} {\bibinfo {title} {Exclusion limits on dark matter-neutrino
  scattering cross section},\ }\bibfield  {journal} {\bibinfo  {journal} {Phys.
  Rev. D}\ }\textbf {\bibinfo {volume} {105}},\ \href
  {https://doi.org/10.1103/physrevd.105.103029} {10.1103/physrevd.105.103029}
  (\bibinfo {year} {2022})\BibitemShut {NoStop}%
\bibitem [{\citenamefont {Bardhan}\ \emph {et~al.}(2023)\citenamefont
  {Bardhan}, \citenamefont {Bhowmick}, \citenamefont {Ghosh}, \citenamefont
  {Guha},\ and\ \citenamefont {Sachdeva}}]{Bardhan_2023}%
  \BibitemOpen
  \bibfield  {author} {\bibinfo {author} {\bibfnamefont {D.}~\bibnamefont
  {Bardhan}}, \bibinfo {author} {\bibfnamefont {S.}~\bibnamefont {Bhowmick}},
  \bibinfo {author} {\bibfnamefont {D.}~\bibnamefont {Ghosh}}, \bibinfo
  {author} {\bibfnamefont {A.}~\bibnamefont {Guha}},\ and\ \bibinfo {author}
  {\bibfnamefont {D.}~\bibnamefont {Sachdeva}},\ }\bibfield  {title} {\bibinfo
  {title} {Bounds on boosted dark matter from direct detection: The role of
  energy-dependent cross sections},\ }\bibfield  {journal} {\bibinfo  {journal}
  {Phys. Rev. D}\ }\textbf {\bibinfo {volume} {107}},\ \href
  {https://doi.org/10.1103/physrevd.107.015010} {10.1103/physrevd.107.015010}
  (\bibinfo {year} {2023})\BibitemShut {NoStop}%
\bibitem [{\citenamefont {{De Romeri}}\ \emph {et~al.}(2023)\citenamefont {{De
  Romeri}}, \citenamefont {Majumdar}, \citenamefont {Papoulias},\ and\
  \citenamefont {Srivastava}}]{DeRomeri_2023}%
  \BibitemOpen
  \bibfield  {author} {\bibinfo {author} {\bibfnamefont {V.}~\bibnamefont {{De
  Romeri}}}, \bibinfo {author} {\bibfnamefont {A.}~\bibnamefont {Majumdar}},
  \bibinfo {author} {\bibfnamefont {D.~K.}\ \bibnamefont {Papoulias}},\ and\
  \bibinfo {author} {\bibfnamefont {R.}~\bibnamefont {Srivastava}},\
  }\href@noop {} {\bibinfo {title} {{XENONnT} and {LUX-ZEPLIN} constraints on
  {DSNB}-boosted dark matter}} (\bibinfo {year} {2023}),\ \Eprint
  {https://arxiv.org/abs/2309.04117} {arXiv:2309.04117 [hep-ph]} \BibitemShut
  {NoStop}%
\bibitem [{\citenamefont {Lin}\ \emph {et~al.}(2023)\citenamefont {Lin},
  \citenamefont {Wu}, \citenamefont {Wu},\ and\ \citenamefont
  {Wong}}]{Lin_2023}%
  \BibitemOpen
  \bibfield  {author} {\bibinfo {author} {\bibfnamefont {Y.-H.}\ \bibnamefont
  {Lin}}, \bibinfo {author} {\bibfnamefont {W.-H.}\ \bibnamefont {Wu}},
  \bibinfo {author} {\bibfnamefont {M.-R.}\ \bibnamefont {Wu}},\ and\ \bibinfo
  {author} {\bibfnamefont {H.~T.-K.}\ \bibnamefont {Wong}},\ }\bibfield
  {title} {\bibinfo {title} {Searching for afterglow: Light dark matter boosted
  by supernova neutrinos},\ }\bibfield  {journal} {\bibinfo  {journal} {Phys.
  Rev. Lett.}\ }\textbf {\bibinfo {volume} {130}},\ \href
  {https://doi.org/10.1103/physrevlett.130.111002}
  {10.1103/physrevlett.130.111002} (\bibinfo {year} {2023})\BibitemShut
  {NoStop}%
\bibitem [{\citenamefont {Bullock}\ and\ \citenamefont
  {Boylan-Kolchin}(2017)}]{Bullock_2017}%
  \BibitemOpen
  \bibfield  {author} {\bibinfo {author} {\bibfnamefont {J.~S.}\ \bibnamefont
  {Bullock}}\ and\ \bibinfo {author} {\bibfnamefont {M.}~\bibnamefont
  {Boylan-Kolchin}},\ }\bibfield  {title} {\bibinfo {title} {Small-scale
  challenges to the {$\Lambda$CDM} paradigm},\ }\href
  {https://doi.org/10.1146/annurev-astro-091916-055313} {\bibfield  {journal}
  {\bibinfo  {journal} {Annual Review of Astronomy and Astrophysics}\ }\textbf
  {\bibinfo {volume} {55}},\ \bibinfo {pages} {343} (\bibinfo {year}
  {2017})}\BibitemShut {NoStop}%
\bibitem [{\citenamefont {Salucci}(2019)}]{Salucci:2018hqu}%
  \BibitemOpen
  \bibfield  {author} {\bibinfo {author} {\bibfnamefont {P.}~\bibnamefont
  {Salucci}},\ }\bibfield  {title} {\bibinfo {title} {{The distribution of dark
  matter in galaxies}},\ }\href {https://doi.org/10.1007/s00159-018-0113-1}
  {\bibfield  {journal} {\bibinfo  {journal} {Astron. Astrophys. Rev.}\
  }\textbf {\bibinfo {volume} {27}},\ \bibinfo {pages} {2} (\bibinfo {year}
  {2019})}\BibitemShut {NoStop}%
\bibitem [{\citenamefont {{Navarro}}\ \emph {et~al.}(1996)\citenamefont
  {{Navarro}}, \citenamefont {{Frenk}},\ and\ \citenamefont {{White}}}]{NFW}%
  \BibitemOpen
  \bibfield  {author} {\bibinfo {author} {\bibfnamefont {J.~F.}\ \bibnamefont
  {{Navarro}}}, \bibinfo {author} {\bibfnamefont {C.~S.}\ \bibnamefont
  {{Frenk}}},\ and\ \bibinfo {author} {\bibfnamefont {S.~D.~M.}\ \bibnamefont
  {{White}}},\ }\bibfield  {title} {\bibinfo {title} {{The Structure of Cold
  Dark Matter Halos}},\ }\href {https://doi.org/10.1086/177173} {\bibfield
  {journal} {\bibinfo  {journal} {Astrophys. J.}\ }\textbf {\bibinfo {volume}
  {462}},\ \bibinfo {pages} {563} (\bibinfo {year} {1996})}\BibitemShut
  {NoStop}%
\bibitem [{\citenamefont {Wang}\ \emph {et~al.}(2020)\citenamefont {Wang},
  \citenamefont {Bose}, \citenamefont {Frenk}, \citenamefont {Gao},
  \citenamefont {Jenkins}, \citenamefont {Springel},\ and\ \citenamefont
  {White}}]{Wang_2020}%
  \BibitemOpen
  \bibfield  {author} {\bibinfo {author} {\bibfnamefont {J.}~\bibnamefont
  {Wang}}, \bibinfo {author} {\bibfnamefont {S.}~\bibnamefont {Bose}}, \bibinfo
  {author} {\bibfnamefont {C.~S.}\ \bibnamefont {Frenk}}, \bibinfo {author}
  {\bibfnamefont {L.}~\bibnamefont {Gao}}, \bibinfo {author} {\bibfnamefont
  {A.}~\bibnamefont {Jenkins}}, \bibinfo {author} {\bibfnamefont
  {V.}~\bibnamefont {Springel}},\ and\ \bibinfo {author} {\bibfnamefont
  {S.~D.~M.}\ \bibnamefont {White}},\ }\bibfield  {title} {\bibinfo {title}
  {Universal structure of dark matter haloes over a mass range of 20 orders of
  magnitude},\ }\href {https://doi.org/10.1038/s41586-020-2642-9} {\bibfield
  {journal} {\bibinfo  {journal} {Nature}\ }\textbf {\bibinfo {volume} {585}},\
  \bibinfo {pages} {39} (\bibinfo {year} {2020})}\BibitemShut {NoStop}%
\bibitem [{\citenamefont {{Flores}}\ and\ \citenamefont
  {{Primack}}(1994)}]{Flores_1994}%
  \BibitemOpen
  \bibfield  {author} {\bibinfo {author} {\bibfnamefont {R.~A.}\ \bibnamefont
  {{Flores}}}\ and\ \bibinfo {author} {\bibfnamefont {J.~R.}\ \bibnamefont
  {{Primack}}},\ }\bibfield  {title} {\bibinfo {title} {{Observational and
  theoretical constraints on singular dark matter halos}},\ }\href
  {https://doi.org/10.1086/187350} {\bibfield  {journal} {\bibinfo  {journal}
  {Astrophys. J. Lett.}\ }\textbf {\bibinfo {volume} {427}},\ \bibinfo {pages}
  {L1} (\bibinfo {year} {1994})}\BibitemShut {NoStop}%
\bibitem [{\citenamefont {Moore}(1994)}]{Moore_1994}%
  \BibitemOpen
  \bibfield  {author} {\bibinfo {author} {\bibfnamefont {B.}~\bibnamefont
  {Moore}},\ }\bibfield  {title} {\bibinfo {title} {Evidence against
  dissipation-less dark matter from observations of galaxy haloes},\ }\href
  {https://doi.org/10.1038/370629a0} {\bibfield  {journal} {\bibinfo  {journal}
  {Nature}\ }\textbf {\bibinfo {volume} {370}},\ \bibinfo {pages} {629}
  (\bibinfo {year} {1994})}\BibitemShut {NoStop}%
\bibitem [{\citenamefont {Spergel}\ and\ \citenamefont
  {Steinhardt}(2000)}]{Spergel_2000}%
  \BibitemOpen
  \bibfield  {author} {\bibinfo {author} {\bibfnamefont {D.~N.}\ \bibnamefont
  {Spergel}}\ and\ \bibinfo {author} {\bibfnamefont {P.~J.}\ \bibnamefont
  {Steinhardt}},\ }\bibfield  {title} {\bibinfo {title} {Observational evidence
  for self-interacting cold dark matter},\ }\href
  {https://doi.org/10.1103/physrevlett.84.3760} {\bibfield  {journal} {\bibinfo
   {journal} {Phys. Rev. Lett.}\ }\textbf {\bibinfo {volume} {84}},\ \bibinfo
  {pages} {3760} (\bibinfo {year} {2000})}\BibitemShut {NoStop}%
\bibitem [{\citenamefont {Oh}\ \emph {et~al.}(2008)\citenamefont {Oh},
  \citenamefont {de~Blok}, \citenamefont {Walter}, \citenamefont {Brinks},\
  and\ \citenamefont {Kennicutt}}]{Oh_2008}%
  \BibitemOpen
  \bibfield  {author} {\bibinfo {author} {\bibfnamefont {S.-H.}\ \bibnamefont
  {Oh}}, \bibinfo {author} {\bibfnamefont {W.~J.~G.}\ \bibnamefont {de~Blok}},
  \bibinfo {author} {\bibfnamefont {F.}~\bibnamefont {Walter}}, \bibinfo
  {author} {\bibfnamefont {E.}~\bibnamefont {Brinks}},\ and\ \bibinfo {author}
  {\bibfnamefont {R.~C.}\ \bibnamefont {Kennicutt}},\ }\bibfield  {title}
  {\bibinfo {title} {High-resolution dark matter density profiles of {THINGS}
  dwarf galaxies: Correcting for noncircular motions},\ }\href
  {https://doi.org/10.1088/0004-6256/136/6/2761} {\bibfield  {journal}
  {\bibinfo  {journal} {Astron. J.}\ }\textbf {\bibinfo {volume} {136}},\
  \bibinfo {pages} {2761} (\bibinfo {year} {2008})}\BibitemShut {NoStop}%
\bibitem [{\citenamefont {Donato}\ \emph {et~al.}(2009)\citenamefont {Donato},
  \citenamefont {Gentile}, \citenamefont {Salucci}, \citenamefont
  {Frigerio~Martins}, \citenamefont {Wilkinson}, \citenamefont {Gilmore},
  \citenamefont {Grebel}, \citenamefont {Koch},\ and\ \citenamefont
  {Wyse}}]{Donato_2009}%
  \BibitemOpen
  \bibfield  {author} {\bibinfo {author} {\bibfnamefont {F.}~\bibnamefont
  {Donato}}, \bibinfo {author} {\bibfnamefont {G.}~\bibnamefont {Gentile}},
  \bibinfo {author} {\bibfnamefont {P.}~\bibnamefont {Salucci}}, \bibinfo
  {author} {\bibfnamefont {C.}~\bibnamefont {Frigerio~Martins}}, \bibinfo
  {author} {\bibfnamefont {M.~I.}\ \bibnamefont {Wilkinson}}, \bibinfo {author}
  {\bibfnamefont {G.}~\bibnamefont {Gilmore}}, \bibinfo {author} {\bibfnamefont
  {E.~K.}\ \bibnamefont {Grebel}}, \bibinfo {author} {\bibfnamefont
  {A.}~\bibnamefont {Koch}},\ and\ \bibinfo {author} {\bibfnamefont
  {R.}~\bibnamefont {Wyse}},\ }\bibfield  {title} {\bibinfo {title} {{A
  constant dark matter halo surface density in galaxies}},\ }\href
  {https://doi.org/10.1111/j.1365-2966.2009.15004.x} {\bibfield  {journal}
  {\bibinfo  {journal} {Mon. Not. R. Astron. Soc.}\ }\textbf {\bibinfo {volume}
  {397}},\ \bibinfo {pages} {1169} (\bibinfo {year} {2009})}\BibitemShut
  {NoStop}%
\bibitem [{\citenamefont {Walker}\ and\ \citenamefont
  {Peñarrubia}(2011)}]{Walker_2011}%
  \BibitemOpen
  \bibfield  {author} {\bibinfo {author} {\bibfnamefont {M.~G.}\ \bibnamefont
  {Walker}}\ and\ \bibinfo {author} {\bibfnamefont {J.}~\bibnamefont
  {Peñarrubia}},\ }\bibfield  {title} {\bibinfo {title} {A method for
  measuring (slopes of) the mass profiles of dwarf spheroidal galaxies},\
  }\href {https://doi.org/10.1088/0004-637X/742/1/20} {\bibfield  {journal}
  {\bibinfo  {journal} {Astrophys. J.}\ }\textbf {\bibinfo {volume} {742}},\
  \bibinfo {pages} {20} (\bibinfo {year} {2011})}\BibitemShut {NoStop}%
\bibitem [{\citenamefont {Salucci}\ \emph {et~al.}(2012)\citenamefont
  {Salucci}, \citenamefont {Wilkinson}, \citenamefont {Walker}, \citenamefont
  {Gilmore}, \citenamefont {Grebel}, \citenamefont {Koch}, \citenamefont
  {Martins},\ and\ \citenamefont {Wyse}}]{Salucci_2012}%
  \BibitemOpen
  \bibfield  {author} {\bibinfo {author} {\bibfnamefont {P.}~\bibnamefont
  {Salucci}}, \bibinfo {author} {\bibfnamefont {M.~I.}\ \bibnamefont
  {Wilkinson}}, \bibinfo {author} {\bibfnamefont {M.~G.}\ \bibnamefont
  {Walker}}, \bibinfo {author} {\bibfnamefont {G.~F.}\ \bibnamefont {Gilmore}},
  \bibinfo {author} {\bibfnamefont {E.~K.}\ \bibnamefont {Grebel}}, \bibinfo
  {author} {\bibfnamefont {A.}~\bibnamefont {Koch}}, \bibinfo {author}
  {\bibfnamefont {C.~F.}\ \bibnamefont {Martins}},\ and\ \bibinfo {author}
  {\bibfnamefont {R.~F.~G.}\ \bibnamefont {Wyse}},\ }\bibfield  {title}
  {\bibinfo {title} {Dwarf spheroidal galaxy kinematics and spiral galaxy
  scaling laws},\ }\href {https://doi.org/10.1111/j.1365-2966.2011.20144.x}
  {\bibfield  {journal} {\bibinfo  {journal} {Mon. Not. R. Astron. Soc.}\
  }\textbf {\bibinfo {volume} {420}},\ \bibinfo {pages} {2034} (\bibinfo {year}
  {2012})}\BibitemShut {NoStop}%
\bibitem [{\citenamefont {Oh}\ \emph {et~al.}(2015)\citenamefont {Oh} \emph
  {et~al.}}]{Oh_2015}%
  \BibitemOpen
  \bibfield  {author} {\bibinfo {author} {\bibfnamefont {S.-H.}\ \bibnamefont
  {Oh}} \emph {et~al.},\ }\bibfield  {title} {\bibinfo {title} {High-resolution
  mass models of dwarf galaxies from {LITTLE} {THINGS}},\ }\href
  {https://doi.org/10.1088/0004-6256/149/6/180} {\bibfield  {journal} {\bibinfo
   {journal} {Astron. J.}\ }\textbf {\bibinfo {volume} {149}},\ \bibinfo
  {pages} {180} (\bibinfo {year} {2015})}\BibitemShut {NoStop}%
\bibitem [{\citenamefont {Navarro}\ \emph {et~al.}(1996)\citenamefont
  {Navarro}, \citenamefont {Eke},\ and\ \citenamefont {Frenk}}]{Navarro_1996}%
  \BibitemOpen
  \bibfield  {author} {\bibinfo {author} {\bibfnamefont {J.~F.}\ \bibnamefont
  {Navarro}}, \bibinfo {author} {\bibfnamefont {V.~R.}\ \bibnamefont {Eke}},\
  and\ \bibinfo {author} {\bibfnamefont {C.~S.}\ \bibnamefont {Frenk}},\
  }\bibfield  {title} {\bibinfo {title} {{The cores of dwarf galaxy haloes}},\
  }\href {https://doi.org/10.1093/mnras/283.3.L72} {\bibfield  {journal}
  {\bibinfo  {journal} {Mon. Not. R. Astron. Soc.}\ }\textbf {\bibinfo {volume}
  {283}},\ \bibinfo {pages} {L72} (\bibinfo {year} {1996})}\BibitemShut
  {NoStop}%
\bibitem [{\citenamefont {Peñarrubia}\ \emph {et~al.}(2012)\citenamefont
  {Peñarrubia}, \citenamefont {Pontzen}, \citenamefont {Walker},\ and\
  \citenamefont {Koposov}}]{Penarrubia_2012}%
  \BibitemOpen
  \bibfield  {author} {\bibinfo {author} {\bibfnamefont {J.}~\bibnamefont
  {Peñarrubia}}, \bibinfo {author} {\bibfnamefont {A.}~\bibnamefont
  {Pontzen}}, \bibinfo {author} {\bibfnamefont {M.~G.}\ \bibnamefont
  {Walker}},\ and\ \bibinfo {author} {\bibfnamefont {S.~E.}\ \bibnamefont
  {Koposov}},\ }\bibfield  {title} {\bibinfo {title} {The coupling between the
  core/cusp and missing satellite problems},\ }\href
  {https://doi.org/10.1088/2041-8205/759/2/L42} {\bibfield  {journal} {\bibinfo
   {journal} {Astrophys. J. Lett.}\ }\textbf {\bibinfo {volume} {759}},\
  \bibinfo {pages} {L42} (\bibinfo {year} {2012})}\BibitemShut {NoStop}%
\bibitem [{\citenamefont {Pontzen}\ and\ \citenamefont
  {Governato}(2012)}]{Pontzen_2012}%
  \BibitemOpen
  \bibfield  {author} {\bibinfo {author} {\bibfnamefont {A.}~\bibnamefont
  {Pontzen}}\ and\ \bibinfo {author} {\bibfnamefont {F.}~\bibnamefont
  {Governato}},\ }\bibfield  {title} {\bibinfo {title} {{How supernova feedback
  turns dark matter cusps into cores}},\ }\href
  {https://doi.org/10.1111/j.1365-2966.2012.20571.x} {\bibfield  {journal}
  {\bibinfo  {journal} {Mon. Not. R. Astron. Soc.}\ }\textbf {\bibinfo {volume}
  {421}},\ \bibinfo {pages} {3464} (\bibinfo {year} {2012})}\BibitemShut
  {NoStop}%
\bibitem [{\citenamefont {{Arnett}}(1966)}]{Arnett_1966}%
  \BibitemOpen
  \bibfield  {author} {\bibinfo {author} {\bibfnamefont {W.~D.}\ \bibnamefont
  {{Arnett}}},\ }\bibfield  {title} {\bibinfo {title} {Gravitational collapse
  and weak interactions},\ }\href {https://doi.org/10.1139/p66-210} {\bibfield
  {journal} {\bibinfo  {journal} {Can. J. Phys.}\ }\textbf {\bibinfo {volume}
  {44}},\ \bibinfo {pages} {2553} (\bibinfo {year} {1966})}\BibitemShut
  {NoStop}%
\bibitem [{\citenamefont {{Colgate}}\ and\ \citenamefont
  {{White}}(1966)}]{Colgate&White_1966}%
  \BibitemOpen
  \bibfield  {author} {\bibinfo {author} {\bibfnamefont {S.~A.}\ \bibnamefont
  {{Colgate}}}\ and\ \bibinfo {author} {\bibfnamefont {R.~H.}\ \bibnamefont
  {{White}}},\ }\bibfield  {title} {\bibinfo {title} {The hydrodynamic behavior
  of supernovae explosions},\ }\href {https://doi.org/10.1086/148549}
  {\bibfield  {journal} {\bibinfo  {journal} {Astrophys. J.}\ }\textbf
  {\bibinfo {volume} {143}},\ \bibinfo {pages} {626} (\bibinfo {year}
  {1966})}\BibitemShut {NoStop}%
\bibitem [{\citenamefont {{Wilson}}(1985)}]{Wilson_1982}%
  \BibitemOpen
  \bibfield  {author} {\bibinfo {author} {\bibfnamefont {J.~R.}\ \bibnamefont
  {{Wilson}}},\ }\bibfield  {title} {\bibinfo {title} {{Supernovae and
  Post-Collapse Behavior}},\ }in\ \href
  {https://ui.adsabs.harvard.edu/abs/1985nuas.conf..422W} {\emph {\bibinfo
  {booktitle} {Numerical Astrophysics}}},\ \bibinfo {editor} {edited by\
  \bibinfo {editor} {\bibfnamefont {J.~M.}\ \bibnamefont {Centrella}}, \bibinfo
  {editor} {\bibfnamefont {J.~M.}\ \bibnamefont {Leblanc}},\ and\ \bibinfo
  {editor} {\bibfnamefont {R.~L.}\ \bibnamefont {Bowers}}}\ (\bibinfo
  {publisher} {Jones \& Bartlett},\ \bibinfo {address} {Boston},\ \bibinfo
  {year} {1985})\ p.\ \bibinfo {pages} {422}\BibitemShut {NoStop}%
\bibitem [{\citenamefont {{Bethe}}\ and\ \citenamefont
  {{Wilson}}(1985)}]{Bethe&Wilson_1985}%
  \BibitemOpen
  \bibfield  {author} {\bibinfo {author} {\bibfnamefont {H.~A.}\ \bibnamefont
  {{Bethe}}}\ and\ \bibinfo {author} {\bibfnamefont {J.~R.}\ \bibnamefont
  {{Wilson}}},\ }\bibfield  {title} {\bibinfo {title} {{Revival of a stalled
  supernova shock by neutrino heating}},\ }\href
  {https://doi.org/10.1086/163343} {\bibfield  {journal} {\bibinfo  {journal}
  {Astrophys. J.}\ }\textbf {\bibinfo {volume} {295}},\ \bibinfo {pages} {14}
  (\bibinfo {year} {1985})}\BibitemShut {NoStop}%
\bibitem [{\citenamefont {Bethe}(1990)}]{Bethe_1990}%
  \BibitemOpen
  \bibfield  {author} {\bibinfo {author} {\bibfnamefont {H.~A.}\ \bibnamefont
  {Bethe}},\ }\bibfield  {title} {\bibinfo {title} {Supernova mechanisms},\
  }\href {https://doi.org/10.1103/RevModPhys.62.801} {\bibfield  {journal}
  {\bibinfo  {journal} {Rev. Mod. Phys.}\ }\textbf {\bibinfo {volume} {62}},\
  \bibinfo {pages} {801} (\bibinfo {year} {1990})}\BibitemShut {NoStop}%
\bibitem [{\citenamefont {Burrows}\ and\ \citenamefont
  {Goshy}(1993)}]{burrows1993theory}%
  \BibitemOpen
  \bibfield  {author} {\bibinfo {author} {\bibfnamefont {A.}~\bibnamefont
  {Burrows}}\ and\ \bibinfo {author} {\bibfnamefont {J.}~\bibnamefont
  {Goshy}},\ }\bibfield  {title} {\bibinfo {title} {A theory of supernova
  explosions},\ }\href@noop {} {\bibfield  {journal} {\bibinfo  {journal}
  {Astrophys. J.}\ }\textbf {\bibinfo {volume} {416}},\ \bibinfo {pages} {L75}
  (\bibinfo {year} {1993})}\BibitemShut {NoStop}%
\bibitem [{\citenamefont {{Kotake}}\ \emph {et~al.}(2006)\citenamefont
  {{Kotake}}, \citenamefont {{Sato}},\ and\ \citenamefont
  {{Takahashi}}}]{Kotake_2006}%
  \BibitemOpen
  \bibfield  {author} {\bibinfo {author} {\bibfnamefont {K.}~\bibnamefont
  {{Kotake}}}, \bibinfo {author} {\bibfnamefont {K.}~\bibnamefont {{Sato}}},\
  and\ \bibinfo {author} {\bibfnamefont {K.}~\bibnamefont {{Takahashi}}},\
  }\bibfield  {title} {\bibinfo {title} {{Explosion mechanism, neutrino burst
  and gravitational wave in core-collapse supernovae}},\ }\href
  {https://doi.org/10.1088/0034-4885/69/4/R03} {\bibfield  {journal} {\bibinfo
  {journal} {Rep. Prog. Phys.}\ }\textbf {\bibinfo {volume} {69}},\ \bibinfo
  {pages} {971} (\bibinfo {year} {2006})}\BibitemShut {NoStop}%
\bibitem [{\citenamefont {Janka}(2012)}]{janka2012b}%
  \BibitemOpen
  \bibfield  {author} {\bibinfo {author} {\bibfnamefont {H.-T.}\ \bibnamefont
  {Janka}},\ }\bibfield  {title} {\bibinfo {title} {Explosion mechanisms of
  core-collapse supernovae},\ }\href
  {https://doi.org/10.1146/annurev-nucl-102711-094901} {\bibfield  {journal}
  {\bibinfo  {journal} {Annu. Rev. Nucl. Part. Sci.}\ }\textbf {\bibinfo
  {volume} {62}},\ \bibinfo {pages} {407} (\bibinfo {year} {2012})}\BibitemShut
  {NoStop}%
\bibitem [{\citenamefont {Mirizzi}\ \emph {et~al.}(2016)\citenamefont
  {Mirizzi}, \citenamefont {Tamborra}, \citenamefont {Janka}, \citenamefont
  {Saviano}, \citenamefont {Scholberg}, \citenamefont {Bollig}, \citenamefont
  {Hudepohl},\ and\ \citenamefont {Chakraborty}}]{Mirizzi_2016}%
  \BibitemOpen
  \bibfield  {author} {\bibinfo {author} {\bibfnamefont {A.}~\bibnamefont
  {Mirizzi}}, \bibinfo {author} {\bibfnamefont {I.}~\bibnamefont {Tamborra}},
  \bibinfo {author} {\bibfnamefont {H.-T.}\ \bibnamefont {Janka}}, \bibinfo
  {author} {\bibfnamefont {N.}~\bibnamefont {Saviano}}, \bibinfo {author}
  {\bibfnamefont {K.}~\bibnamefont {Scholberg}}, \bibinfo {author}
  {\bibfnamefont {R.}~\bibnamefont {Bollig}}, \bibinfo {author} {\bibfnamefont
  {L.}~\bibnamefont {Hudepohl}},\ and\ \bibinfo {author} {\bibfnamefont
  {S.}~\bibnamefont {Chakraborty}},\ }\bibfield  {title} {\bibinfo {title}
  {Supernova neutrinos: Production, oscillations and detection},\ }\bibfield
  {journal} {\bibinfo  {journal} {Riv. Nuovo Cimento}\ }\textbf {\bibinfo
  {volume} {39}},\ \href {https://doi.org/10.1393/ncr/i2016-10120-8}
  {10.1393/ncr/i2016-10120-8} (\bibinfo {year} {2016})\BibitemShut {NoStop}%
\bibitem [{\citenamefont {Hayashi}\ \emph {et~al.}(2021)\citenamefont
  {Hayashi}, \citenamefont {Ibe}, \citenamefont {Kobayashi}, \citenamefont
  {Nakayama},\ and\ \citenamefont {Shirai}}]{Hayashi:2020syu}%
  \BibitemOpen
  \bibfield  {author} {\bibinfo {author} {\bibfnamefont {K.}~\bibnamefont
  {Hayashi}}, \bibinfo {author} {\bibfnamefont {M.}~\bibnamefont {Ibe}},
  \bibinfo {author} {\bibfnamefont {S.}~\bibnamefont {Kobayashi}}, \bibinfo
  {author} {\bibfnamefont {Y.}~\bibnamefont {Nakayama}},\ and\ \bibinfo
  {author} {\bibfnamefont {S.}~\bibnamefont {Shirai}},\ }\bibfield  {title}
  {\bibinfo {title} {{Probing dark matter self-interaction with ultrafaint
  dwarf galaxies}},\ }\href {https://doi.org/10.1103/PhysRevD.103.023017}
  {\bibfield  {journal} {\bibinfo  {journal} {Phys. Rev. D}\ }\textbf {\bibinfo
  {volume} {103}},\ \bibinfo {pages} {023017} (\bibinfo {year}
  {2021})}\BibitemShut {NoStop}%
\bibitem [{\citenamefont {{Mu{\~n}oz}}\ \emph {et~al.}(2018)\citenamefont
  {{Mu{\~n}oz}}, \citenamefont {{C{\^o}t{\'e}}}, \citenamefont {{Santana}},
  \citenamefont {{Geha}}, \citenamefont {{Simon}}, \citenamefont
  {{Oyarz{\'u}n}}, \citenamefont {{Stetson}},\ and\ \citenamefont
  {{Djorgovski}}}]{2018ApJ...860...66M}%
  \BibitemOpen
  \bibfield  {author} {\bibinfo {author} {\bibfnamefont {R.~R.}\ \bibnamefont
  {{Mu{\~n}oz}}}, \bibinfo {author} {\bibfnamefont {P.}~\bibnamefont
  {{C{\^o}t{\'e}}}}, \bibinfo {author} {\bibfnamefont {F.~A.}\ \bibnamefont
  {{Santana}}}, \bibinfo {author} {\bibfnamefont {M.}~\bibnamefont {{Geha}}},
  \bibinfo {author} {\bibfnamefont {J.~D.}\ \bibnamefont {{Simon}}}, \bibinfo
  {author} {\bibfnamefont {G.~A.}\ \bibnamefont {{Oyarz{\'u}n}}}, \bibinfo
  {author} {\bibfnamefont {P.~B.}\ \bibnamefont {{Stetson}}},\ and\ \bibinfo
  {author} {\bibfnamefont {S.~G.}\ \bibnamefont {{Djorgovski}}},\ }\bibfield
  {title} {\bibinfo {title} {{A MegaCam Survey of Outer Halo Satellites. III.
  Photometric and Structural Parameters}},\ }\href
  {https://doi.org/10.3847/1538-4357/aac16b} {\bibfield  {journal} {\bibinfo
  {journal} {Astrophys J.}\ }\textbf {\bibinfo {volume} {860}},\ \bibinfo {eid}
  {66} (\bibinfo {year} {2018})}\BibitemShut {NoStop}%
\bibitem [{\citenamefont {Simon}\ \emph
  {et~al.}(2011{\natexlab{a}})\citenamefont {Simon} \emph
  {et~al.}}]{Simon:2010ek}%
  \BibitemOpen
  \bibfield  {author} {\bibinfo {author} {\bibfnamefont {J.~D.}\ \bibnamefont
  {Simon}} \emph {et~al.},\ }\bibfield  {title} {\bibinfo {title} {{A Complete
  Spectroscopic Survey of the Milky Way Satellite Segue 1: The Darkest
  Galaxy}},\ }\href {https://doi.org/10.1088/0004-637X/733/1/46} {\bibfield
  {journal} {\bibinfo  {journal} {Astrophys. J.}\ }\textbf {\bibinfo {volume}
  {733}},\ \bibinfo {pages} {46} (\bibinfo {year}
  {2011}{\natexlab{a}})}\BibitemShut {NoStop}%
\bibitem [{\citenamefont {Kirby}\ \emph {et~al.}(2013)\citenamefont {Kirby},
  \citenamefont {Boylan-Kolchin}, \citenamefont {Cohen}, \citenamefont {Geha},
  \citenamefont {Bullock},\ and\ \citenamefont {Kaplinghat}}]{Kirby:2013isa}%
  \BibitemOpen
  \bibfield  {author} {\bibinfo {author} {\bibfnamefont {E.~N.}\ \bibnamefont
  {Kirby}}, \bibinfo {author} {\bibfnamefont {M.}~\bibnamefont
  {Boylan-Kolchin}}, \bibinfo {author} {\bibfnamefont {J.~G.}\ \bibnamefont
  {Cohen}}, \bibinfo {author} {\bibfnamefont {M.}~\bibnamefont {Geha}},
  \bibinfo {author} {\bibfnamefont {J.~S.}\ \bibnamefont {Bullock}},\ and\
  \bibinfo {author} {\bibfnamefont {M.}~\bibnamefont {Kaplinghat}},\ }\bibfield
   {title} {\bibinfo {title} {{Segue 2: The Least Massive Galaxy}},\ }\href
  {https://doi.org/10.1088/0004-637X/770/1/16} {\bibfield  {journal} {\bibinfo
  {journal} {Astrophys. J.}\ }\textbf {\bibinfo {volume} {770}},\ \bibinfo
  {pages} {16} (\bibinfo {year} {2013})}\BibitemShut {NoStop}%
\bibitem [{\citenamefont {Koposov}\ \emph
  {et~al.}(2011{\natexlab{a}})\citenamefont {Koposov} \emph
  {et~al.}}]{Koposov:2011zi}%
  \BibitemOpen
  \bibfield  {author} {\bibinfo {author} {\bibfnamefont {S.~E.}\ \bibnamefont
  {Koposov}} \emph {et~al.},\ }\bibfield  {title} {\bibinfo {title} {{Accurate
  Stellar Kinematics at Faint Magnitudes: application to the
  Bootes\textasciitilde{}I dwarf spheroidal galaxy}},\ }\href
  {https://doi.org/10.1088/0004-637X/736/2/146} {\bibfield  {journal} {\bibinfo
   {journal} {Astrophys. J.}\ }\textbf {\bibinfo {volume} {736}},\ \bibinfo
  {pages} {146} (\bibinfo {year} {2011}{\natexlab{a}})}\BibitemShut {NoStop}%
\bibitem [{\citenamefont {Simon}\ and\ \citenamefont
  {Geha}(2007{\natexlab{a}})}]{Simon:2007dq}%
  \BibitemOpen
  \bibfield  {author} {\bibinfo {author} {\bibfnamefont {J.~D.}\ \bibnamefont
  {Simon}}\ and\ \bibinfo {author} {\bibfnamefont {M.}~\bibnamefont {Geha}},\
  }\bibfield  {title} {\bibinfo {title} {{The Kinematics of the Ultra-Faint
  Milky Way Satellites: Solving the Missing Satellite Problem}},\ }\href
  {https://doi.org/10.1086/521816} {\bibfield  {journal} {\bibinfo  {journal}
  {Astrophys. J.}\ }\textbf {\bibinfo {volume} {670}},\ \bibinfo {pages} {313}
  (\bibinfo {year} {2007}{\natexlab{a}})}\BibitemShut {NoStop}%
\bibitem [{\citenamefont {Simon}\ \emph {et~al.}(2015)\citenamefont {Simon}
  \emph {et~al.}}]{DES:2015tfc}%
  \BibitemOpen
  \bibfield  {author} {\bibinfo {author} {\bibfnamefont {J.~D.}\ \bibnamefont
  {Simon}} \emph {et~al.} (\bibinfo {collaboration} {DES}),\ }\bibfield
  {title} {\bibinfo {title} {{Stellar Kinematics and Metallicities in the
  Ultra-Faint Dwarf Galaxy Reticulum II}},\ }\href
  {https://doi.org/10.1088/0004-637X/808/1/95} {\bibfield  {journal} {\bibinfo
  {journal} {Astrophys. J.}\ }\textbf {\bibinfo {volume} {808}},\ \bibinfo
  {pages} {95} (\bibinfo {year} {2015})}\BibitemShut {NoStop}%
\bibitem [{\citenamefont {{Martin}}\ \emph {et~al.}(2016)\citenamefont
  {{Martin}} \emph {et~al.}}]{2016MNRAS.458L..59M}%
  \BibitemOpen
  \bibfield  {author} {\bibinfo {author} {\bibfnamefont {N.~F.}\ \bibnamefont
  {{Martin}}} \emph {et~al.},\ }\bibfield  {title} {\bibinfo {title} {{Is Draco
  II one of the faintest dwarf galaxies? First study from Keck/DEIMOS
  spectroscopy}},\ }\href {https://doi.org/10.1093/mnrasl/slw013} {\bibfield
  {journal} {\bibinfo  {journal} {Mon. Not. R. Astron. Soc.}\ }\textbf
  {\bibinfo {volume} {458}},\ \bibinfo {pages} {L59} (\bibinfo {year}
  {2016})}\BibitemShut {NoStop}%
\bibitem [{\citenamefont {{Kirby}}\ \emph {et~al.}(2017)\citenamefont
  {{Kirby}}, \citenamefont {{Cohen}}, \citenamefont {{Simon}}, \citenamefont
  {{Guhathakurta}}, \citenamefont {{Thygesen}},\ and\ \citenamefont
  {{Duggan}}}]{2017ApJ...838...83K}%
  \BibitemOpen
  \bibfield  {author} {\bibinfo {author} {\bibfnamefont {E.~N.}\ \bibnamefont
  {{Kirby}}}, \bibinfo {author} {\bibfnamefont {J.~G.}\ \bibnamefont
  {{Cohen}}}, \bibinfo {author} {\bibfnamefont {J.~D.}\ \bibnamefont
  {{Simon}}}, \bibinfo {author} {\bibfnamefont {P.}~\bibnamefont
  {{Guhathakurta}}}, \bibinfo {author} {\bibfnamefont {A.~O.}\ \bibnamefont
  {{Thygesen}}},\ and\ \bibinfo {author} {\bibfnamefont {G.~E.}\ \bibnamefont
  {{Duggan}}},\ }\bibfield  {title} {\bibinfo {title} {{Triangulum II. Not
  Especially Dense After All}},\ }\href
  {https://doi.org/10.3847/1538-4357/aa6570} {\bibfield  {journal} {\bibinfo
  {journal} {Astrophys. J.}\ }\textbf {\bibinfo {volume} {838}},\ \bibinfo
  {eid} {83} (\bibinfo {year} {2017})}\BibitemShut {NoStop}%
\bibitem [{\citenamefont {Kirby}\ \emph {et~al.}(2015)\citenamefont {Kirby},
  \citenamefont {Simon},\ and\ \citenamefont {Cohen}}]{Kirby:2015ija}%
  \BibitemOpen
  \bibfield  {author} {\bibinfo {author} {\bibfnamefont {E.~N.}\ \bibnamefont
  {Kirby}}, \bibinfo {author} {\bibfnamefont {J.~D.}\ \bibnamefont {Simon}},\
  and\ \bibinfo {author} {\bibfnamefont {J.~G.}\ \bibnamefont {Cohen}},\
  }\bibfield  {title} {\bibinfo {title} {{Spectroscopic Confirmation of the
  Dwarf Galaxies Hydra II and Pisces II and the Globular Cluster Laevens 1}},\
  }\href {https://doi.org/10.1088/0004-637X/810/1/56} {\bibfield  {journal}
  {\bibinfo  {journal} {Astrophys. J.}\ }\textbf {\bibinfo {volume} {810}},\
  \bibinfo {pages} {56} (\bibinfo {year} {2015})}\BibitemShut {NoStop}%
\bibitem [{\citenamefont {Walker}\ \emph {et~al.}(2016)\citenamefont {Walker}
  \emph {et~al.}}]{Walker:2015twz}%
  \BibitemOpen
  \bibfield  {author} {\bibinfo {author} {\bibfnamefont {M.~G.}\ \bibnamefont
  {Walker}} \emph {et~al.},\ }\bibfield  {title} {\bibinfo {title}
  {{Magellan/M2FS spectroscopy of Tucana 2 and Grus 1}},\ }\href
  {https://doi.org/10.3847/0004-637X/819/1/53} {\bibfield  {journal} {\bibinfo
  {journal} {Astrophys. J.}\ }\textbf {\bibinfo {volume} {819}},\ \bibinfo
  {pages} {53} (\bibinfo {year} {2016})}\BibitemShut {NoStop}%
\bibitem [{\citenamefont {Simon}\ \emph
  {et~al.}(2020{\natexlab{a}})\citenamefont {Simon} \emph
  {et~al.}}]{DES:2019ncb}%
  \BibitemOpen
  \bibfield  {author} {\bibinfo {author} {\bibfnamefont {J.~D.}\ \bibnamefont
  {Simon}} \emph {et~al.} (\bibinfo {collaboration} {DES}),\ }\bibfield
  {title} {\bibinfo {title} {{Birds of a Feather? Magellan/IMACS Spectroscopy
  of the Ultra-Faint Satellites Grus II, Tucana IV, and Tucana V}},\ }\href
  {https://doi.org/10.3847/1538-4357/ab7ccb} {\bibfield  {journal} {\bibinfo
  {journal} {Astrophys. J.}\ }\textbf {\bibinfo {volume} {892}},\ \bibinfo
  {pages} {137} (\bibinfo {year} {2020}{\natexlab{a}})}\BibitemShut {NoStop}%
\bibitem [{\citenamefont {Koposov}\ \emph
  {et~al.}(2015{\natexlab{a}})\citenamefont {Koposov} \emph
  {et~al.}}]{Koposov:2015jla}%
  \BibitemOpen
  \bibfield  {author} {\bibinfo {author} {\bibfnamefont {S.~E.}\ \bibnamefont
  {Koposov}} \emph {et~al.},\ }\bibfield  {title} {\bibinfo {title}
  {{Kinematics and chemistry of recently discovered Reticulum 2 and Horologium
  1 dwarf galaxies}},\ }\href {https://doi.org/10.1088/0004-637X/811/1/62}
  {\bibfield  {journal} {\bibinfo  {journal} {Astrophys. J.}\ }\textbf
  {\bibinfo {volume} {811}},\ \bibinfo {pages} {62} (\bibinfo {year}
  {2015}{\natexlab{a}})}\BibitemShut {NoStop}%
\bibitem [{\citenamefont {Simon}\ \emph {et~al.}(2017)\citenamefont {Simon}
  \emph {et~al.}}]{DES:2016fyd}%
  \BibitemOpen
  \bibfield  {author} {\bibinfo {author} {\bibfnamefont {J.~D.}\ \bibnamefont
  {Simon}} \emph {et~al.} (\bibinfo {collaboration} {DES}),\ }\bibfield
  {title} {\bibinfo {title} {{Nearest Neighbor: The Low-Mass Milky Way
  Satellite Tucana III}},\ }\href {https://doi.org/10.3847/1538-4357/aa5be7}
  {\bibfield  {journal} {\bibinfo  {journal} {Astrophys. J.}\ }\textbf
  {\bibinfo {volume} {838}},\ \bibinfo {pages} {11} (\bibinfo {year}
  {2017})}\BibitemShut {NoStop}%
\bibitem [{\citenamefont {{Walker}}\ \emph {et~al.}(2015)\citenamefont
  {{Walker}}, \citenamefont {{Olszewski}},\ and\ \citenamefont
  {{Mateo}}}]{2015MNRAS.448.2717W}%
  \BibitemOpen
  \bibfield  {author} {\bibinfo {author} {\bibfnamefont {M.~G.}\ \bibnamefont
  {{Walker}}}, \bibinfo {author} {\bibfnamefont {E.~W.}\ \bibnamefont
  {{Olszewski}}},\ and\ \bibinfo {author} {\bibfnamefont {M.}~\bibnamefont
  {{Mateo}}},\ }\bibfield  {title} {\bibinfo {title} {{Bayesian analysis of
  resolved stellar spectra: application to MMT/Hectochelle observations of the
  Draco dwarf spheroidal}},\ }\href {https://doi.org/10.1093/mnras/stv099}
  {\bibfield  {journal} {\bibinfo  {journal} {Mon. Not. R. Astron. Soc.}\
  }\textbf {\bibinfo {volume} {448}},\ \bibinfo {pages} {2717} (\bibinfo {year}
  {2015})}\BibitemShut {NoStop}%
\bibitem [{\citenamefont {Walker}\ \emph
  {et~al.}(2008{\natexlab{a}})\citenamefont {Walker}, \citenamefont {Mateo},\
  and\ \citenamefont {Olszewski}}]{Walker:2008ji}%
  \BibitemOpen
  \bibfield  {author} {\bibinfo {author} {\bibfnamefont {M.~G.}\ \bibnamefont
  {Walker}}, \bibinfo {author} {\bibfnamefont {M.}~\bibnamefont {Mateo}},\ and\
  \bibinfo {author} {\bibfnamefont {E.~W.}\ \bibnamefont {Olszewski}},\
  }\bibfield  {title} {\bibinfo {title} {{Systemic Proper Motions of Milky Way
  Satellites from Stellar Redshifts: the Carina, Fornax, Sculptor and Sextans
  Dwarf Spheroidals}},\ }\href {https://doi.org/10.1086/595586} {\bibfield
  {journal} {\bibinfo  {journal} {Astrophys. J. Lett.}\ }\textbf {\bibinfo
  {volume} {688}},\ \bibinfo {pages} {L75} (\bibinfo {year}
  {2008}{\natexlab{a}})}\BibitemShut {NoStop}%
\bibitem [{\citenamefont {Walker}\ \emph
  {et~al.}(2009{\natexlab{a}})\citenamefont {Walker}, \citenamefont {Mateo},\
  and\ \citenamefont {Olszewski}}]{Walker:2008ax}%
  \BibitemOpen
  \bibfield  {author} {\bibinfo {author} {\bibfnamefont {M.~G.}\ \bibnamefont
  {Walker}}, \bibinfo {author} {\bibfnamefont {M.}~\bibnamefont {Mateo}},\ and\
  \bibinfo {author} {\bibfnamefont {E.}~\bibnamefont {Olszewski}},\ }\bibfield
  {title} {\bibinfo {title} {{Stellar Velocities in the Carina, Fornax,
  Sculptor and Sextans dSph Galaxies: Data from the Magellan/MMFS Survey}},\
  }\href {https://doi.org/10.1088/0004-6256/137/2/3100} {\bibfield  {journal}
  {\bibinfo  {journal} {Astron. J.}\ }\textbf {\bibinfo {volume} {137}},\
  \bibinfo {pages} {3100} (\bibinfo {year} {2009}{\natexlab{a}})}\BibitemShut
  {NoStop}%
\bibitem [{\citenamefont {{Fabrizio}}\ \emph {et~al.}(2016)\citenamefont
  {{Fabrizio}} \emph {et~al.}}]{2016ApJ...830..126F}%
  \BibitemOpen
  \bibfield  {author} {\bibinfo {author} {\bibfnamefont {M.}~\bibnamefont
  {{Fabrizio}}} \emph {et~al.},\ }\bibfield  {title} {\bibinfo {title} {{The
  Carina Project. X. On the Kinematics of Old and Intermediate-age Stellar
  Populations1,2}},\ }\href {https://doi.org/10.3847/0004-637X/830/2/126}
  {\bibfield  {journal} {\bibinfo  {journal} {Astrophys. J.}\ }\textbf
  {\bibinfo {volume} {830}},\ \bibinfo {eid} {126} (\bibinfo {year}
  {2016})}\BibitemShut {NoStop}%
\bibitem [{\citenamefont {{Spencer}}\ \emph {et~al.}(2018)\citenamefont
  {{Spencer}}, \citenamefont {{Mateo}}, \citenamefont {{Olszewski}},
  \citenamefont {{Walker}}, \citenamefont {{McConnachie}},\ and\ \citenamefont
  {{Kirby}}}]{2018AJ....156..257S}%
  \BibitemOpen
  \bibfield  {author} {\bibinfo {author} {\bibfnamefont {M.~E.}\ \bibnamefont
  {{Spencer}}}, \bibinfo {author} {\bibfnamefont {M.}~\bibnamefont {{Mateo}}},
  \bibinfo {author} {\bibfnamefont {E.~W.}\ \bibnamefont {{Olszewski}}},
  \bibinfo {author} {\bibfnamefont {M.~G.}\ \bibnamefont {{Walker}}}, \bibinfo
  {author} {\bibfnamefont {A.~W.}\ \bibnamefont {{McConnachie}}},\ and\
  \bibinfo {author} {\bibfnamefont {E.~N.}\ \bibnamefont {{Kirby}}},\
  }\bibfield  {title} {\bibinfo {title} {{The Binary Fraction of Stars in Dwarf
  Galaxies: The Cases of Draco and Ursa Minor}},\ }\href
  {https://doi.org/10.3847/1538-3881/aae3e4} {\bibfield  {journal} {\bibinfo
  {journal} {Astron. J.}\ }\textbf {\bibinfo {volume} {156}},\ \bibinfo {eid}
  {257} (\bibinfo {year} {2018})}\BibitemShut {NoStop}%
\bibitem [{\citenamefont {Mateo}\ \emph
  {et~al.}(2008{\natexlab{a}})\citenamefont {Mateo}, \citenamefont
  {Olszewski},\ and\ \citenamefont {Walker}}]{Mateo:2007xh}%
  \BibitemOpen
  \bibfield  {author} {\bibinfo {author} {\bibfnamefont {M.}~\bibnamefont
  {Mateo}}, \bibinfo {author} {\bibfnamefont {E.~W.}\ \bibnamefont
  {Olszewski}},\ and\ \bibinfo {author} {\bibfnamefont {M.~G.}\ \bibnamefont
  {Walker}},\ }\bibfield  {title} {\bibinfo {title} {{The Velocity Dispersion
  Profile of the Remote Dwarf Spheroidal Galaxy Leo. 1. A Tidal Hit and
  Run?}},\ }\href {https://doi.org/10.1086/522326} {\bibfield  {journal}
  {\bibinfo  {journal} {Astrophys. J.}\ }\textbf {\bibinfo {volume} {675}},\
  \bibinfo {pages} {201} (\bibinfo {year} {2008}{\natexlab{a}})}\BibitemShut
  {NoStop}%
\bibitem [{\citenamefont {{Spencer}}\ \emph {et~al.}(2017)\citenamefont
  {{Spencer}}, \citenamefont {{Mateo}}, \citenamefont {{Walker}},\ and\
  \citenamefont {{Olszewski}}}]{2017ApJ...836..202S}%
  \BibitemOpen
  \bibfield  {author} {\bibinfo {author} {\bibfnamefont {M.~E.}\ \bibnamefont
  {{Spencer}}}, \bibinfo {author} {\bibfnamefont {M.}~\bibnamefont {{Mateo}}},
  \bibinfo {author} {\bibfnamefont {M.~G.}\ \bibnamefont {{Walker}}},\ and\
  \bibinfo {author} {\bibfnamefont {E.~W.}\ \bibnamefont {{Olszewski}}},\
  }\bibfield  {title} {\bibinfo {title} {{A Multi-epoch Kinematic Study of the
  Remote Dwarf Spheroidal Galaxy Leo II}},\ }\href
  {https://doi.org/10.3847/1538-4357/836/2/202} {\bibfield  {journal} {\bibinfo
   {journal} {\apj}\ }\textbf {\bibinfo {volume} {836}},\ \bibinfo {eid} {202}
  (\bibinfo {year} {2017})}\BibitemShut {NoStop}%
\bibitem [{\citenamefont {Molin\'e}\ \emph {et~al.}(2017)\citenamefont
  {Molin\'e}, \citenamefont {S\'anchez-Conde}, \citenamefont {Palomares-Ruiz},\
  and\ \citenamefont {Prada}}]{Moline:2016pbm}%
  \BibitemOpen
  \bibfield  {author} {\bibinfo {author} {\bibfnamefont {A.}~\bibnamefont
  {Molin\'e}}, \bibinfo {author} {\bibfnamefont {M.~A.}\ \bibnamefont
  {S\'anchez-Conde}}, \bibinfo {author} {\bibfnamefont {S.}~\bibnamefont
  {Palomares-Ruiz}},\ and\ \bibinfo {author} {\bibfnamefont {F.}~\bibnamefont
  {Prada}},\ }\bibfield  {title} {\bibinfo {title} {{Characterization of
  subhalo structural properties and implications for dark matter annihilation
  signals}},\ }\href {https://doi.org/10.1093/mnras/stx026} {\bibfield
  {journal} {\bibinfo  {journal} {Mon. Not. R. Astron. Soc.}\ }\textbf
  {\bibinfo {volume} {466}},\ \bibinfo {pages} {4974} (\bibinfo {year}
  {2017})}\BibitemShut {NoStop}%
\bibitem [{\citenamefont {Grcevich}\ and\ \citenamefont
  {Putman}(2009)}]{Grcevich_2009}%
  \BibitemOpen
  \bibfield  {author} {\bibinfo {author} {\bibfnamefont {J.}~\bibnamefont
  {Grcevich}}\ and\ \bibinfo {author} {\bibfnamefont {M.~E.}\ \bibnamefont
  {Putman}},\ }\bibfield  {title} {\bibinfo {title} {H {I} in the local group
  dwarf galaxies and stripping by the galactic halo},\ }\href
  {https://doi.org/10.1088/0004-637X/696/1/385} {\bibfield  {journal} {\bibinfo
   {journal} {Astrophys. J.}\ }\textbf {\bibinfo {volume} {696}},\ \bibinfo
  {pages} {385} (\bibinfo {year} {2009})}\BibitemShut {NoStop}%
\bibitem [{\citenamefont {Simon}\ \emph
  {et~al.}(2011{\natexlab{b}})\citenamefont {Simon} \emph
  {et~al.}}]{Simon_2011}%
  \BibitemOpen
  \bibfield  {author} {\bibinfo {author} {\bibfnamefont {J.~D.}\ \bibnamefont
  {Simon}} \emph {et~al.},\ }\bibfield  {title} {\bibinfo {title} {A complete
  spectroscopic survey of the {Milky} {Way} satellite {Segue} 1: the darkest
  $\mathrm{galaxy}^*$},\ }\href {https://doi.org/10.1088/0004-637X/733/1/46}
  {\bibfield  {journal} {\bibinfo  {journal} {Astrophys. J.}\ }\textbf
  {\bibinfo {volume} {733}},\ \bibinfo {pages} {46} (\bibinfo {year}
  {2011}{\natexlab{b}})}\BibitemShut {NoStop}%
\bibitem [{\citenamefont {Belokurov}\ \emph {et~al.}(2009)\citenamefont
  {Belokurov}, \citenamefont {Walker}, \citenamefont {Evans}, \citenamefont
  {Gilmore}, \citenamefont {Irwin}, \citenamefont {Mateo}, \citenamefont
  {Mayer}, \citenamefont {Olszewski}, \citenamefont {Bechtold},\ and\
  \citenamefont {Pickering}}]{Belokurov_2009}%
  \BibitemOpen
  \bibfield  {author} {\bibinfo {author} {\bibfnamefont {V.}~\bibnamefont
  {Belokurov}}, \bibinfo {author} {\bibfnamefont {M.~G.}\ \bibnamefont
  {Walker}}, \bibinfo {author} {\bibfnamefont {N.~W.}\ \bibnamefont {Evans}},
  \bibinfo {author} {\bibfnamefont {G.}~\bibnamefont {Gilmore}}, \bibinfo
  {author} {\bibfnamefont {M.~J.}\ \bibnamefont {Irwin}}, \bibinfo {author}
  {\bibfnamefont {M.}~\bibnamefont {Mateo}}, \bibinfo {author} {\bibfnamefont
  {L.}~\bibnamefont {Mayer}}, \bibinfo {author} {\bibfnamefont
  {E.}~\bibnamefont {Olszewski}}, \bibinfo {author} {\bibfnamefont
  {J.}~\bibnamefont {Bechtold}},\ and\ \bibinfo {author} {\bibfnamefont
  {T.}~\bibnamefont {Pickering}},\ }\bibfield  {title} {\bibinfo {title} {{The
  discovery of Segue 2: a prototype of the population of satellites of
  satellites}},\ }\href {https://doi.org/10.1111/j.1365-2966.2009.15106.x}
  {\bibfield  {journal} {\bibinfo  {journal} {Mon. Not. R. Astron. Soc.}\
  }\textbf {\bibinfo {volume} {397}},\ \bibinfo {pages} {1748} (\bibinfo {year}
  {2009})}\BibitemShut {NoStop}%
\bibitem [{\citenamefont {Koposov}\ \emph
  {et~al.}(2011{\natexlab{b}})\citenamefont {Koposov} \emph
  {et~al.}}]{Koposov_2011}%
  \BibitemOpen
  \bibfield  {author} {\bibinfo {author} {\bibfnamefont {S.~E.}\ \bibnamefont
  {Koposov}} \emph {et~al.},\ }\bibfield  {title} {\bibinfo {title} {Accurate
  stellar kinematics at faint magnitudes: Application to the {B}o\"otes {I}
  dwarf spheroidal galaxy},\ }\href
  {https://doi.org/10.1088/0004-637X/736/2/146} {\bibfield  {journal} {\bibinfo
   {journal} {Astrophys. J.}\ }\textbf {\bibinfo {volume} {736}},\ \bibinfo
  {pages} {146} (\bibinfo {year} {2011}{\natexlab{b}})}\BibitemShut {NoStop}%
\bibitem [{\citenamefont {{Ad\'en, D.}}\ \emph {et~al.}(2009)\citenamefont
  {{Ad\'en, D.}}, \citenamefont {{Feltzing, S.}}, \citenamefont {{Koch, A.}},
  \citenamefont {{Wilkinson, M. I.}}, \citenamefont {{Grebel, E. K.}},
  \citenamefont {{Lundstr\"om, I.}}, \citenamefont {{Gilmore, G. F.}},
  \citenamefont {{Zucker, D. B.}}, \citenamefont {{Belokurov, V.}},
  \citenamefont {{Evans, N. W.}},\ and\ \citenamefont {{Faria,
  D.}}}]{Aden_2009a}%
  \BibitemOpen
  \bibfield  {author} {\bibinfo {author} {\bibnamefont {{Ad\'en, D.}}},
  \bibinfo {author} {\bibnamefont {{Feltzing, S.}}}, \bibinfo {author}
  {\bibnamefont {{Koch, A.}}}, \bibinfo {author} {\bibnamefont {{Wilkinson, M.
  I.}}}, \bibinfo {author} {\bibnamefont {{Grebel, E. K.}}}, \bibinfo {author}
  {\bibnamefont {{Lundstr\"om, I.}}}, \bibinfo {author} {\bibnamefont
  {{Gilmore, G. F.}}}, \bibinfo {author} {\bibnamefont {{Zucker, D. B.}}},
  \bibinfo {author} {\bibnamefont {{Belokurov, V.}}}, \bibinfo {author}
  {\bibnamefont {{Evans, N. W.}}},\ and\ \bibinfo {author} {\bibnamefont
  {{Faria, D.}}},\ }\bibfield  {title} {\bibinfo {title} {A photometric and
  spectroscopic study of the new dwarf spheroidal galaxy in
  $\mathrm{Hercules}^{*,**}$ - metallicity, velocities, and a clean list of
  {RGB} members},\ }\href {https://doi.org/10.1051/0004-6361/200912718}
  {\bibfield  {journal} {\bibinfo  {journal} {Astron. Astrophys.}\ }\textbf
  {\bibinfo {volume} {506}},\ \bibinfo {pages} {1147} (\bibinfo {year}
  {2009})}\BibitemShut {NoStop}%
\bibitem [{\citenamefont {Simon}\ and\ \citenamefont
  {Geha}(2007{\natexlab{b}})}]{Simon_2007}%
  \BibitemOpen
  \bibfield  {author} {\bibinfo {author} {\bibfnamefont {J.~D.}\ \bibnamefont
  {Simon}}\ and\ \bibinfo {author} {\bibfnamefont {M.}~\bibnamefont {Geha}},\
  }\bibfield  {title} {\bibinfo {title} {The kinematics of the ultra-faint
  {Milky Way} satellites: Solving the missing satellite problem},\ }\href
  {https://doi.org/10.1086/521816} {\bibfield  {journal} {\bibinfo  {journal}
  {Astrophys. J.}\ }\textbf {\bibinfo {volume} {670}},\ \bibinfo {pages} {313}
  (\bibinfo {year} {2007}{\natexlab{b}})}\BibitemShut {NoStop}%
\bibitem [{\citenamefont {Walker}\ \emph
  {et~al.}(2009{\natexlab{b}})\citenamefont {Walker}, \citenamefont
  {Belokurov}, \citenamefont {Evans}, \citenamefont {Irwin}, \citenamefont
  {Mateo}, \citenamefont {Olszewski},\ and\ \citenamefont
  {Gilmore}}]{Walker_2009a}%
  \BibitemOpen
  \bibfield  {author} {\bibinfo {author} {\bibfnamefont {M.~G.}\ \bibnamefont
  {Walker}}, \bibinfo {author} {\bibfnamefont {V.}~\bibnamefont {Belokurov}},
  \bibinfo {author} {\bibfnamefont {N.~W.}\ \bibnamefont {Evans}}, \bibinfo
  {author} {\bibfnamefont {M.~J.}\ \bibnamefont {Irwin}}, \bibinfo {author}
  {\bibfnamefont {M.}~\bibnamefont {Mateo}}, \bibinfo {author} {\bibfnamefont
  {E.~W.}\ \bibnamefont {Olszewski}},\ and\ \bibinfo {author} {\bibfnamefont
  {G.}~\bibnamefont {Gilmore}},\ }\bibfield  {title} {\bibinfo {title} {Leo v:
  Spectroscopy of a distant and disturbed satellite*},\ }\href
  {https://doi.org/10.1088/0004-637X/694/2/L144} {\bibfield  {journal}
  {\bibinfo  {journal} {Astrophys. J.}\ }\textbf {\bibinfo {volume} {694}},\
  \bibinfo {pages} {L144} (\bibinfo {year} {2009}{\natexlab{b}})}\BibitemShut
  {NoStop}%
\bibitem [{\citenamefont {Belokurov}\ \emph {et~al.}(2008)\citenamefont
  {Belokurov}, \citenamefont {Walker}, \citenamefont {Evans}, \citenamefont
  {Faria}, \citenamefont {Gilmore}, \citenamefont {Irwin}, \citenamefont
  {Koposov}, \citenamefont {Mateo}, \citenamefont {Olszewski},\ and\
  \citenamefont {Zucker}}]{Belokurov_2008}%
  \BibitemOpen
  \bibfield  {author} {\bibinfo {author} {\bibfnamefont {V.}~\bibnamefont
  {Belokurov}}, \bibinfo {author} {\bibfnamefont {M.~G.}\ \bibnamefont
  {Walker}}, \bibinfo {author} {\bibfnamefont {N.~W.}\ \bibnamefont {Evans}},
  \bibinfo {author} {\bibfnamefont {D.~C.}\ \bibnamefont {Faria}}, \bibinfo
  {author} {\bibfnamefont {G.}~\bibnamefont {Gilmore}}, \bibinfo {author}
  {\bibfnamefont {M.~J.}\ \bibnamefont {Irwin}}, \bibinfo {author}
  {\bibfnamefont {S.}~\bibnamefont {Koposov}}, \bibinfo {author} {\bibfnamefont
  {M.}~\bibnamefont {Mateo}}, \bibinfo {author} {\bibfnamefont
  {E.}~\bibnamefont {Olszewski}},\ and\ \bibinfo {author} {\bibfnamefont
  {D.~B.}\ \bibnamefont {Zucker}},\ }\bibfield  {title} {\bibinfo {title} {Leo
  v: A companion of a companion of the milky way galaxy?},\ }\href
  {https://doi.org/10.1086/592962} {\bibfield  {journal} {\bibinfo  {journal}
  {Astrophys. J.}\ }\textbf {\bibinfo {volume} {686}},\ \bibinfo {pages} {L83}
  (\bibinfo {year} {2008})}\BibitemShut {NoStop}%
\bibitem [{\citenamefont {Koposov}\ \emph
  {et~al.}(2015{\natexlab{b}})\citenamefont {Koposov}, \citenamefont
  {Belokurov}, \citenamefont {Torrealba},\ and\ \citenamefont
  {Evans}}]{Koposov_2015a}%
  \BibitemOpen
  \bibfield  {author} {\bibinfo {author} {\bibfnamefont {S.~E.}\ \bibnamefont
  {Koposov}}, \bibinfo {author} {\bibfnamefont {V.}~\bibnamefont {Belokurov}},
  \bibinfo {author} {\bibfnamefont {G.}~\bibnamefont {Torrealba}},\ and\
  \bibinfo {author} {\bibfnamefont {N.~W.}\ \bibnamefont {Evans}},\ }\bibfield
  {title} {\bibinfo {title} {Beasts of the southern wild: Discovery of nine
  ultra faint satellites in the vicinity of the {M}agellanic clouds},\ }\href
  {https://doi.org/10.1088/0004-637X/805/2/130} {\bibfield  {journal} {\bibinfo
   {journal} {Astrophys. J.}\ }\textbf {\bibinfo {volume} {805}},\ \bibinfo
  {pages} {130} (\bibinfo {year} {2015}{\natexlab{b}})}\BibitemShut {NoStop}%
\bibitem [{\citenamefont {Koposov}\ \emph
  {et~al.}(2015{\natexlab{c}})\citenamefont {Koposov} \emph
  {et~al.}}]{Koposov_2015b}%
  \BibitemOpen
  \bibfield  {author} {\bibinfo {author} {\bibfnamefont {S.~E.}\ \bibnamefont
  {Koposov}} \emph {et~al.},\ }\bibfield  {title} {\bibinfo {title} {Kinematics
  and chemistry of recently discovered {R}eticulum 2 and {H}orologium 1 dwarf
  galaxies},\ }\href {https://doi.org/10.1088/0004-637X/811/1/62} {\bibfield
  {journal} {\bibinfo  {journal} {Astrophys. J.}\ }\textbf {\bibinfo {volume}
  {811}},\ \bibinfo {pages} {62} (\bibinfo {year}
  {2015}{\natexlab{c}})}\BibitemShut {NoStop}%
\bibitem [{\citenamefont {Muñoz}\ \emph {et~al.}(2018)\citenamefont {Muñoz},
  \citenamefont {Côté}, \citenamefont {Santana}, \citenamefont {Geha},
  \citenamefont {Simon}, \citenamefont {Oyarzún}, \citenamefont {Stetson},\
  and\ \citenamefont {Djorgovski}}]{Muñoz_2018}%
  \BibitemOpen
  \bibfield  {author} {\bibinfo {author} {\bibfnamefont {R.~R.}\ \bibnamefont
  {Muñoz}}, \bibinfo {author} {\bibfnamefont {P.}~\bibnamefont {Côté}},
  \bibinfo {author} {\bibfnamefont {F.~A.}\ \bibnamefont {Santana}}, \bibinfo
  {author} {\bibfnamefont {M.}~\bibnamefont {Geha}}, \bibinfo {author}
  {\bibfnamefont {J.~D.}\ \bibnamefont {Simon}}, \bibinfo {author}
  {\bibfnamefont {G.~A.}\ \bibnamefont {Oyarzún}}, \bibinfo {author}
  {\bibfnamefont {P.~B.}\ \bibnamefont {Stetson}},\ and\ \bibinfo {author}
  {\bibfnamefont {S.~G.}\ \bibnamefont {Djorgovski}},\ }\bibfield  {title}
  {\bibinfo {title} {A {MegaCam} survey of outer halo satellites. {III}.
  photometric and structural
  $\mathrm{parameters}^{*}$\textsuperscript{\Cross}},\ }\href
  {https://doi.org/10.3847/1538-4357/aac16b} {\bibfield  {journal} {\bibinfo
  {journal} {Astrophys. J.}\ }\textbf {\bibinfo {volume} {860}},\ \bibinfo
  {pages} {66} (\bibinfo {year} {2018})}\BibitemShut {NoStop}%
\bibitem [{\citenamefont {McConnachie}(2012)}]{McConnachie_2012}%
  \BibitemOpen
  \bibfield  {author} {\bibinfo {author} {\bibfnamefont {A.~W.}\ \bibnamefont
  {McConnachie}},\ }\bibfield  {title} {\bibinfo {title} {The observed
  properties of dwarf galaxies in and around the local group},\ }\href
  {https://doi.org/10.1088/0004-6256/144/1/4} {\bibfield  {journal} {\bibinfo
  {journal} {Astron. J.}\ }\textbf {\bibinfo {volume} {144}},\ \bibinfo {pages}
  {4} (\bibinfo {year} {2012})}\BibitemShut {NoStop}%
\bibitem [{\citenamefont {Chiti}\ \emph {et~al.}(2022)\citenamefont {Chiti},
  \citenamefont {Simon}, \citenamefont {Frebel}, \citenamefont {Pace},
  \citenamefont {Ji},\ and\ \citenamefont {Li}}]{Chiti_2022a}%
  \BibitemOpen
  \bibfield  {author} {\bibinfo {author} {\bibfnamefont {A.}~\bibnamefont
  {Chiti}}, \bibinfo {author} {\bibfnamefont {J.~D.}\ \bibnamefont {Simon}},
  \bibinfo {author} {\bibfnamefont {A.}~\bibnamefont {Frebel}}, \bibinfo
  {author} {\bibfnamefont {A.~B.}\ \bibnamefont {Pace}}, \bibinfo {author}
  {\bibfnamefont {A.~P.}\ \bibnamefont {Ji}},\ and\ \bibinfo {author}
  {\bibfnamefont {T.~S.}\ \bibnamefont {Li}},\ }\bibfield  {title} {\bibinfo
  {title} {{Magellan/IMACS} spectroscopy of {Grus} i: A low metallicity
  ultra-faint dwarf galaxy$^*$},\ }\href
  {https://doi.org/10.3847/1538-4357/ac96ed} {\bibfield  {journal} {\bibinfo
  {journal} {Astrophys. J.}\ }\textbf {\bibinfo {volume} {939}},\ \bibinfo
  {pages} {41} (\bibinfo {year} {2022})}\BibitemShut {NoStop}%
\bibitem [{\citenamefont {Chiti}\ \emph {et~al.}(2023)\citenamefont {Chiti},
  \citenamefont {Frebel}, \citenamefont {Ji}, \citenamefont {Mardini},
  \citenamefont {Ou}, \citenamefont {Simon}, \citenamefont {Jerjen},
  \citenamefont {Kim},\ and\ \citenamefont {Norris}}]{Chiti_2023}%
  \BibitemOpen
  \bibfield  {author} {\bibinfo {author} {\bibfnamefont {A.}~\bibnamefont
  {Chiti}}, \bibinfo {author} {\bibfnamefont {A.}~\bibnamefont {Frebel}},
  \bibinfo {author} {\bibfnamefont {A.~P.}\ \bibnamefont {Ji}}, \bibinfo
  {author} {\bibfnamefont {M.~K.}\ \bibnamefont {Mardini}}, \bibinfo {author}
  {\bibfnamefont {X.}~\bibnamefont {Ou}}, \bibinfo {author} {\bibfnamefont
  {J.~D.}\ \bibnamefont {Simon}}, \bibinfo {author} {\bibfnamefont
  {H.}~\bibnamefont {Jerjen}}, \bibinfo {author} {\bibfnamefont
  {D.}~\bibnamefont {Kim}},\ and\ \bibinfo {author} {\bibfnamefont {J.~E.}\
  \bibnamefont {Norris}},\ }\bibfield  {title} {\bibinfo {title} {Detailed
  chemical abundances of stars in the outskirts of the {Tucana} {II} ultrafaint
  dwarf galaxy*},\ }\href {https://doi.org/10.3847/1538-3881/aca416} {\bibfield
   {journal} {\bibinfo  {journal} {Astron. J.}\ }\textbf {\bibinfo {volume}
  {165}},\ \bibinfo {pages} {55} (\bibinfo {year} {2023})}\BibitemShut
  {NoStop}%
\bibitem [{\citenamefont {Drlica-Wagner}\ \emph {et~al.}(2015)\citenamefont
  {Drlica-Wagner} \emph {et~al.}}]{Drlica-Wagner_2015}%
  \BibitemOpen
  \bibfield  {author} {\bibinfo {author} {\bibfnamefont {A.}~\bibnamefont
  {Drlica-Wagner}} \emph {et~al.} (\bibinfo {collaboration} {The DES
  Collaboration}),\ }\bibfield  {title} {\bibinfo {title} {Eight ultra-faint
  galaxy candidates discovered in year two of the dark energy survey},\ }\href
  {https://doi.org/10.1088/0004-637X/813/2/109} {\bibfield  {journal} {\bibinfo
   {journal} {Astrophys. J.}\ }\textbf {\bibinfo {volume} {813}},\ \bibinfo
  {pages} {109} (\bibinfo {year} {2015})}\BibitemShut {NoStop}%
\bibitem [{\citenamefont {Simon}\ \emph
  {et~al.}(2020{\natexlab{b}})\citenamefont {Simon} \emph
  {et~al.}}]{Simon_2020}%
  \BibitemOpen
  \bibfield  {author} {\bibinfo {author} {\bibfnamefont {J.~D.}\ \bibnamefont
  {Simon}} \emph {et~al.} (\bibinfo {collaboration} {The DES Collaboration}),\
  }\bibfield  {title} {\bibinfo {title} {Birds of a feather? {Magellan/IMACS}
  spectroscopy of the ultra-faint satellites {Grus II}, {Tucana IV}, and
  {Tucana V}$^*$},\ }\href {https://doi.org/10.3847/1538-4357/ab7ccb}
  {\bibfield  {journal} {\bibinfo  {journal} {Astrophys. J.}\ }\textbf
  {\bibinfo {volume} {892}},\ \bibinfo {pages} {137} (\bibinfo {year}
  {2020}{\natexlab{b}})}\BibitemShut {NoStop}%
\bibitem [{\citenamefont {Martin}\ \emph {et~al.}(2007)\citenamefont {Martin},
  \citenamefont {Ibata}, \citenamefont {Chapman}, \citenamefont {Irwin},\ and\
  \citenamefont {Lewis}}]{Martin_2007}%
  \BibitemOpen
  \bibfield  {author} {\bibinfo {author} {\bibfnamefont {N.~F.}\ \bibnamefont
  {Martin}}, \bibinfo {author} {\bibfnamefont {R.~A.}\ \bibnamefont {Ibata}},
  \bibinfo {author} {\bibfnamefont {S.~C.}\ \bibnamefont {Chapman}}, \bibinfo
  {author} {\bibfnamefont {M.}~\bibnamefont {Irwin}},\ and\ \bibinfo {author}
  {\bibfnamefont {G.~F.}\ \bibnamefont {Lewis}},\ }\bibfield  {title} {\bibinfo
  {title} {{A Keck/DEIMOS spectroscopic survey of faint Galactic satellites:
  searching for the least massive dwarf galaxies$^*$}},\ }\href
  {https://doi.org/10.1111/j.1365-2966.2007.12055.x} {\bibfield  {journal}
  {\bibinfo  {journal} {Mon. Not. R. Astron. Soc.}\ }\textbf {\bibinfo {volume}
  {380}},\ \bibinfo {pages} {281} (\bibinfo {year} {2007})}\BibitemShut
  {NoStop}%
\bibitem [{\citenamefont {Walker}\ \emph
  {et~al.}(2008{\natexlab{b}})\citenamefont {Walker}, \citenamefont {Mateo},\
  and\ \citenamefont {Olszewski}}]{Walker_2008}%
  \BibitemOpen
  \bibfield  {author} {\bibinfo {author} {\bibfnamefont {M.~G.}\ \bibnamefont
  {Walker}}, \bibinfo {author} {\bibfnamefont {M.}~\bibnamefont {Mateo}},\ and\
  \bibinfo {author} {\bibfnamefont {E.~W.}\ \bibnamefont {Olszewski}},\
  }\bibfield  {title} {\bibinfo {title} {Systemic proper motions of {Milky}
  {Way} satellites from stellar redshifts: The {Carina}, {Fornax}, {Sculpto}r,
  and {Sextans} dwarf spheroidals$^*$},\ }\href
  {https://doi.org/10.1086/595586} {\bibfield  {journal} {\bibinfo  {journal}
  {Astrophys. J.}\ }\textbf {\bibinfo {volume} {688}},\ \bibinfo {pages} {L75}
  (\bibinfo {year} {2008}{\natexlab{b}})}\BibitemShut {NoStop}%
\bibitem [{\citenamefont {Walker}\ \emph
  {et~al.}(2009{\natexlab{c}})\citenamefont {Walker}, \citenamefont {Mateo},\
  and\ \citenamefont {Olszewski}}]{Walker_2009b}%
  \BibitemOpen
  \bibfield  {author} {\bibinfo {author} {\bibfnamefont {M.~G.}\ \bibnamefont
  {Walker}}, \bibinfo {author} {\bibfnamefont {M.}~\bibnamefont {Mateo}},\ and\
  \bibinfo {author} {\bibfnamefont {E.~W.}\ \bibnamefont {Olszewski}},\
  }\bibfield  {title} {\bibinfo {title} {Stellar velocities in the {C}arina,
  {F}ornax, {S}culptor, and {S}extans {dSph} galaxies: Data from the
  {M}agellan/{MMFS} survey$^*$},\ }\href
  {https://doi.org/10.1088/0004-6256/137/2/3100} {\bibfield  {journal}
  {\bibinfo  {journal} {Astron. J.}\ }\textbf {\bibinfo {volume} {137}},\
  \bibinfo {pages} {3100} (\bibinfo {year} {2009}{\natexlab{c}})}\BibitemShut
  {NoStop}%
\bibitem [{\citenamefont {Wilkinson}\ \emph {et~al.}(2004)\citenamefont
  {Wilkinson}, \citenamefont {Kleyna}, \citenamefont {Evans}, \citenamefont
  {Gilmore}, \citenamefont {Irwin},\ and\ \citenamefont
  {Grebel}}]{Wilkinson_2004}%
  \BibitemOpen
  \bibfield  {author} {\bibinfo {author} {\bibfnamefont {M.~I.}\ \bibnamefont
  {Wilkinson}}, \bibinfo {author} {\bibfnamefont {J.~T.}\ \bibnamefont
  {Kleyna}}, \bibinfo {author} {\bibfnamefont {N.~W.}\ \bibnamefont {Evans}},
  \bibinfo {author} {\bibfnamefont {G.~F.}\ \bibnamefont {Gilmore}}, \bibinfo
  {author} {\bibfnamefont {M.~J.}\ \bibnamefont {Irwin}},\ and\ \bibinfo
  {author} {\bibfnamefont {E.~K.}\ \bibnamefont {Grebel}},\ }\bibfield  {title}
  {\bibinfo {title} {Kinematically cold populations at large radii in the
  {Draco} and {Ursa} {Minor} dwarf spheroidal galaxies},\ }\href
  {https://doi.org/10.1086/423619} {\bibfield  {journal} {\bibinfo  {journal}
  {Astrophys. J.}\ }\textbf {\bibinfo {volume} {611}},\ \bibinfo {pages} {L21}
  (\bibinfo {year} {2004})}\BibitemShut {NoStop}%
\bibitem [{\citenamefont {Walker}\ \emph {et~al.}(2007)\citenamefont {Walker},
  \citenamefont {Mateo}, \citenamefont {Olszewski}, \citenamefont {Gnedin},
  \citenamefont {Wang}, \citenamefont {Sen},\ and\ \citenamefont
  {Woodroofe}}]{Walker_2007}%
  \BibitemOpen
  \bibfield  {author} {\bibinfo {author} {\bibfnamefont {M.~G.}\ \bibnamefont
  {Walker}}, \bibinfo {author} {\bibfnamefont {M.}~\bibnamefont {Mateo}},
  \bibinfo {author} {\bibfnamefont {E.~W.}\ \bibnamefont {Olszewski}}, \bibinfo
  {author} {\bibfnamefont {O.~Y.}\ \bibnamefont {Gnedin}}, \bibinfo {author}
  {\bibfnamefont {X.}~\bibnamefont {Wang}}, \bibinfo {author} {\bibfnamefont
  {B.}~\bibnamefont {Sen}},\ and\ \bibinfo {author} {\bibfnamefont
  {M.}~\bibnamefont {Woodroofe}},\ }\bibfield  {title} {\bibinfo {title}
  {Velocity dispersion profiles of seven dwarf spheroidal galaxies$^*$},\
  }\href {https://doi.org/10.1086/521998} {\bibfield  {journal} {\bibinfo
  {journal} {Astrophys. J.}\ }\textbf {\bibinfo {volume} {667}},\ \bibinfo
  {pages} {L53} (\bibinfo {year} {2007})}\BibitemShut {NoStop}%
\bibitem [{\citenamefont {Bouchard}\ \emph {et~al.}(2006)\citenamefont
  {Bouchard}, \citenamefont {Carignan},\ and\ \citenamefont
  {Staveley-Smith}}]{Bouchard_2006}%
  \BibitemOpen
  \bibfield  {author} {\bibinfo {author} {\bibfnamefont {A.}~\bibnamefont
  {Bouchard}}, \bibinfo {author} {\bibfnamefont {C.}~\bibnamefont {Carignan}},\
  and\ \bibinfo {author} {\bibfnamefont {L.}~\bibnamefont {Staveley-Smith}},\
  }\bibfield  {title} {\bibinfo {title} {Neutral hydrogen clouds near
  early-type dwarf galaxies of the local group},\ }\href
  {https://doi.org/10.1086/503629} {\bibfield  {journal} {\bibinfo  {journal}
  {Astron. J.}\ }\textbf {\bibinfo {volume} {131}},\ \bibinfo {pages} {2913}
  (\bibinfo {year} {2006})}\BibitemShut {NoStop}%
\bibitem [{\citenamefont {Mateo}\ \emph
  {et~al.}(2008{\natexlab{b}})\citenamefont {Mateo}, \citenamefont
  {Olszewski},\ and\ \citenamefont {Walker}}]{Mateo_2008}%
  \BibitemOpen
  \bibfield  {author} {\bibinfo {author} {\bibfnamefont {M.}~\bibnamefont
  {Mateo}}, \bibinfo {author} {\bibfnamefont {E.~W.}\ \bibnamefont
  {Olszewski}},\ and\ \bibinfo {author} {\bibfnamefont {M.~G.}\ \bibnamefont
  {Walker}},\ }\bibfield  {title} {\bibinfo {title} {The velocity dispersion
  profile of the remote dwarf spheroidal galaxy {Leo} i: A tidal hit and
  run?},\ }\href {https://doi.org/10.1086/522326} {\bibfield  {journal}
  {\bibinfo  {journal} {Astrophys. J.}\ }\textbf {\bibinfo {volume} {675}},\
  \bibinfo {pages} {201} (\bibinfo {year} {2008}{\natexlab{b}})}\BibitemShut
  {NoStop}%
\bibitem [{\citenamefont {Carignan}\ \emph {et~al.}(1998)\citenamefont
  {Carignan}, \citenamefont {Beaulieu}, \citenamefont {Côté}, \citenamefont
  {Demers},\ and\ \citenamefont {Mateo}}]{Carignan_1998}%
  \BibitemOpen
  \bibfield  {author} {\bibinfo {author} {\bibfnamefont {C.}~\bibnamefont
  {Carignan}}, \bibinfo {author} {\bibfnamefont {S.}~\bibnamefont {Beaulieu}},
  \bibinfo {author} {\bibfnamefont {S.}~\bibnamefont {Côté}}, \bibinfo
  {author} {\bibfnamefont {S.}~\bibnamefont {Demers}},\ and\ \bibinfo {author}
  {\bibfnamefont {M.}~\bibnamefont {Mateo}},\ }\bibfield  {title} {\bibinfo
  {title} {Detection of h i associated with the {Sculptor} dwarf spheroidal
  galaxy},\ }\href {https://doi.org/10.1086/300540} {\bibfield  {journal}
  {\bibinfo  {journal} {Astron. J.}\ }\textbf {\bibinfo {volume} {116}},\
  \bibinfo {pages} {1690} (\bibinfo {year} {1998})}\BibitemShut {NoStop}%
\bibitem [{\citenamefont {Walker}\ \emph
  {et~al.}(2009{\natexlab{d}})\citenamefont {Walker}, \citenamefont {Mateo},
  \citenamefont {Olszewski}, \citenamefont {Peñarrubia}, \citenamefont
  {Evans},\ and\ \citenamefont {Gilmore}}]{Walker_2009c}%
  \BibitemOpen
  \bibfield  {author} {\bibinfo {author} {\bibfnamefont {M.~G.}\ \bibnamefont
  {Walker}}, \bibinfo {author} {\bibfnamefont {M.}~\bibnamefont {Mateo}},
  \bibinfo {author} {\bibfnamefont {E.~W.}\ \bibnamefont {Olszewski}}, \bibinfo
  {author} {\bibfnamefont {J.}~\bibnamefont {Peñarrubia}}, \bibinfo {author}
  {\bibfnamefont {N.~W.}\ \bibnamefont {Evans}},\ and\ \bibinfo {author}
  {\bibfnamefont {G.}~\bibnamefont {Gilmore}},\ }\bibfield  {title} {\bibinfo
  {title} {A universal mass profile for dwarf spheroidal galaxies?$^*$},\
  }\href {https://doi.org/10.1088/0004-637X/704/2/1274} {\bibfield  {journal}
  {\bibinfo  {journal} {Astrophys. J.}\ }\textbf {\bibinfo {volume} {704}},\
  \bibinfo {pages} {1274} (\bibinfo {year} {2009}{\natexlab{d}})}\BibitemShut
  {NoStop}%
\bibitem [{\citenamefont {El-Zant}\ \emph {et~al.}(2001)\citenamefont
  {El-Zant}, \citenamefont {Shlosman},\ and\ \citenamefont
  {Hoffman}}]{El-Zant_2001}%
  \BibitemOpen
  \bibfield  {author} {\bibinfo {author} {\bibfnamefont {A.}~\bibnamefont
  {El-Zant}}, \bibinfo {author} {\bibfnamefont {I.}~\bibnamefont {Shlosman}},\
  and\ \bibinfo {author} {\bibfnamefont {Y.}~\bibnamefont {Hoffman}},\
  }\bibfield  {title} {\bibinfo {title} {Dark halos: The flattening of the
  density cusp by dynamical friction},\ }\href {https://doi.org/10.1086/322516}
  {\bibfield  {journal} {\bibinfo  {journal} {Astrophys. J.}\ }\textbf
  {\bibinfo {volume} {560}},\ \bibinfo {pages} {636} (\bibinfo {year}
  {2001})}\BibitemShut {NoStop}%
\bibitem [{\citenamefont {Mayer}\ \emph {et~al.}(2001)\citenamefont {Mayer},
  \citenamefont {Governato}, \citenamefont {Colpi}, \citenamefont {Moore},
  \citenamefont {Quinn}, \citenamefont {Wadsley}, \citenamefont {Stadel},\ and\
  \citenamefont {Lake}}]{Mayer_2001}%
  \BibitemOpen
  \bibfield  {author} {\bibinfo {author} {\bibfnamefont {L.}~\bibnamefont
  {Mayer}}, \bibinfo {author} {\bibfnamefont {F.}~\bibnamefont {Governato}},
  \bibinfo {author} {\bibfnamefont {M.}~\bibnamefont {Colpi}}, \bibinfo
  {author} {\bibfnamefont {B.}~\bibnamefont {Moore}}, \bibinfo {author}
  {\bibfnamefont {T.}~\bibnamefont {Quinn}}, \bibinfo {author} {\bibfnamefont
  {J.}~\bibnamefont {Wadsley}}, \bibinfo {author} {\bibfnamefont
  {J.}~\bibnamefont {Stadel}},\ and\ \bibinfo {author} {\bibfnamefont
  {G.}~\bibnamefont {Lake}},\ }\bibfield  {title} {\bibinfo {title} {Tidal
  stirring and the origin of dwarf spheroidals in the local group},\ }\href
  {https://doi.org/10.1086/318898} {\bibfield  {journal} {\bibinfo  {journal}
  {Astrophys. J.}\ }\textbf {\bibinfo {volume} {547}},\ \bibinfo {pages} {L123}
  (\bibinfo {year} {2001})}\BibitemShut {NoStop}%
\bibitem [{\citenamefont {Read}\ and\ \citenamefont
  {Gilmore}(2005)}]{Read&Gilmore_2005}%
  \BibitemOpen
  \bibfield  {author} {\bibinfo {author} {\bibfnamefont {J.~I.}\ \bibnamefont
  {Read}}\ and\ \bibinfo {author} {\bibfnamefont {G.}~\bibnamefont {Gilmore}},\
  }\bibfield  {title} {\bibinfo {title} {{Mass loss from dwarf spheroidal
  galaxies: the origins of shallow dark matter cores and exponential surface
  brightness profiles}},\ }\href
  {https://doi.org/10.1111/j.1365-2966.2004.08424.x} {\bibfield  {journal}
  {\bibinfo  {journal} {Mon. Not. R. Astron. Soc.}\ }\textbf {\bibinfo {volume}
  {356}},\ \bibinfo {pages} {107} (\bibinfo {year} {2005})}\BibitemShut
  {NoStop}%
\bibitem [{\citenamefont {Mashchenko}\ \emph {et~al.}(2008)\citenamefont
  {Mashchenko}, \citenamefont {Wadsley},\ and\ \citenamefont
  {Couchman}}]{Mashchenko_2008}%
  \BibitemOpen
  \bibfield  {author} {\bibinfo {author} {\bibfnamefont {S.}~\bibnamefont
  {Mashchenko}}, \bibinfo {author} {\bibfnamefont {J.}~\bibnamefont
  {Wadsley}},\ and\ \bibinfo {author} {\bibfnamefont {H.~M.~P.}\ \bibnamefont
  {Couchman}},\ }\bibfield  {title} {\bibinfo {title} {Stellar feedback in
  dwarf galaxy formation},\ }\href {https://doi.org/10.1126/science.1148666}
  {\bibfield  {journal} {\bibinfo  {journal} {Science}\ }\textbf {\bibinfo
  {volume} {319}},\ \bibinfo {pages} {174} (\bibinfo {year}
  {2008})}\BibitemShut {NoStop}%
\bibitem [{\citenamefont {Goerdt}\ \emph {et~al.}(2010)\citenamefont {Goerdt},
  \citenamefont {Moore}, \citenamefont {Read},\ and\ \citenamefont
  {Stadel}}]{Goerdt_2010}%
  \BibitemOpen
  \bibfield  {author} {\bibinfo {author} {\bibfnamefont {T.}~\bibnamefont
  {Goerdt}}, \bibinfo {author} {\bibfnamefont {B.}~\bibnamefont {Moore}},
  \bibinfo {author} {\bibfnamefont {J.~I.}\ \bibnamefont {Read}},\ and\
  \bibinfo {author} {\bibfnamefont {J.}~\bibnamefont {Stadel}},\ }\bibfield
  {title} {\bibinfo {title} {Core creation in galaxies and halos via sinking
  massive objects},\ }\href {https://doi.org/10.1088/0004-637X/725/2/1707}
  {\bibfield  {journal} {\bibinfo  {journal} {Astrophys. J.}\ }\textbf
  {\bibinfo {volume} {725}},\ \bibinfo {pages} {1707} (\bibinfo {year}
  {2010})}\BibitemShut {NoStop}%
\bibitem [{\citenamefont {Governato}\ \emph {et~al.}(2010)\citenamefont
  {Governato} \emph {et~al.}}]{Governato_2010}%
  \BibitemOpen
  \bibfield  {author} {\bibinfo {author} {\bibfnamefont {F.}~\bibnamefont
  {Governato}} \emph {et~al.},\ }\bibfield  {title} {\bibinfo {title}
  {Bulgeless dwarf galaxies and dark matter cores from supernova-driven
  outflows},\ }\href {https://doi.org/10.1038/nature08640} {\bibfield
  {journal} {\bibinfo  {journal} {Nature}\ }\textbf {\bibinfo {volume} {463}},\
  \bibinfo {pages} {203} (\bibinfo {year} {2010})}\BibitemShut {NoStop}%
\bibitem [{\citenamefont {Kazantzidis}\ \emph {et~al.}(2011)\citenamefont
  {Kazantzidis}, \citenamefont {Lokas}, \citenamefont {Callegari},
  \citenamefont {Mayer},\ and\ \citenamefont {Moustakas}}]{Kazantzidis:2010cw}%
  \BibitemOpen
  \bibfield  {author} {\bibinfo {author} {\bibfnamefont {S.}~\bibnamefont
  {Kazantzidis}}, \bibinfo {author} {\bibfnamefont {E.~L.}\ \bibnamefont
  {Lokas}}, \bibinfo {author} {\bibfnamefont {S.}~\bibnamefont {Callegari}},
  \bibinfo {author} {\bibfnamefont {L.}~\bibnamefont {Mayer}},\ and\ \bibinfo
  {author} {\bibfnamefont {L.~A.}\ \bibnamefont {Moustakas}},\ }\bibfield
  {title} {\bibinfo {title} {{On the Efficiency of the Tidal Stirring Mechanism
  for the Origin of Dwarf Spheroidals: Dependence on the Orbital and Structural
  Parameters of the Progenitor Disky Dwarfs}},\ }\href
  {https://doi.org/10.1088/0004-637X/726/2/98} {\bibfield  {journal} {\bibinfo
  {journal} {Astrophys. J.}\ }\textbf {\bibinfo {volume} {726}},\ \bibinfo
  {pages} {98} (\bibinfo {year} {2011})}\BibitemShut {NoStop}%
\bibitem [{\citenamefont {Cole}\ \emph {et~al.}(2011)\citenamefont {Cole},
  \citenamefont {Dehnen},\ and\ \citenamefont {Wilkinson}}]{Cole_2011}%
  \BibitemOpen
  \bibfield  {author} {\bibinfo {author} {\bibfnamefont {D.~R.}\ \bibnamefont
  {Cole}}, \bibinfo {author} {\bibfnamefont {W.}~\bibnamefont {Dehnen}},\ and\
  \bibinfo {author} {\bibfnamefont {M.~I.}\ \bibnamefont {Wilkinson}},\
  }\bibfield  {title} {\bibinfo {title} {{Weakening dark matter cusps by clumpy
  baryonic infall}},\ }\href {https://doi.org/10.1111/j.1365-2966.2011.19110.x}
  {\bibfield  {journal} {\bibinfo  {journal} {Mon. Not. R. Astron. Soc.}\
  }\textbf {\bibinfo {volume} {416}},\ \bibinfo {pages} {1118} (\bibinfo {year}
  {2011})}\BibitemShut {NoStop}%
\bibitem [{\citenamefont {Brooks}\ and\ \citenamefont
  {Zolotov}(2014)}]{Brooks:2012vi}%
  \BibitemOpen
  \bibfield  {author} {\bibinfo {author} {\bibfnamefont {A.~M.}\ \bibnamefont
  {Brooks}}\ and\ \bibinfo {author} {\bibfnamefont {A.}~\bibnamefont
  {Zolotov}},\ }\bibfield  {title} {\bibinfo {title} {{Why Baryons Matter: The
  Kinematics of Dwarf Spheroidal Satellites}},\ }\href
  {https://doi.org/10.1088/0004-637X/786/2/87} {\bibfield  {journal} {\bibinfo
  {journal} {Astrophys. J.}\ }\textbf {\bibinfo {volume} {786}},\ \bibinfo
  {pages} {87} (\bibinfo {year} {2014})}\BibitemShut {NoStop}%
\bibitem [{\citenamefont {Chang}\ \emph {et~al.}(2013)\citenamefont {Chang},
  \citenamefont {Maccio'},\ and\ \citenamefont {Kang}}]{Chang:2012rx}%
  \BibitemOpen
  \bibfield  {author} {\bibinfo {author} {\bibfnamefont {J.}~\bibnamefont
  {Chang}}, \bibinfo {author} {\bibfnamefont {A.~V.}\ \bibnamefont {Maccio'}},\
  and\ \bibinfo {author} {\bibfnamefont {X.}~\bibnamefont {Kang}},\ }\bibfield
  {title} {\bibinfo {title} {{The dependence of tidal stripping efficiency on
  the satellite and host galaxy morphology}},\ }\href
  {https://doi.org/10.1093/mnras/stt434} {\bibfield  {journal} {\bibinfo
  {journal} {Mon. Not. R. Astron. Soc.}\ }\textbf {\bibinfo {volume} {431}},\
  \bibinfo {pages} {3533} (\bibinfo {year} {2013})}\BibitemShut {NoStop}%
\bibitem [{\citenamefont {Di~Cintio}\ \emph {et~al.}(2013)\citenamefont
  {Di~Cintio}, \citenamefont {Brook}, \citenamefont {Macciò}, \citenamefont
  {Stinson}, \citenamefont {Knebe}, \citenamefont {Dutton},\ and\ \citenamefont
  {Wadsley}}]{DiCintio_2013}%
  \BibitemOpen
  \bibfield  {author} {\bibinfo {author} {\bibfnamefont {A.}~\bibnamefont
  {Di~Cintio}}, \bibinfo {author} {\bibfnamefont {C.~B.}\ \bibnamefont
  {Brook}}, \bibinfo {author} {\bibfnamefont {A.~V.}\ \bibnamefont {Macciò}},
  \bibinfo {author} {\bibfnamefont {G.~S.}\ \bibnamefont {Stinson}}, \bibinfo
  {author} {\bibfnamefont {A.}~\bibnamefont {Knebe}}, \bibinfo {author}
  {\bibfnamefont {A.~A.}\ \bibnamefont {Dutton}},\ and\ \bibinfo {author}
  {\bibfnamefont {J.}~\bibnamefont {Wadsley}},\ }\bibfield  {title} {\bibinfo
  {title} {{The dependence of dark matter profiles on the stellar-to-halo mass
  ratio: a prediction for cusps versus cores}},\ }\href
  {https://doi.org/10.1093/mnras/stt1891} {\bibfield  {journal} {\bibinfo
  {journal} {Mon. Not. R. Astron. Soc.}\ }\textbf {\bibinfo {volume} {437}},\
  \bibinfo {pages} {415} (\bibinfo {year} {2013})}\BibitemShut {NoStop}%
\bibitem [{\citenamefont {Garrison-Kimmel}\ \emph {et~al.}(2013)\citenamefont
  {Garrison-Kimmel}, \citenamefont {Rocha}, \citenamefont {Boylan-Kolchin},
  \citenamefont {Bullock},\ and\ \citenamefont
  {Lally}}]{Garrison-Kimmel:2013yys}%
  \BibitemOpen
  \bibfield  {author} {\bibinfo {author} {\bibfnamefont {S.}~\bibnamefont
  {Garrison-Kimmel}}, \bibinfo {author} {\bibfnamefont {M.}~\bibnamefont
  {Rocha}}, \bibinfo {author} {\bibfnamefont {M.}~\bibnamefont
  {Boylan-Kolchin}}, \bibinfo {author} {\bibfnamefont {J.}~\bibnamefont
  {Bullock}},\ and\ \bibinfo {author} {\bibfnamefont {J.}~\bibnamefont
  {Lally}},\ }\bibfield  {title} {\bibinfo {title} {{Can Feedback Solve the Too
  Big to Fail Problem?}},\ }\href {https://doi.org/10.1093/mnras/stt984}
  {\bibfield  {journal} {\bibinfo  {journal} {Mon. Not. R. Astron. Soc.}\
  }\textbf {\bibinfo {volume} {433}},\ \bibinfo {pages} {3539} (\bibinfo {year}
  {2013})}\BibitemShut {NoStop}%
\bibitem [{\citenamefont {Kazantzidis}\ \emph {et~al.}(2013)\citenamefont
  {Kazantzidis}, \citenamefont {Lokas},\ and\ \citenamefont
  {Mayer}}]{Kazantzidis:2013wi}%
  \BibitemOpen
  \bibfield  {author} {\bibinfo {author} {\bibfnamefont {S.}~\bibnamefont
  {Kazantzidis}}, \bibinfo {author} {\bibfnamefont {E.~L.}\ \bibnamefont
  {Lokas}},\ and\ \bibinfo {author} {\bibfnamefont {L.}~\bibnamefont {Mayer}},\
  }\bibfield  {title} {\bibinfo {title} {{Tidal Stirring of Disky Dwarfs with
  Shallow Dark Matter Density Profiles: Enhanced Transformation into Dwarf
  Spheroidals}},\ }\href {https://doi.org/10.1088/2041-8205/764/2/L29}
  {\bibfield  {journal} {\bibinfo  {journal} {Astrophys. J. Lett.}\ }\textbf
  {\bibinfo {volume} {764}},\ \bibinfo {pages} {L29} (\bibinfo {year}
  {2013})}\BibitemShut {NoStop}%
\bibitem [{\citenamefont {Teyssier}\ \emph {et~al.}(2013)\citenamefont
  {Teyssier}, \citenamefont {Pontzen}, \citenamefont {Dubois},\ and\
  \citenamefont {Read}}]{Teyssier_2013}%
  \BibitemOpen
  \bibfield  {author} {\bibinfo {author} {\bibfnamefont {R.}~\bibnamefont
  {Teyssier}}, \bibinfo {author} {\bibfnamefont {A.}~\bibnamefont {Pontzen}},
  \bibinfo {author} {\bibfnamefont {Y.}~\bibnamefont {Dubois}},\ and\ \bibinfo
  {author} {\bibfnamefont {J.~I.}\ \bibnamefont {Read}},\ }\bibfield  {title}
  {\bibinfo {title} {{Cusp-core transformations in dwarf galaxies:
  observational predictions}},\ }\href {https://doi.org/10.1093/mnras/sts563}
  {\bibfield  {journal} {\bibinfo  {journal} {Mon. Not. R. Astron. Soc.}\
  }\textbf {\bibinfo {volume} {429}},\ \bibinfo {pages} {3068} (\bibinfo {year}
  {2013})}\BibitemShut {NoStop}%
\bibitem [{\citenamefont {Nipoti}\ and\ \citenamefont
  {Binney}(2015)}]{Nipoti:2014xha}%
  \BibitemOpen
  \bibfield  {author} {\bibinfo {author} {\bibfnamefont {C.}~\bibnamefont
  {Nipoti}}\ and\ \bibinfo {author} {\bibfnamefont {J.}~\bibnamefont
  {Binney}},\ }\bibfield  {title} {\bibinfo {title} {{Early flattening of dark
  matter cusps in dwarf spheroidal galaxies}},\ }\href
  {https://doi.org/10.1093/mnras/stu2217} {\bibfield  {journal} {\bibinfo
  {journal} {Mon. Not. R. Astron. Soc.}\ }\textbf {\bibinfo {volume} {446}},\
  \bibinfo {pages} {1820} (\bibinfo {year} {2015})}\BibitemShut {NoStop}%
\bibitem [{\citenamefont {Chan}\ \emph {et~al.}(2015)\citenamefont {Chan},
  \citenamefont {Kereš}, \citenamefont {Oñorbe}, \citenamefont {Hopkins},
  \citenamefont {Muratov}, \citenamefont {Faucher-Giguère},\ and\
  \citenamefont {Quataert}}]{Chan_2015}%
  \BibitemOpen
  \bibfield  {author} {\bibinfo {author} {\bibfnamefont {T.~K.}\ \bibnamefont
  {Chan}}, \bibinfo {author} {\bibfnamefont {D.}~\bibnamefont {Kereš}},
  \bibinfo {author} {\bibfnamefont {J.}~\bibnamefont {Oñorbe}}, \bibinfo
  {author} {\bibfnamefont {P.~F.}\ \bibnamefont {Hopkins}}, \bibinfo {author}
  {\bibfnamefont {A.~L.}\ \bibnamefont {Muratov}}, \bibinfo {author}
  {\bibfnamefont {C.-A.}\ \bibnamefont {Faucher-Giguère}},\ and\ \bibinfo
  {author} {\bibfnamefont {E.}~\bibnamefont {Quataert}},\ }\bibfield  {title}
  {\bibinfo {title} {{The impact of baryonic physics on the structure of dark
  matter haloes: the view from the FIRE cosmological simulations}},\ }\href
  {https://doi.org/10.1093/mnras/stv2165} {\bibfield  {journal} {\bibinfo
  {journal} {Mon. Not. R. Astron. Soc.}\ }\textbf {\bibinfo {volume} {454}},\
  \bibinfo {pages} {2981} (\bibinfo {year} {2015})}\BibitemShut {NoStop}%
\bibitem [{\citenamefont {Read}\ \emph {et~al.}(2016)\citenamefont {Read},
  \citenamefont {Agertz},\ and\ \citenamefont {Collins}}]{Read:2015sta}%
  \BibitemOpen
  \bibfield  {author} {\bibinfo {author} {\bibfnamefont {J.~I.}\ \bibnamefont
  {Read}}, \bibinfo {author} {\bibfnamefont {O.}~\bibnamefont {Agertz}},\ and\
  \bibinfo {author} {\bibfnamefont {M.~L.~M.}\ \bibnamefont {Collins}},\
  }\bibfield  {title} {\bibinfo {title} {{Dark matter cores all the way
  down}},\ }\href {https://doi.org/10.1093/mnras/stw713} {\bibfield  {journal}
  {\bibinfo  {journal} {Mon. Not. R. Astron. Soc.}\ }\textbf {\bibinfo {volume}
  {459}},\ \bibinfo {pages} {2573} (\bibinfo {year} {2016})}\BibitemShut
  {NoStop}%
\bibitem [{\citenamefont {Dutton}\ \emph {et~al.}(2016)\citenamefont {Dutton},
  \citenamefont {Macciò}, \citenamefont {Dekel}, \citenamefont {Wang},
  \citenamefont {Stinson}, \citenamefont {Obreja}, \citenamefont {Di~Cintio},
  \citenamefont {Brook}, \citenamefont {Buck},\ and\ \citenamefont
  {Kang}}]{Dutton_2016}%
  \BibitemOpen
  \bibfield  {author} {\bibinfo {author} {\bibfnamefont {A.~A.}\ \bibnamefont
  {Dutton}}, \bibinfo {author} {\bibfnamefont {A.~V.}\ \bibnamefont {Macciò}},
  \bibinfo {author} {\bibfnamefont {A.}~\bibnamefont {Dekel}}, \bibinfo
  {author} {\bibfnamefont {L.}~\bibnamefont {Wang}}, \bibinfo {author}
  {\bibfnamefont {G.}~\bibnamefont {Stinson}}, \bibinfo {author} {\bibfnamefont
  {A.}~\bibnamefont {Obreja}}, \bibinfo {author} {\bibfnamefont
  {A.}~\bibnamefont {Di~Cintio}}, \bibinfo {author} {\bibfnamefont
  {C.}~\bibnamefont {Brook}}, \bibinfo {author} {\bibfnamefont
  {T.}~\bibnamefont {Buck}},\ and\ \bibinfo {author} {\bibfnamefont
  {X.}~\bibnamefont {Kang}},\ }\bibfield  {title} {\bibinfo {title} {{NIHAO IX:
  the role of gas inflows and outflows in driving the contraction and expansion
  of cold dark matter haloes}},\ }\href {https://doi.org/10.1093/mnras/stw1537}
  {\bibfield  {journal} {\bibinfo  {journal} {Mon. Not. R. Astron. Soc.}\
  }\textbf {\bibinfo {volume} {461}},\ \bibinfo {pages} {2658} (\bibinfo {year}
  {2016})}\BibitemShut {NoStop}%
\bibitem [{\citenamefont {Tollet}\ \emph {et~al.}(2016)\citenamefont {Tollet}
  \emph {et~al.}}]{Tollet_2016}%
  \BibitemOpen
  \bibfield  {author} {\bibinfo {author} {\bibfnamefont {E.}~\bibnamefont
  {Tollet}} \emph {et~al.},\ }\bibfield  {title} {\bibinfo {title} {{NIHAO –
  IV: core creation and destruction in dark matter density profiles across
  cosmic time}},\ }\href {https://doi.org/10.1093/mnras/stv2856} {\bibfield
  {journal} {\bibinfo  {journal} {Mon. Not. R. Astron. Soc.}\ }\textbf
  {\bibinfo {volume} {456}},\ \bibinfo {pages} {3542} (\bibinfo {year}
  {2016})}\BibitemShut {NoStop}%
\bibitem [{\citenamefont {Wang}\ \emph {et~al.}(2017)\citenamefont {Wang},
  \citenamefont {Fattahi}, \citenamefont {Cooper}, \citenamefont {Sawala},
  \citenamefont {Strigari}, \citenamefont {Frenk}, \citenamefont {Navarro},
  \citenamefont {Oman},\ and\ \citenamefont {Schaller}}]{Wang:2016qol}%
  \BibitemOpen
  \bibfield  {author} {\bibinfo {author} {\bibfnamefont {M.~Y.}\ \bibnamefont
  {Wang}}, \bibinfo {author} {\bibfnamefont {A.}~\bibnamefont {Fattahi}},
  \bibinfo {author} {\bibfnamefont {A.~P.}\ \bibnamefont {Cooper}}, \bibinfo
  {author} {\bibfnamefont {T.}~\bibnamefont {Sawala}}, \bibinfo {author}
  {\bibfnamefont {L.~E.}\ \bibnamefont {Strigari}}, \bibinfo {author}
  {\bibfnamefont {C.~S.}\ \bibnamefont {Frenk}}, \bibinfo {author}
  {\bibfnamefont {J.~F.}\ \bibnamefont {Navarro}}, \bibinfo {author}
  {\bibfnamefont {K.}~\bibnamefont {Oman}},\ and\ \bibinfo {author}
  {\bibfnamefont {M.}~\bibnamefont {Schaller}},\ }\bibfield  {title} {\bibinfo
  {title} {{Tidal features of classical Milky Way satellites in a $\Lambda$
  cold dark matter universe}},\ }\href {https://doi.org/10.1093/mnras/stx742}
  {\bibfield  {journal} {\bibinfo  {journal} {Mon. Not. R. Astron. Soc.}\
  }\textbf {\bibinfo {volume} {468}},\ \bibinfo {pages} {4887} (\bibinfo {year}
  {2017})}\BibitemShut {NoStop}%
\bibitem [{\citenamefont {Fitts}\ \emph {et~al.}(2017)\citenamefont {Fitts}
  \emph {et~al.}}]{Fitts_2017}%
  \BibitemOpen
  \bibfield  {author} {\bibinfo {author} {\bibfnamefont {A.}~\bibnamefont
  {Fitts}} \emph {et~al.},\ }\bibfield  {title} {\bibinfo {title} {{fire in the
  field: simulating the threshold of galaxy formation}},\ }\href
  {https://doi.org/10.1093/mnras/stx1757} {\bibfield  {journal} {\bibinfo
  {journal} {Mon. Not. R. Astron. Soc.}\ }\textbf {\bibinfo {volume} {471}},\
  \bibinfo {pages} {3547} (\bibinfo {year} {2017})}\BibitemShut {NoStop}%
\bibitem [{\citenamefont {Hiroshima}\ \emph {et~al.}(2018)\citenamefont
  {Hiroshima}, \citenamefont {Ando},\ and\ \citenamefont
  {Ishiyama}}]{Hiroshima:2018kfv}%
  \BibitemOpen
  \bibfield  {author} {\bibinfo {author} {\bibfnamefont {N.}~\bibnamefont
  {Hiroshima}}, \bibinfo {author} {\bibfnamefont {S.}~\bibnamefont {Ando}},\
  and\ \bibinfo {author} {\bibfnamefont {T.}~\bibnamefont {Ishiyama}},\
  }\bibfield  {title} {\bibinfo {title} {{Modeling evolution of dark matter
  substructure and annihilation boost}},\ }\href
  {https://doi.org/10.1103/PhysRevD.97.123002} {\bibfield  {journal} {\bibinfo
  {journal} {Phys. Rev. D}\ }\textbf {\bibinfo {volume} {97}},\ \bibinfo
  {pages} {123002} (\bibinfo {year} {2018})}\BibitemShut {NoStop}%
\bibitem [{\citenamefont {Freundlich}\ \emph {et~al.}(2019)\citenamefont
  {Freundlich}, \citenamefont {Dekel}, \citenamefont {Jiang}, \citenamefont
  {Ishai}, \citenamefont {Cornuault}, \citenamefont {Lapiner}, \citenamefont
  {Dutton},\ and\ \citenamefont {Macciò}}]{Freundlich_2019}%
  \BibitemOpen
  \bibfield  {author} {\bibinfo {author} {\bibfnamefont {J.}~\bibnamefont
  {Freundlich}}, \bibinfo {author} {\bibfnamefont {A.}~\bibnamefont {Dekel}},
  \bibinfo {author} {\bibfnamefont {F.}~\bibnamefont {Jiang}}, \bibinfo
  {author} {\bibfnamefont {G.}~\bibnamefont {Ishai}}, \bibinfo {author}
  {\bibfnamefont {N.}~\bibnamefont {Cornuault}}, \bibinfo {author}
  {\bibfnamefont {S.}~\bibnamefont {Lapiner}}, \bibinfo {author} {\bibfnamefont
  {A.~A.}\ \bibnamefont {Dutton}},\ and\ \bibinfo {author} {\bibfnamefont
  {A.~V.}\ \bibnamefont {Macciò}},\ }\bibfield  {title} {\bibinfo {title} {{A
  model for core formation in dark matter haloes and ultra-diffuse galaxies by
  outflow episodes}},\ }\href {https://doi.org/10.1093/mnras/stz3306}
  {\bibfield  {journal} {\bibinfo  {journal} {Mon. Not. R. Astron. Soc.}\
  }\textbf {\bibinfo {volume} {491}},\ \bibinfo {pages} {4523} (\bibinfo {year}
  {2019})}\BibitemShut {NoStop}%
\bibitem [{\citenamefont {Lazar}\ \emph {et~al.}(2020)\citenamefont {Lazar}
  \emph {et~al.}}]{Lazar_2020}%
  \BibitemOpen
  \bibfield  {author} {\bibinfo {author} {\bibfnamefont {A.}~\bibnamefont
  {Lazar}} \emph {et~al.},\ }\bibfield  {title} {\bibinfo {title} {{A dark
  matter profile to model diverse feedback-induced core sizes of
  $\mathrm{\Lambda}$CDM haloes}},\ }\href
  {https://doi.org/10.1093/mnras/staa2101} {\bibfield  {journal} {\bibinfo
  {journal} {Mon. Not. R. Astron. Soc.}\ }\textbf {\bibinfo {volume} {497}},\
  \bibinfo {pages} {2393} (\bibinfo {year} {2020})}\BibitemShut {NoStop}%
\bibitem [{\citenamefont {Burger}\ and\ \citenamefont
  {Zavala}(2021)}]{Burger_2021}%
  \BibitemOpen
  \bibfield  {author} {\bibinfo {author} {\bibfnamefont {J.~D.}\ \bibnamefont
  {Burger}}\ and\ \bibinfo {author} {\bibfnamefont {J.}~\bibnamefont
  {Zavala}},\ }\bibfield  {title} {\bibinfo {title} {Supernova-driven mechanism
  of cusp-core transformation: an appraisal},\ }\href
  {https://doi.org/10.3847/1538-4357/ac1a0f} {\bibfield  {journal} {\bibinfo
  {journal} {Astrophys. J.}\ }\textbf {\bibinfo {volume} {921}},\ \bibinfo
  {pages} {126} (\bibinfo {year} {2021})}\BibitemShut {NoStop}%
\bibitem [{\citenamefont {Keil}\ \emph {et~al.}(2003)\citenamefont {Keil},
  \citenamefont {Raffelt},\ and\ \citenamefont {Janka}}]{PinchedSpectrum}%
  \BibitemOpen
  \bibfield  {author} {\bibinfo {author} {\bibfnamefont {M.~T.}\ \bibnamefont
  {Keil}}, \bibinfo {author} {\bibfnamefont {G.~G.}\ \bibnamefont {Raffelt}},\
  and\ \bibinfo {author} {\bibfnamefont {H.-T.}\ \bibnamefont {Janka}},\
  }\bibfield  {title} {\bibinfo {title} {Monte carlo study of supernova
  neutrino spectra formation},\ }\href {https://doi.org/10.1086/375130}
  {\bibfield  {journal} {\bibinfo  {journal} {Astrophys. J.}\ }\textbf
  {\bibinfo {volume} {590}},\ \bibinfo {pages} {971} (\bibinfo {year}
  {2003})}\BibitemShut {NoStop}%
\bibitem [{\citenamefont {Totani}\ \emph {et~al.}(1998)\citenamefont {Totani},
  \citenamefont {Sato}, \citenamefont {Dalhed},\ and\ \citenamefont
  {Wilson}}]{Totani_1998}%
  \BibitemOpen
  \bibfield  {author} {\bibinfo {author} {\bibfnamefont {T.}~\bibnamefont
  {Totani}}, \bibinfo {author} {\bibfnamefont {K.}~\bibnamefont {Sato}},
  \bibinfo {author} {\bibfnamefont {H.~E.}\ \bibnamefont {Dalhed}},\ and\
  \bibinfo {author} {\bibfnamefont {J.~R.}\ \bibnamefont {Wilson}},\ }\bibfield
   {title} {\bibinfo {title} {Future detection of supernova neutrino burst and
  explosion mechanism},\ }\href {https://doi.org/10.1086/305364} {\bibfield
  {journal} {\bibinfo  {journal} {Astrophys. J.}\ }\textbf {\bibinfo {volume}
  {496}},\ \bibinfo {pages} {216} (\bibinfo {year} {1998})}\BibitemShut
  {NoStop}%
\bibitem [{\citenamefont {Thompson}\ \emph {et~al.}(2003)\citenamefont
  {Thompson}, \citenamefont {Burrows},\ and\ \citenamefont
  {Pinto}}]{Thompson_2003}%
  \BibitemOpen
  \bibfield  {author} {\bibinfo {author} {\bibfnamefont {T.~A.}\ \bibnamefont
  {Thompson}}, \bibinfo {author} {\bibfnamefont {A.}~\bibnamefont {Burrows}},\
  and\ \bibinfo {author} {\bibfnamefont {P.~A.}\ \bibnamefont {Pinto}},\
  }\bibfield  {title} {\bibinfo {title} {Shock breakout in core-collapse
  supernovae and its neutrino signature},\ }\href
  {https://doi.org/10.1086/375701} {\bibfield  {journal} {\bibinfo  {journal}
  {Astrophys. J.}\ }\textbf {\bibinfo {volume} {592}},\ \bibinfo {pages} {434}
  (\bibinfo {year} {2003})}\BibitemShut {NoStop}%
\bibitem [{\citenamefont {Sumiyoshi}\ \emph {et~al.}(2005)\citenamefont
  {Sumiyoshi}, \citenamefont {Yamada}, \citenamefont {Suzuki}, \citenamefont
  {Shen}, \citenamefont {Chiba},\ and\ \citenamefont {Toki}}]{Sumiyoshi_2005}%
  \BibitemOpen
  \bibfield  {author} {\bibinfo {author} {\bibfnamefont {K.}~\bibnamefont
  {Sumiyoshi}}, \bibinfo {author} {\bibfnamefont {S.}~\bibnamefont {Yamada}},
  \bibinfo {author} {\bibfnamefont {H.}~\bibnamefont {Suzuki}}, \bibinfo
  {author} {\bibfnamefont {H.}~\bibnamefont {Shen}}, \bibinfo {author}
  {\bibfnamefont {S.}~\bibnamefont {Chiba}},\ and\ \bibinfo {author}
  {\bibfnamefont {H.}~\bibnamefont {Toki}},\ }\bibfield  {title} {\bibinfo
  {title} {Postbounce evolution of core-collapse supernovae: Long-term effects
  of the equation of state},\ }\href {https://doi.org/10.1086/431788}
  {\bibfield  {journal} {\bibinfo  {journal} {Astrophys. J.}\ }\textbf
  {\bibinfo {volume} {629}},\ \bibinfo {pages} {922} (\bibinfo {year}
  {2005})}\BibitemShut {NoStop}%
\bibitem [{\citenamefont {Nakazato}\ \emph {et~al.}(2013)\citenamefont
  {Nakazato}, \citenamefont {Sumiyoshi}, \citenamefont {Suzuki}, \citenamefont
  {Totani}, \citenamefont {Umeda},\ and\ \citenamefont
  {Yamada}}]{Nakazato_2013}%
  \BibitemOpen
  \bibfield  {author} {\bibinfo {author} {\bibfnamefont {K.}~\bibnamefont
  {Nakazato}}, \bibinfo {author} {\bibfnamefont {K.}~\bibnamefont {Sumiyoshi}},
  \bibinfo {author} {\bibfnamefont {H.}~\bibnamefont {Suzuki}}, \bibinfo
  {author} {\bibfnamefont {T.}~\bibnamefont {Totani}}, \bibinfo {author}
  {\bibfnamefont {H.}~\bibnamefont {Umeda}},\ and\ \bibinfo {author}
  {\bibfnamefont {S.}~\bibnamefont {Yamada}},\ }\bibfield  {title} {\bibinfo
  {title} {Supernova neutrino light curves and spectra for various progenitor
  stars: from core collapse to proto-neutrino star cooling},\ }\href
  {https://doi.org/10.1088/0067-0049/205/1/2} {\bibfield  {journal} {\bibinfo
  {journal} {Astrophys. J. Suppl. Ser.}\ }\textbf {\bibinfo {volume} {205}},\
  \bibinfo {pages} {2} (\bibinfo {year} {2013})}\BibitemShut {NoStop}%
\bibitem [{\citenamefont {Suwa}\ \emph {et~al.}(2019)\citenamefont {Suwa},
  \citenamefont {Sumiyoshi}, \citenamefont {Nakazato}, \citenamefont
  {Takahira}, \citenamefont {Koshio}, \citenamefont {Mori},\ and\ \citenamefont
  {Wendell}}]{Suwa_2019}%
  \BibitemOpen
  \bibfield  {author} {\bibinfo {author} {\bibfnamefont {Y.}~\bibnamefont
  {Suwa}}, \bibinfo {author} {\bibfnamefont {K.}~\bibnamefont {Sumiyoshi}},
  \bibinfo {author} {\bibfnamefont {K.}~\bibnamefont {Nakazato}}, \bibinfo
  {author} {\bibfnamefont {Y.}~\bibnamefont {Takahira}}, \bibinfo {author}
  {\bibfnamefont {Y.}~\bibnamefont {Koshio}}, \bibinfo {author} {\bibfnamefont
  {M.}~\bibnamefont {Mori}},\ and\ \bibinfo {author} {\bibfnamefont {R.~A.}\
  \bibnamefont {Wendell}},\ }\bibfield  {title} {\bibinfo {title} {Observing
  supernova neutrino light curves with super-kamiokande: Expected event number
  over 10 s},\ }\href {https://doi.org/10.3847/1538-4357/ab2e05} {\bibfield
  {journal} {\bibinfo  {journal} {Astrophys. J.}\ }\textbf {\bibinfo {volume}
  {881}},\ \bibinfo {pages} {139} (\bibinfo {year} {2019})}\BibitemShut
  {NoStop}%
\bibitem [{\citenamefont {Bollig}\ \emph {et~al.}(2021)\citenamefont {Bollig},
  \citenamefont {Yadav}, \citenamefont {Kresse}, \citenamefont {Janka},
  \citenamefont {Müller},\ and\ \citenamefont {Heger}}]{Bollig_2021}%
  \BibitemOpen
  \bibfield  {author} {\bibinfo {author} {\bibfnamefont {R.}~\bibnamefont
  {Bollig}}, \bibinfo {author} {\bibfnamefont {N.}~\bibnamefont {Yadav}},
  \bibinfo {author} {\bibfnamefont {D.}~\bibnamefont {Kresse}}, \bibinfo
  {author} {\bibfnamefont {H.-T.}\ \bibnamefont {Janka}}, \bibinfo {author}
  {\bibfnamefont {B.}~\bibnamefont {Müller}},\ and\ \bibinfo {author}
  {\bibfnamefont {A.}~\bibnamefont {Heger}},\ }\bibfield  {title} {\bibinfo
  {title} {Self-consistent 3d supernova models from -7 minutes to $+$7 s: A
  1-bethe explosion of a $\sim$19 $\mathrm{M}_\odot$ progenitor},\ }\href
  {https://doi.org/10.3847/1538-4357/abf82e} {\bibfield  {journal} {\bibinfo
  {journal} {Astrophys. J.}\ }\textbf {\bibinfo {volume} {915}},\ \bibinfo
  {pages} {28} (\bibinfo {year} {2021})}\BibitemShut {NoStop}%
\bibitem [{\citenamefont {Kroupa}(2002)}]{Kroupa_2002}%
  \BibitemOpen
  \bibfield  {author} {\bibinfo {author} {\bibfnamefont {P.}~\bibnamefont
  {Kroupa}},\ }\bibfield  {title} {\bibinfo {title} {The initial mass function
  of stars: Evidence for uniformity in variable systems},\ }\href
  {https://doi.org/10.1126/science.1067524} {\bibfield  {journal} {\bibinfo
  {journal} {Science}\ }\textbf {\bibinfo {volume} {295}},\ \bibinfo {pages}
  {82} (\bibinfo {year} {2002})}\BibitemShut {NoStop}%
\bibitem [{\citenamefont {Smartt}(2009)}]{Smartt_2009}%
  \BibitemOpen
  \bibfield  {author} {\bibinfo {author} {\bibfnamefont {S.~J.}\ \bibnamefont
  {Smartt}},\ }\bibfield  {title} {\bibinfo {title} {Progenitors of
  core-collapse supernovae},\ }\href
  {https://doi.org/10.1146/annurev-astro-082708-101737} {\bibfield  {journal}
  {\bibinfo  {journal} {Annu. Rev. Astron. Astrophys.}\ }\textbf {\bibinfo
  {volume} {47}},\ \bibinfo {pages} {63} (\bibinfo {year} {2009})}\BibitemShut
  {NoStop}%
\bibitem [{\citenamefont {Smartt}\ \emph {et~al.}(2009)\citenamefont {Smartt},
  \citenamefont {Eldridge}, \citenamefont {Crockett},\ and\ \citenamefont
  {Maund}}]{Smartt_2015}%
  \BibitemOpen
  \bibfield  {author} {\bibinfo {author} {\bibfnamefont {S.~J.}\ \bibnamefont
  {Smartt}}, \bibinfo {author} {\bibfnamefont {J.~J.}\ \bibnamefont
  {Eldridge}}, \bibinfo {author} {\bibfnamefont {R.~M.}\ \bibnamefont
  {Crockett}},\ and\ \bibinfo {author} {\bibfnamefont {J.~R.}\ \bibnamefont
  {Maund}},\ }\bibfield  {title} {\bibinfo {title} {{The death of massive stars
  – I. Observational constraints on the progenitors of Type II-P
  supernovae}},\ }\href {https://doi.org/10.1111/j.1365-2966.2009.14506.x}
  {\bibfield  {journal} {\bibinfo  {journal} {Mon. Not. R. Astron. Soc.}\
  }\textbf {\bibinfo {volume} {395}},\ \bibinfo {pages} {1409} (\bibinfo {year}
  {2009})}\BibitemShut {NoStop}%
\bibitem [{\citenamefont {Díaz-Rodríguez}\ \emph {et~al.}(2018)\citenamefont
  {Díaz-Rodríguez}, \citenamefont {Murphy}, \citenamefont {Rubin},
  \citenamefont {Dolphin}, \citenamefont {Williams},\ and\ \citenamefont
  {Dalcanton}}]{Diaz-Rodríguez_2018}%
  \BibitemOpen
  \bibfield  {author} {\bibinfo {author} {\bibfnamefont {M.}~\bibnamefont
  {Díaz-Rodríguez}}, \bibinfo {author} {\bibfnamefont {J.~W.}\ \bibnamefont
  {Murphy}}, \bibinfo {author} {\bibfnamefont {D.~A.}\ \bibnamefont {Rubin}},
  \bibinfo {author} {\bibfnamefont {A.~E.}\ \bibnamefont {Dolphin}}, \bibinfo
  {author} {\bibfnamefont {B.~F.}\ \bibnamefont {Williams}},\ and\ \bibinfo
  {author} {\bibfnamefont {J.~J.}\ \bibnamefont {Dalcanton}},\ }\bibfield
  {title} {\bibinfo {title} {Progenitor mass distribution for core-collapse
  supernova remnants in {M31} and {M33}},\ }\href
  {https://doi.org/10.3847/1538-4357/aac6e1} {\bibfield  {journal} {\bibinfo
  {journal} {Astrophys. J.}\ }\textbf {\bibinfo {volume} {861}},\ \bibinfo
  {pages} {92} (\bibinfo {year} {2018})}\BibitemShut {NoStop}%
\bibitem [{\citenamefont {Díaz-Rodríguez}\ \emph {et~al.}(2021)\citenamefont
  {Díaz-Rodríguez}, \citenamefont {Murphy}, \citenamefont {Williams},
  \citenamefont {Dalcanton},\ and\ \citenamefont
  {Dolphin}}]{Diaz-Rodriguez_2021}%
  \BibitemOpen
  \bibfield  {author} {\bibinfo {author} {\bibfnamefont {M.}~\bibnamefont
  {Díaz-Rodríguez}}, \bibinfo {author} {\bibfnamefont {J.~W.}\ \bibnamefont
  {Murphy}}, \bibinfo {author} {\bibfnamefont {B.~F.}\ \bibnamefont
  {Williams}}, \bibinfo {author} {\bibfnamefont {J.~J.}\ \bibnamefont
  {Dalcanton}},\ and\ \bibinfo {author} {\bibfnamefont {A.~E.}\ \bibnamefont
  {Dolphin}},\ }\bibfield  {title} {\bibinfo {title} {{Progenitor mass
  distribution for 22 historic core-collapse supernovae}},\ }\href
  {https://doi.org/10.1093/mnras/stab1800} {\bibfield  {journal} {\bibinfo
  {journal} {Mon. Not. R. Astron. Soc.}\ }\textbf {\bibinfo {volume} {506}},\
  \bibinfo {pages} {781} (\bibinfo {year} {2021})}\BibitemShut {NoStop}%
\bibitem [{\citenamefont {Hayashi}\ \emph {et~al.}(2023)\citenamefont
  {Hayashi}, \citenamefont {Hirai}, \citenamefont {Chiba},\ and\ \citenamefont
  {Ishiyama}}]{Hayashi_2023}%
  \BibitemOpen
  \bibfield  {author} {\bibinfo {author} {\bibfnamefont {K.}~\bibnamefont
  {Hayashi}}, \bibinfo {author} {\bibfnamefont {Y.}~\bibnamefont {Hirai}},
  \bibinfo {author} {\bibfnamefont {M.}~\bibnamefont {Chiba}},\ and\ \bibinfo
  {author} {\bibfnamefont {T.}~\bibnamefont {Ishiyama}},\ }\bibfield  {title}
  {\bibinfo {title} {Dark matter halo properties of the galactic dwarf
  satellites: Implication for chemo-dynamical evolution of the satellites and a
  challenge to lambda cold dark matter},\ }\href
  {https://doi.org/10.3847/1538-4357/ace33e} {\bibfield  {journal} {\bibinfo
  {journal} {Astrophys. J.}\ }\textbf {\bibinfo {volume} {953}},\ \bibinfo
  {pages} {185} (\bibinfo {year} {2023})}\BibitemShut {NoStop}%
\bibitem [{\citenamefont {Hirata}\ \emph {et~al.}(1987)\citenamefont {Hirata},
  \citenamefont {Kajita}, \citenamefont {Koshiba}, \citenamefont {Nakahata},
  \citenamefont {Oyama}, \citenamefont {Sato} \emph {et~al.}}]{1987A1}%
  \BibitemOpen
  \bibfield  {author} {\bibinfo {author} {\bibfnamefont {K.}~\bibnamefont
  {Hirata}}, \bibinfo {author} {\bibfnamefont {T.}~\bibnamefont {Kajita}},
  \bibinfo {author} {\bibfnamefont {M.}~\bibnamefont {Koshiba}}, \bibinfo
  {author} {\bibfnamefont {M.}~\bibnamefont {Nakahata}}, \bibinfo {author}
  {\bibfnamefont {Y.}~\bibnamefont {Oyama}}, \bibinfo {author} {\bibfnamefont
  {N.}~\bibnamefont {Sato}}, \emph {et~al.},\ }\bibfield  {title} {\bibinfo
  {title} {Observation of a neutrino burst from the supernova {SN1987A}},\
  }\href {https://doi.org/10.1103/PhysRevLett.58.1490} {\bibfield  {journal}
  {\bibinfo  {journal} {Phys. Rev. Lett.}\ }\textbf {\bibinfo {volume} {58}},\
  \bibinfo {pages} {1490} (\bibinfo {year} {1987})}\BibitemShut {NoStop}%
\bibitem [{\citenamefont {Bionta}\ \emph {et~al.}(1987)\citenamefont {Bionta},
  \citenamefont {Blewitt}, \citenamefont {Bratton}, \citenamefont {Casper},
  \citenamefont {Ciocio}, \citenamefont {Claus} \emph {et~al.}}]{1987A2}%
  \BibitemOpen
  \bibfield  {author} {\bibinfo {author} {\bibfnamefont {R.~M.}\ \bibnamefont
  {Bionta}}, \bibinfo {author} {\bibfnamefont {G.}~\bibnamefont {Blewitt}},
  \bibinfo {author} {\bibfnamefont {C.~B.}\ \bibnamefont {Bratton}}, \bibinfo
  {author} {\bibfnamefont {D.}~\bibnamefont {Casper}}, \bibinfo {author}
  {\bibfnamefont {A.}~\bibnamefont {Ciocio}}, \bibinfo {author} {\bibfnamefont
  {R.}~\bibnamefont {Claus}}, \emph {et~al.},\ }\bibfield  {title} {\bibinfo
  {title} {Observation of a neutrino burst in coincidence with supernova 1987a
  in the large magellanic cloud},\ }\href
  {https://doi.org/10.1103/PhysRevLett.58.1494} {\bibfield  {journal} {\bibinfo
   {journal} {Phys. Rev. Lett.}\ }\textbf {\bibinfo {volume} {58}},\ \bibinfo
  {pages} {1494} (\bibinfo {year} {1987})}\BibitemShut {NoStop}%
\bibitem [{\citenamefont {Alexeyev}\ \emph {et~al.}(1988)\citenamefont
  {Alexeyev}, \citenamefont {Alexeyeva}, \citenamefont {Krivosheina},\ and\
  \citenamefont {Volchenko}}]{1987A3}%
  \BibitemOpen
  \bibfield  {author} {\bibinfo {author} {\bibfnamefont {E.}~\bibnamefont
  {Alexeyev}}, \bibinfo {author} {\bibfnamefont {L.}~\bibnamefont {Alexeyeva}},
  \bibinfo {author} {\bibfnamefont {I.}~\bibnamefont {Krivosheina}},\ and\
  \bibinfo {author} {\bibfnamefont {V.}~\bibnamefont {Volchenko}},\ }\bibfield
  {title} {\bibinfo {title} {Detection of the neutrino signal from {SN} 1987a
  in the lmc using the inr baksan underground scintillation telescope},\
  }\bibfield  {journal} {\bibinfo  {journal} {Phys. Lett. B}\ }\textbf
  {\bibinfo {volume} {205}},\ \href
  {https://doi.org/https://doi.org/10.1016/0370-2693(88)91651-6}
  {https://doi.org/10.1016/0370-2693(88)91651-6} (\bibinfo {year}
  {1988})\BibitemShut {NoStop}%
\bibitem [{\citenamefont {Aartsen}\ \emph {et~al.}(2018)\citenamefont {Aartsen}
  \emph {et~al.}}]{IceCube2018}%
  \BibitemOpen
  \bibfield  {author} {\bibinfo {author} {\bibfnamefont {M.}~\bibnamefont
  {Aartsen}} \emph {et~al.} (\bibinfo {collaboration} {The IceCube
  Collaboration and Fermi-LAT and MAGIC and AGILE and ASAS-SN and HAWC and
  H.E.S.S. and INTEGRAL and Kanata and Kiso and Kapteyn and Liverpool Telescope
  and Subaru and Swift/NuSTAR and VERITAS and VLA/17B-403 Teams}),\ }\bibfield
  {title} {\bibinfo {title} {Multimessenger observations of a flaring blazar
  coincident with high-energy neutrino {IceCube}-{170922A}},\ }\href
  {https://doi.org/10.1126/science.aat1378} {\bibfield  {journal} {\bibinfo
  {journal} {Science}\ }\textbf {\bibinfo {volume} {361}},\ \bibinfo {pages}
  {eaat1378} (\bibinfo {year} {2018})}\BibitemShut {NoStop}%
\bibitem [{\citenamefont {Stein}\ \emph {et~al.}(2021)\citenamefont {Stein}
  \emph {et~al.}}]{Stein:2020xhk}%
  \BibitemOpen
  \bibfield  {author} {\bibinfo {author} {\bibfnamefont {R.}~\bibnamefont
  {Stein}} \emph {et~al.},\ }\bibfield  {title} {\bibinfo {title} {{A tidal
  disruption event coincident with a high-energy neutrino}},\ }\href
  {https://doi.org/10.1038/s41550-020-01295-8} {\bibfield  {journal} {\bibinfo
  {journal} {Nature Astron.}\ }\textbf {\bibinfo {volume} {5}},\ \bibinfo
  {pages} {510} (\bibinfo {year} {2021})}\BibitemShut {NoStop}%
\bibitem [{\citenamefont {Olivares-Del~Campo}\ \emph
  {et~al.}(2018)\citenamefont {Olivares-Del~Campo}, \citenamefont {B\oe{}hm},
  \citenamefont {Palomares-Ruiz},\ and\ \citenamefont
  {Pascoli}}]{Olivares-Del-Campo_2018}%
  \BibitemOpen
  \bibfield  {author} {\bibinfo {author} {\bibfnamefont {A.}~\bibnamefont
  {Olivares-Del~Campo}}, \bibinfo {author} {\bibfnamefont {C.}~\bibnamefont
  {B\oe{}hm}}, \bibinfo {author} {\bibfnamefont {S.}~\bibnamefont
  {Palomares-Ruiz}},\ and\ \bibinfo {author} {\bibfnamefont {S.}~\bibnamefont
  {Pascoli}},\ }\bibfield  {title} {\bibinfo {title} {Dark matter-neutrino
  interactions through the lens of their cosmological implications},\ }\href
  {https://doi.org/10.1103/PhysRevD.97.075039} {\bibfield  {journal} {\bibinfo
  {journal} {Phys. Rev. D}\ }\textbf {\bibinfo {volume} {97}},\ \bibinfo
  {pages} {075039} (\bibinfo {year} {2018})}\BibitemShut {NoStop}%
\bibitem [{\citenamefont {Palanque-Delabrouille}\ \emph
  {et~al.}(2020)\citenamefont {Palanque-Delabrouille}, \citenamefont {Yèche},
  \citenamefont {Schöneberg}, \citenamefont {Lesgourgues}, \citenamefont
  {Walther}, \citenamefont {Chabanier},\ and\ \citenamefont
  {Armengaud}}]{Palanque-Delabrouille_2020}%
  \BibitemOpen
  \bibfield  {author} {\bibinfo {author} {\bibfnamefont {N.}~\bibnamefont
  {Palanque-Delabrouille}}, \bibinfo {author} {\bibfnamefont {C.}~\bibnamefont
  {Yèche}}, \bibinfo {author} {\bibfnamefont {N.}~\bibnamefont {Schöneberg}},
  \bibinfo {author} {\bibfnamefont {J.}~\bibnamefont {Lesgourgues}}, \bibinfo
  {author} {\bibfnamefont {M.}~\bibnamefont {Walther}}, \bibinfo {author}
  {\bibfnamefont {S.}~\bibnamefont {Chabanier}},\ and\ \bibinfo {author}
  {\bibfnamefont {E.}~\bibnamefont {Armengaud}},\ }\bibfield  {title} {\bibinfo
  {title} {Hints, neutrino bounds, and wdm constraints from {SDSS DR14
  Lyman}-$\alpha$ and {Planck} full-survey data},\ }\href
  {https://doi.org/10.1088/1475-7516/2020/04/038} {\bibfield  {journal}
  {\bibinfo  {journal} {J. Cosmol. Astropart. Phys.}\ }\textbf {\bibinfo
  {volume} {2020}}\bibinfo  {number} { (04)},\ \bibinfo {pages}
  {038}}\BibitemShut {NoStop}%
\bibitem [{\citenamefont {Arg\"uelles}\ \emph {et~al.}(2017)\citenamefont
  {Arg\"uelles}, \citenamefont {Kheirandish},\ and\ \citenamefont
  {Vincent}}]{Arguelles_2017}%
  \BibitemOpen
\bibfield  {number} {  }\bibfield  {author} {\bibinfo {author} {\bibfnamefont
  {C.~A.}\ \bibnamefont {Arg\"uelles}}, \bibinfo {author} {\bibfnamefont
  {A.}~\bibnamefont {Kheirandish}},\ and\ \bibinfo {author} {\bibfnamefont
  {A.~C.}\ \bibnamefont {Vincent}},\ }\bibfield  {title} {\bibinfo {title}
  {Imaging galactic dark matter with high-energy cosmic neutrinos},\ }\href
  {https://doi.org/10.1103/PhysRevLett.119.201801} {\bibfield  {journal}
  {\bibinfo  {journal} {Phys. Rev. Lett.}\ }\textbf {\bibinfo {volume} {119}},\
  \bibinfo {pages} {201801} (\bibinfo {year} {2017})}\BibitemShut {NoStop}%
\bibitem [{\citenamefont {Weisz}\ \emph {et~al.}(2014)\citenamefont {Weisz},
  \citenamefont {Dolphin}, \citenamefont {Skillman}, \citenamefont {Holtzman},
  \citenamefont {Gilbert}, \citenamefont {Dalcanton},\ and\ \citenamefont
  {Williams}}]{Weisz:2014qra}%
  \BibitemOpen
  \bibfield  {author} {\bibinfo {author} {\bibfnamefont {D.~R.}\ \bibnamefont
  {Weisz}}, \bibinfo {author} {\bibfnamefont {A.~E.}\ \bibnamefont {Dolphin}},
  \bibinfo {author} {\bibfnamefont {E.~D.}\ \bibnamefont {Skillman}}, \bibinfo
  {author} {\bibfnamefont {J.}~\bibnamefont {Holtzman}}, \bibinfo {author}
  {\bibfnamefont {K.~M.}\ \bibnamefont {Gilbert}}, \bibinfo {author}
  {\bibfnamefont {J.~J.}\ \bibnamefont {Dalcanton}},\ and\ \bibinfo {author}
  {\bibfnamefont {B.~F.}\ \bibnamefont {Williams}},\ }\bibfield  {title}
  {\bibinfo {title} {{The Star Formation Histories of Local Group Dwarf
  Galaxies I. Hubble Space Telescope / Wide Field Planetary Camera 2
  Observations}},\ }\href {https://doi.org/10.1088/0004-637X/789/2/147}
  {\bibfield  {journal} {\bibinfo  {journal} {Astrophys. J.}\ }\textbf
  {\bibinfo {volume} {789}},\ \bibinfo {pages} {147} (\bibinfo {year}
  {2014})}\BibitemShut {NoStop}%
\bibitem [{\citenamefont {Li}\ \emph {et~al.}(2022)\citenamefont {Li},
  \citenamefont {Vagins},\ and\ \citenamefont {Wurm}}]{Li_2022}%
  \BibitemOpen
  \bibfield  {author} {\bibinfo {author} {\bibfnamefont {Y.-F.}\ \bibnamefont
  {Li}}, \bibinfo {author} {\bibfnamefont {M.}~\bibnamefont {Vagins}},\ and\
  \bibinfo {author} {\bibfnamefont {M.}~\bibnamefont {Wurm}},\ }\bibfield
  {title} {\bibinfo {title} {Prospects for the detection of the diffuse
  supernova neutrino background with the experiments {SK}-{Gd} and {JUNO}},\
  }\href {https://doi.org/10.3390/universe8030181} {\bibfield  {journal}
  {\bibinfo  {journal} {Universe}\ }\textbf {\bibinfo {volume} {8}},\ \bibinfo
  {pages} {181} (\bibinfo {year} {2022})}\BibitemShut {NoStop}%
\bibitem [{\citenamefont {Ekanger}\ \emph {et~al.}(2024)\citenamefont
  {Ekanger}, \citenamefont {Horiuchi}, \citenamefont {Nagakura},\ and\
  \citenamefont {Reitz}}]{Ekanger_2023:DSNB}%
  \BibitemOpen
  \bibfield  {author} {\bibinfo {author} {\bibfnamefont {N.}~\bibnamefont
  {Ekanger}}, \bibinfo {author} {\bibfnamefont {S.}~\bibnamefont {Horiuchi}},
  \bibinfo {author} {\bibfnamefont {H.}~\bibnamefont {Nagakura}},\ and\
  \bibinfo {author} {\bibfnamefont {S.}~\bibnamefont {Reitz}},\ }\bibfield
  {title} {\bibinfo {title} {{Diffuse supernova neutrino background with
  up-to-date star formation rate measurements and long-term multidimensional
  supernova simulations}},\ }\href
  {https://doi.org/10.1103/PhysRevD.109.023024} {\bibfield  {journal} {\bibinfo
   {journal} {Phys. Rev. D}\ }\textbf {\bibinfo {volume} {109}},\ \bibinfo
  {pages} {023024} (\bibinfo {year} {2024})}\BibitemShut {NoStop}%
\bibitem [{\citenamefont {Skidmore}\ \emph {et~al.}(2015)\citenamefont
  {Skidmore} \emph {et~al.}}]{TMT:2015pvw}%
  \BibitemOpen
  \bibfield  {author} {\bibinfo {author} {\bibfnamefont {W.}~\bibnamefont
  {Skidmore}} \emph {et~al.} (\bibinfo {collaboration} {TMT International
  Science Development Teams \& TMT Science Advisory Committee}),\ }\bibfield
  {title} {\bibinfo {title} {{Thirty Meter Telescope Detailed Science Case:
  2015}},\ }\href {https://doi.org/10.1088/1674-4527/15/12/001} {\bibfield
  {journal} {\bibinfo  {journal} {Res. Astron. Astrophys.}\ }\textbf {\bibinfo
  {volume} {15}},\ \bibinfo {pages} {1945} (\bibinfo {year}
  {2015})}\BibitemShut {NoStop}%
\bibitem [{\citenamefont {Željko Ivezić}\ \emph {et~al.}(2019)\citenamefont
  {Željko Ivezić} \emph {et~al.}}]{LSST}%
  \BibitemOpen
  \bibfield  {author} {\bibinfo {author} {\bibnamefont {Željko Ivezić}} \emph
  {et~al.},\ }\bibfield  {title} {\bibinfo {title} {Lsst: From science drivers
  to reference design and anticipated data products},\ }\href
  {https://doi.org/10.3847/1538-4357/ab042c} {\bibfield  {journal} {\bibinfo
  {journal} {Astrophys. J.}\ }\textbf {\bibinfo {volume} {873}},\ \bibinfo
  {pages} {111} (\bibinfo {year} {2019})}\BibitemShut {NoStop}%
\bibitem [{\citenamefont {{de Martino}}\ \emph {et~al.}(2022)\citenamefont {{de
  Martino}}, \citenamefont {Diaferio},\ and\ \citenamefont
  {Ostorero}}]{deMartino_2022}%
  \BibitemOpen
  \bibfield  {author} {\bibinfo {author} {\bibfnamefont {I.}~\bibnamefont {{de
  Martino}}}, \bibinfo {author} {\bibfnamefont {A.}~\bibnamefont {Diaferio}},\
  and\ \bibinfo {author} {\bibfnamefont {L.}~\bibnamefont {Ostorero}},\
  }\bibfield  {title} {\bibinfo {title} {{The proper motion of stars in dwarf
  galaxies: distinguishing central density cusps from cores}},\ }\href
  {https://doi.org/10.1093/mnras/stac2336} {\bibfield  {journal} {\bibinfo
  {journal} {Mon. Not. R. Astron. Soc.}\ }\textbf {\bibinfo {volume} {516}},\
  \bibinfo {pages} {3556} (\bibinfo {year} {2022})}\BibitemShut {NoStop}%
\bibitem [{\citenamefont {Guerra}\ \emph {et~al.}(2023)\citenamefont {Guerra},
  \citenamefont {Geha},\ and\ \citenamefont {Strigari}}]{Guerra_2023}%
  \BibitemOpen
  \bibfield  {author} {\bibinfo {author} {\bibfnamefont {J.}~\bibnamefont
  {Guerra}}, \bibinfo {author} {\bibfnamefont {M.}~\bibnamefont {Geha}},\ and\
  \bibinfo {author} {\bibfnamefont {L.~E.}\ \bibnamefont {Strigari}},\
  }\bibfield  {title} {\bibinfo {title} {Forecasts on the dark matter density
  profiles of dwarf spheroidal galaxies with current and future kinematic
  observations},\ }\href {https://doi.org/10.3847/1538-4357/aca8a5} {\bibfield
  {journal} {\bibinfo  {journal} {Astrophys. J.}\ }\textbf {\bibinfo {volume}
  {943}},\ \bibinfo {pages} {121} (\bibinfo {year} {2023})}\BibitemShut
  {NoStop}%
\end{thebibliography}%

\end{document}